\documentclass[preprint,12pt]{elsarticle}

\usepackage[T1]{fontenc}
\usepackage{xcolor}
\usepackage{etex}
\usepackage{url}
\usepackage{breakurl}
\usepackage{latexsym}
\usepackage{amsmath}
\usepackage{amssymb}
\usepackage{stmaryrd}
\usepackage{color}
\usepackage{graphicx}
\usepackage{multirow}
\usepackage{soul}
\usepackage{xspace}
\usepackage{mathtools}
\usepackage{proof}
\usepackage{accsupp} 
\usepackage{wrapfig}
\usepackage{tikz}
\usetikzlibrary{calc,arrows,shapes,chains,matrix,positioning,scopes,decorations.pathmorphing,shadows,fit,mindmap}
\definecolor{myDarkGray}{rgb}{.75,.75,.75}
\definecolor{light-gray}{gray}{0.80}
\definecolor{light-blue}{rgb}{0.94, 0.97, 1.0}

\DeclareMathAlphabet{\mathpzc}{OT1}{pzc}{m}{it}

\usepackage{todonotes}

\usepackage{listings}

\lstdefinestyle{mystyle}{
	flexiblecolumns=true,
	keepspaces=true,
	tabsize=4,
	showstringspaces=false,
	basewidth={0em,0em},,
	basicstyle=\footnotesize,
	commentstyle=\itshape,
	stringstyle=\ttfamily
}
\lstdefinestyle{mystyleSmall}{
	flexiblecolumns=true,
	keepspaces=true,
	tabsize=4,
	showstringspaces=false,
	basewidth={0em,0em},,
	basicstyle=\scriptsize,
	commentstyle=\itshape,
	stringstyle=\ttfamily
}

\newcommand{\locv}{ u }
\newcommand{\sito}{ l }
\newcommand{\llv}{ \ell }
\newcommand{\mmnil}{{\bf nil}}
\newcommand{\oap}[2]{\mmout( #2 ) @ #1}
\newcommand{\iap}[2]{\mmin( #2 ) @ #1}
\newcommand{\rap}[2]{\mmread( #2 ) @ #1}
\newcommand{\goap}[2]{\mmeval( #2 ) @ #1}
\newcommand{\nnew}[1]{\mmnew( #1 )}
\newcommand{\mmout}{{\bf out}}
\newcommand{\mmin}{{\bf in}}
\newcommand{\mmread}{{\bf read}}
\newcommand{\mmeval}{{\bf eval}}
\newcommand{\mmnew}{{\bf newloc}}
\newcommand{\fl}[1]{{\mathit{fv}}(#1)}    

\newcommand{\netconf}[2]{ #1 \vdash #2 }
\newcommand{\prlaw}[1]{(\mbox{{\sc{#1}}})}  
\newcommand{\redarrow}[1]{\succ\!\longrightarrow}
\newcommand{\musub}[3]{ #1 [\raisebox{.5ex}{\small$#2$}\! / \!\mbox{\small $#3$}]}
\newcommand{\outrule}{\prlaw{Out}}
\newcommand{\inrule}{\prlaw{In}}
\newcommand{\readrule}{\prlaw{Read}}
\newcommand{\evalrule}{\prlaw{Eval}}
\newcommand{\newrule}{\prlaw{New}}

\newcommand{\valt}[2]{\mbox{$[\![ \: #1 \: ]\!]_{#2}$}}
\newcommand{\emptytuple}{\mmnil}
\newcommand{\mumatch}{match}

\newcommand{\Klava}{\textsc{Klava}\xspace}
\newcommand{\XKlaim}{\textsc{X-Klaim}\xspace}
\newcommand{\xtext}{\textsc{Xtext}\xspace}

\newcommand{\MetaKlaim}{\textsc{MetaKlaim}\xspace}
\newcommand{\MetaML}{\textsc{MetaML}\xspace}

\newcommand{\F}{\textsc{F}}

\newcommand{\ie}{i.e.~}
\newcommand{\eg}{e.g.~}

\newcommand{\pic}{\ensuremath{\pi}-calculus} 
\newcommand{\metaklaim}{\textsc{Meta}\textsc{Klaim}\xspace}

\newcommand{\lwsbpel}{\textsf{B}$\mathit{lite}$\xspace}
\newcommand{\wsbpel}{\textsc{WS-BPEL}\xspace}
\newcommand{\socLogic}{\textsf{SocL}\xspace}
\newcommand{\cowsmc}{\textsf{CMC}\xspace}
\newcommand{\jresp}{\textsf{j}\textsc{RESP}\xspace}
\newcommand{\klaim}{\textsc{Klaim}\xspace}
\newcommand{\cows}{\textsc{Cows}\xspace}
\newcommand{\SCEL}{\textsc{Scel}\xspace}
\newcommand{\SCELlight}{\textsc{Scelight}\xspace}
\newcommand{\PSCEL}{\textsc{Pscel}\xspace}
\newcommand{\abc}{\textsc{A}\texttt{b}\textsc{C}\xspace}

\newcommand{\java}{Java}
\newcommand{\maudel}[0]{Maude}
\newcommand{\multivesta}[0]{\textsc{MultiVeStA}}
\newcommand{\spin}{Spin}
\newcommand{\misscel}{\textsc{Misscel}}

\newcommand{\defas}{\triangleq}

\newcommand{\define}{\defas}

\newcommand{\res}[1]{(\nu #1)}                  
\newcommand{\choice}{\: + \:}
\newcommand{\parcomp}{\parallel}             
\newcommand{\enspred}[1]{\mathpzc{#1}}
\newcommand{\auth}[2]{#1\mid#2}
\newcommand{\knw}{\mathcal{K}}      
\newcommand{\pols}{\Pi}          
\newcommand{\loccomp}[2]{ \interf{#1}[#2] }                  
\newcommand{\loccompAlt}[2]{#1 [#2] }                  
\newcommand{\interf}[1]{\ensuremath{\mathbf{\cal #1}}}
\newcommand{\naddr}{n}              
\newcommand{\vaddr}{x}              
\newcommand{\nvaddr}{c}          
\newcommand{\self}{\mathsf{self}}
\newcommand{\this}{\mathsf{this}}
\newcommand{\bexpr}{e}               
\newcommand{\procnil}{\ensuremath{\textbf{nil}}}
\newcommand{\acput}{\textbf{put}}                           
\newcommand{\acget}{\textbf{get}}
\newcommand{\acread}{\textbf{qry}}
\newcommand{\acputp}[2]{\ensuremath{\mathbf{put}(#1)@#2}}  
\newcommand{\acgetp}[2]{\ensuremath{\mathbf{get}(#1)@#2}}
\newcommand{\acreadp}[2]{\ensuremath{\mathbf{qry}(#1)@#2}}
\newcommand{\form}{?}
\newcommand{\tuple}[1]{\langle #1 \rangle}
\newcommand{\acnew}{\textbf{new}}
\newcommand{\acnewp}[1]{\ensuremath{\mathbf{new}(#1)}}
\newcommand{\acfresh}{\textbf{fresh}}
\newcommand{\acfreshp}[1]{\ensuremath{\mathbf{fresh}(#1)}}
\newcommand{\partner}{p} 
\newcommand{\op}{o} 
\newcommand{\expr}{\epsilon} 
 
\newcommand{\epc}[2]{#1 \!\stackrel{{}_\bullet}{} \!#2}

\newcommand{\var}{x} 
 
\newcommand{\name}{n} 
 
\newcommand{\val}{v} 
\newcommand{\kl}{k} 
\newcommand{\xorn}{u} 
\newcommand{\xorv}{w} 
\newcommand{\xnk}{e} 
\newcommand{\guard}{g} 
\newcommand{\killing}[1]{\mathbf{kill}(#1)} 
\newcommand{\out}[3]{\epc{#1}{#2} ! #3} 
\newcommand{\spar}{\mid} 
\newcommand{\prot}[1]{\{\!| #1 |\!\}} 
\newcommand{\scope}[1]{[#1]\,} 
\newcommand{\rep}{\ast\,} 
\newcommand{\nil}{\mathbf{0}} 
\newcommand{\inp}[3]{\epc{#1}{#2} ? #3} 
\newcommand{\arr}[1]{\langle #1 \rangle}

\newcommand{\und}{\_} 
\newcommand{\myfont}[1]{\mathrm {#1}} 
\newcommand{\pTA}{\partner_{br}}
\definecolor{mygray}{rgb}{.90,.90,.90}
\definecolor{light-gray}{gray}{0.80}
\definecolor{light-blue}{rgb}{0.85, 0.92, 1.0}
\newcommand{\bgInstanceA}[1]{\colorbox{mygray}{#1}} 
\newcommand{\bgInstanceB}[1]{\colorbox{light-blue}{#1}} 
\newcommand{\rMapsto}[2]{\xmapsto{#1}\hspace{-1mm}\textsubscript{${#2}$}}
\newcommand{\rTo}[1]{\xrightarrow{#1}}
\newcommand\pto{\mathrel{\ooalign{\hfil$\mapstochar$\hfil\cr$\to$\cr}}}
\newcommand{\COMP}[2]{\Gamma_{#1}\!:\!{#2}}
\newcommand{\Attributes}{\mathcal{A}}
\newcommand{\Values}{\mathcal{V}}
\newcommand{\abcmatch}[1]{\langle{#1}\rangle}
\newcommand{\abcnil}{{\sf 0}}
\newcommand{\abcout}[2]{({#1})@{#2}}
\newcommand{\abcin}[2]{{#1}({#2})}
\newcommand{\abcupdate}[2]{[{#1}:={#2}]}

\newcommand{\abclabout}[3]{#1\triangleright\overline{#3}(#2)}
\newcommand{\abclabin}[3]{#1\triangleright #3(#2)}
\newcommand{\abclabdin}[3]{\widetilde{\abclabin{#1}{#2}{#3}}}

\newcommand*{\llbrace}{%
  \BeginAccSupp{method=hex,unicode,ActualText=2983}%
    \textnormal{\usefont{OMS}{lmr}{m}{n}\char102}%
    \mathchoice{\mkern-4.05mu}{\mkern-4.05mu}{\mkern-4.3mu}{\mkern-4.8mu}%
    \textnormal{\usefont{OMS}{lmr}{m}{n}\char106}%
  \EndAccSupp{}%
}
\newcommand*{\rrbrace}{%
  \BeginAccSupp{method=hex,unicode,ActualText=2984}%
    \textnormal{\usefont{OMS}{lmr}{m}{n}\char106}%
    \mathchoice{\mkern-4.05mu}{\mkern-4.05mu}{\mkern-4.3mu}{\mkern-4.8mu}%
    \textnormal{\usefont{OMS}{lmr}{m}{n}\char103}%
  \EndAccSupp{}%
}
\newcommand{\goat}{{\rm{\it \cal{G}\kern-.13em o\cal{A}t}}\xspace}

\newcommand{\COMPAbC}[3]{#1\hspace{-1.5mm}:_{#2}\hspace{-1mm}{#3}}
\definecolor{lgray}{gray}{0.9}

\newcommand{\x}[1]{{\sf #1}}
\newcommand{\match}[3]{#1{\bf(}#2{\bf,}#3{\bf)}}
\newcommand{\allow}{\vdash}

\newcommand{\rulelabel}[1]{\textit{\small{(#1)}}} 

\newcommand{\procarrow}[1]{\downarrow_{#1}}

\newcommand{\doPut}[2]{#1\triangleright #2}

\newcommand{\acceptPut}[2]{#1\,\bar{\triangleright}\, #2}

\newcommand{\ensarrow}[1]{\rTo{#1}}

\newcommand{\evtarg}{\gamma}          
\newcommand{\replace}[2]{[#2 / #1]}

\newcommand{\valp}[2]{\mbox{$[\![ \: #1 \: ]\!]_{#2}$}}

\newcommand{\valn}[2]{\mbox{$[\![ \: #1 \: ]\!]_{#2}$}}

\makeatletter
\def \rightarrowfill{\m@th\mathord{\smash-}\mkern-6mu%
  \cleaders\hbox{$\mkern-2mu\mathord{\smash-}\mkern-2mu$}\hfill
  \mkern-6mu\mathord\rightarrow}
\makeatother
\newcommand{\transition}[1]{\overstackrel{\rightarrowfill}{\ \ #1\ \ \phantom{\dag}}} 

\def \overstackrel#1#2{\mathrel{\mathop{#1}\limits^{#2}}}

\newcommand{\eval}[1]{[\![ #1 ]\!]}

\newcommand{\halt}[1]{halt(#1)} 
\newcommand{\taukill}{\dag} 
\newcommand{\dset}[1]{\myfont{e}(#1)} 
\newcommand{\nokill}[2]{\myfont{noKill}(#1,#2)} 
\newcommand{\outl}[3]{\epcl{#1}{#2}\lhd #3} 
\newcommand{\epcl}[2]{#1\, {}_{^\bullet} #2} 
\newcommand{\inpl}[3]{\epcl{#1}{#2}\rhd #3} 
\newcommand{\matchCows}[2]{{\cal M}(#1,#2)} 

\newdimen\proofrulebreadth \proofrulebreadth=.04em
\newdimen\proofdotseparation \proofdotseparation=1.25ex
\newdimen\proofrulebaseline \proofrulebaseline=2ex
\newcount\proofdotnumber \proofdotnumber=3
\let\then\relax
\def\hfi{\hskip0pt plus.0001fil}
\mathchardef\squigto="3A3B
\newif\ifinsideprooftree\insideprooftreefalse
\newif\ifonleftofproofrule\onleftofproofrulefalse
\newif\ifproofdots\proofdotsfalse
\newif\ifdoubleproof\doubleprooffalse
\let\wereinproofbit\relax
\newdimen\shortenproofleft
\newdimen\shortenproofright
\newdimen\proofbelowshift
\newbox\proofabove
\newbox\proofbelow
\newbox\proofrulename
\def\shiftproofbelow{\let\next\relax\afterassignment\setshiftproofbelow\dimen0 }
\def\shiftproofbelowneg{\def\next{\multiply\dimen0 by-1 }%
\afterassignment\setshiftproofbelow\dimen0 }
\def\setshiftproofbelow{\next\proofbelowshift=\dimen0 }
\def\setproofrulebreadth{\proofrulebreadth}

\def\prooftree{
\ifnum  \lastpenalty=1 \then   \unpenalty \else
\onleftofproofrulefalse \fi
\ifonleftofproofrule \else   \ifinsideprooftree
        \then   \hskip.5em plus1fil
        \fi
\fi
\bgroup
\setbox\proofbelow=\hbox{}\setbox\proofrulename=\hbox{}%
\let\justifies\proofover\let\leadsto\proofoverdots\let\Justifies\proofoverdbl
\let\using\proofusing\let\[\prooftree
\ifinsideprooftree\let\]\endprooftree\fi
\proofdotsfalse\doubleprooffalse
\let\thickness\setproofrulebreadth
\let\shiftright\shiftproofbelow \let\shift\shiftproofbelow
\let\shiftleft\shiftproofbelowneg
\let\ifwasinsideprooftree\ifinsideprooftree
\insideprooftreetrue
\setbox\proofabove=\hbox\bgroup$\displaystyle 
\let\wereinproofbit\prooftree
\shortenproofleft=0pt \shortenproofright=0pt \proofbelowshift=0pt
\onleftofproofruletrue\penalty1 }

\def\eproofbit{
\ifx    \wereinproofbit\prooftree \then   \ifcase \lastpenalty
        \then   \shortenproofright=0pt  
        \or     \unpenalty\hfil         
        \or     \unpenalty\unskip       
        \else   \shortenproofright=0pt  
        \fi
\fi
\global\dimen0=\shortenproofleft \global\dimen1=\shortenproofright
\global\dimen2=\proofrulebreadth \global\dimen3=\proofbelowshift
\global\dimen4=\proofdotseparation
$\egroup  
\shortenproofleft=\dimen0 \shortenproofright=\dimen1
\proofrulebreadth=\dimen2 \proofbelowshift=\dimen3
\proofdotseparation=\dimen4
}

\def\proofover{
\eproofbit 
\setbox\proofbelow=\hbox\bgroup 
\let\wereinproofbit\proofover
$\displaystyle
}%
\def\proofoverdbl{
\eproofbit 
\doubleprooftrue
\setbox\proofbelow=\hbox\bgroup 
\let\wereinproofbit\proofoverdbl
$\displaystyle

}%
\def\proofoverdots{
\eproofbit 
\proofdotstrue
\setbox\proofbelow=\hbox\bgroup 
\let\wereinproofbit\proofoverdots

$\displaystyle
}%
\def\proofusing{
\eproofbit 
\setbox\proofrulename=\hbox\bgroup 
\let\wereinproofbit\proofusing
\kern0.3em$ }

\def\endprooftree{
\eproofbit 
  \dimen5 =0pt
\dimen0=\wd\proofabove \advance\dimen0-\shortenproofleft
\advance\dimen0-\shortenproofright
\dimen1=.5\dimen0 \advance\dimen1-.5\wd\proofbelow \dimen4=\dimen1
\advance\dimen1\proofbelowshift \advance\dimen4-\proofbelowshift
\ifdim  \dimen1<0pt \then   \advance\shortenproofleft\dimen1
        \advance\dimen0-\dimen1
        \dimen1=0pt
        \ifdim  \shortenproofleft<0pt
        \then   \setbox\proofabove=\hbox{%
                        \kern-\shortenproofleft\unhbox\proofabove}%
                \shortenproofleft=0pt
        \fi
\fi
\ifdim  \dimen4<0pt \then   \advance\shortenproofright\dimen4
        \advance\dimen0-\dimen4
        \dimen4=0pt
\fi
\ifdim  \shortenproofright<\wd\proofrulename \then
\shortenproofright=\wd\proofrulename \fi
\dimen2=\shortenproofleft \advance\dimen2 by\dimen1
\dimen3=\shortenproofright\advance\dimen3 by\dimen4
\ifproofdots \then
        \dimen6=\shortenproofleft \advance\dimen6 .5\dimen0
        \setbox1=\vbox to\proofdotseparation{\vss\hbox{$\cdot$}\vss}
        \setbox0=\hbox{%
                \kern\dimen6
                $\vcenter to\proofdotnumber\proofdotseparation
                        {\leaders\box1\vfill}$%
                \unhbox\proofrulename}%
\else   \dimen6=\fontdimen22\the\textfont2 
        \dimen7=\dimen6
        \advance\dimen6by.5\proofrulebreadth
        \advance\dimen7by-.5\proofrulebreadth
        \setbox0=\hbox{%
                \kern\shortenproofleft
                \ifdoubleproof
                \then   \hbox to\dimen0{%
                        $\mathsurround0pt\mathord=\mkern-6mu%
                        \cleaders\hbox{$\mkern-2mu=\mkern-2mu$}\hfill
                        \mkern-6mu\mathord=$}%
                \else   \vrule height\dimen6 depth-\dimen7 width\dimen0
                \fi
                \unhbox\proofrulename}%
        \ht0=\dimen6 \dp0=-\dimen7
\fi
\let\doll\relax
\ifwasinsideprooftree \then   \let\VBOX\vbox \else
\ifmmode\else$\let\doll=$\fi
        \let\VBOX\vcenter
\fi
\VBOX   {\baselineskip\proofrulebaseline \lineskip.2ex
        \expandafter\lineskiplimit\ifproofdots0ex\else-0.6ex\fi
        \hbox   spread\dimen5   {\hfi\unhbox\proofabove\hfi}%
        \hbox{\box0}%
        \hbox   {\kern\dimen2 \box\proofbelow}}\doll%
\global\dimen2=\dimen2 \global\dimen3=\dimen3
\egroup 
\ifonleftofproofrule \then   \shortenproofleft=\dimen2 \fi
\shortenproofright=\dimen3
\onleftofproofrulefalse \ifinsideprooftree \then   \hskip.5em plus
1fil \penalty2 \fi }

\newcommand{\substl}{\sigma} 
\newcommand{\substi}[1]{\{#1\}} 
\newcommand{\assoc}[2]{#1 \mapsto #2} 

\newcommand{\denyUnless}{\x{deny}\textrm{-}\x{unless}\textrm{-}\x{permit}}

\newcommand{\ruleOpt}[1]{{\bf{(}}#1{\bf{)}}}

\newcommand{\permit}{\x{permit}}

\newcommand{\contextRespPDPBegin}[1]{\x{AD:} \,  }
\newcommand{\contextRespPEPBegin}[1]{\x{ED:} \,  }

\newcommand{\abel}{\emph{ABEL}\xspace}
\newcommand{\kvalt}[1]{\mbox{$[\![ \: #1 \: ]\!]$}}

\newcommand{\modif}[1]{#1}

\journal{JLAMP}

\begin{document}

\begin{frontmatter}



\title{A Formal Approach to the Engineering of Domain-Specific Distributed Systems\tnoteref{full}}
 \tnotetext[full]{This work is a revised and extended version of~\cite{DFPT:coord18}. It appeared in the Proceedings of the 20th International Conference on Coordination Models and Languages (COORDINATION2018).}

\author[IMT,CINI]{Rocco De Nicola}
\author[UNIPI]{Gianluigi Ferrari}
\author[UNIFI]{Rosario Pugliese}
\author[UNICAM]{Francesco Tiezzi}

\address[IMT]{IMT School for Advanced Studies Lucca, Italy}
\address[UNIPI]{Universit\`a di Pisa, Italy}
\address[UNIFI]{Universit\`a degli Studi di Firenze, Italy}
\address[UNICAM]{Universit\`a di Camerino, Italy}
\address[CINI]{National Cybersecurity Laboratory, CINI, Italy}

\begin{abstract}
We review some results regarding specification, programming and verification of different classes of distributed systems which stemmed from the research of the Concurrency and Mobility Group at University of Firenze. More specifically, we examine the distinguishing features of network-aware programming, service-oriented computing, autonomic computing, and collective adaptive systems programming. We then present an overview of four different languages, namely \klaim, \cows, \SCEL\ and \abc. For each language, we discuss design choices, present syntax and semantics, show how the different formalisms can be used to model and program a travel booking scenario, and describe programming environments and verification techniques. 
\end{abstract}


\end{frontmatter}
\tableofcontents
\renewcommand{\sep}{\ \ | \ \ } 
\sloppy

\section{Introduction}
\label{sec:introduction}

Since the mid-90s, we have witnessed an evolution of distributed computing  towards increasingly complex systems formed by several software components featuring asynchronous interactions and operating in open-ended and non-deterministic environments. Such transformation, initially induced by the spreading of internetworking technologies, led to a paradigm shift making software components \emph{aware} of the underlying network infrastructure. Network awareness, on the one hand, constrained the remote access to distributed resources and, on the other hand, enabled computation mobility, to support different kinds of optimisations.

On top of these networked systems, software components have been then deployed to provide \emph{services} accessible by end-users and other system components through communication endpoints. This fostered the development of sophisticated applications built by reusing and composing simpler elements. \modif{Such service-based compositional approach abstracted from the actual distribution of the involved components over the underlying network, but required to deal with the interaction challenges posed by their heterogeneity. Interoperability was then achieved through the definition of standard protocols and suitable run-time support for programming languages that were taking into account also the failures that could occur in long-term interactions. Moreover, the absence of network awareness meant that there was no need for code mobility.}

Later on, the need arose of reducing the maintenance cost of these web-based systems, whose size was becoming bigger and bigger, and of extending their applicability to interact with and control the physical world, possibly in scenarios where human intervention was difficult or even impossible. It was then advocated to rely on \emph{autonomic} components, which are capable of continuously monitoring their internal status and the working environment, and to adapt their behaviour accordingly. 
\modif{In addition to point-to-point interactions, typical of client-server protocols, more sophisticated forms of interaction could occur that simultaneously involve an \emph{ensemble} of components dynamically determined. Ensembles are to be intended as collections of task-oriented or dedicated components that pool their resources and capabilities together to create a more complex system, which offers more functionalities and higher performance than simply the sum of the constituent elements.}

More recently, in some classes of autonomic computational systems we have witnessed the tremendous growth in the number of interacting components that are usually distributed, heterogeneous, decentralised and interdependent, and operate in dynamic and possibly unpredictable environments. The components form \emph{collectives} by combining their behaviours to achieve specific goals or to contribute to an emerging behaviour of the global system. Collectives abstract from the identity of the single components to guarantee scalability.

The evolution of distributed computing described above corresponds to the emergence of classes of systems that characterise specific programming domains.
Correspondingly, dedicated programming paradigms have been proposed, namely:
\begin{description}
\item[network-aware programming] to exploit the knowledge of the underlying infrastructure for better using network facilities and moving programs closer to the resources they want to use~\cite{CodeMobility:98};
\item[service-oriented computing] to allow the exploitation of loosely-coupled services as fundamental resources for developing applications and support the rapid and automatic development of open distributed systems~\cite{PG-SOC:2003};
\item[autonomic computing] to guarantee the self-managing characteristics of distributed computing resources, adapting to unpredictable changes while hiding intrinsic complexity to operators and users~\cite{KC03};
\item[collective adaptive systems programming] to model complex systems with large numbers of heterogeneous entities interacting without a specific central control, and adapting to environmental settings in pursuit of an individual or collective goal~\cite{anderson2013adaptive}.
\end{description}
Besides dealing with the distinctive aspects of each of such domains, the main challenge in engineering these classes of distributed systems is to coordinate the overall behaviour resulting from the involved distributed components while ensuring trustworthiness of the whole system.
To meet this goal, many researchers have adopted a language-based approach that combines the use of formal methods techniques with model-driven software engineering. 
The key ingredients of the resulting methodology, that can be applied to all classes of systems described above, may be summarised as relying on:
\begin{enumerate}
\item a specification language equipped with a formal semantics, 
which associates mathematical models 
to each term of the language to precisely establish the expected behaviour of systems;
\item a set of techniques and tools, built on top of the models, to express and verify properties of interest;
\item a programming framework together with an associated runtime environment, to actually execute the specified systems.
\end{enumerate}

When specialising this methodology, a major challenge for (specification or programming) language designers is to devise appropriate abstractions and linguistic primitives to deal with the specificities of the domain under investigation. Indeed, including the distinctive aspects of the domain as first-class elements of the language makes systems design more intuitive and concise, and their analysis more effective. In fact, when the outcome of a verification activity is expressed by considering the high level features of a system, and not its low-level representation, system designers can be provided  with a more direct feedback. 

This paper reviews some of the efforts, to which the authors have contributed, in applying the outlined methodology to the classes of distributed systems mentioned above by taking as starting point process algebras and some of the verification techniques and tools developed for them. 
%
\modif{The approach was initially applied to network-aware programming and the main result was the definition of the \klaim~language \cite{Klaim98} \modif{that had explicit localities, processes mobility and network connections as primitive notions} (Section~\ref{sec:klaim}). Afterwards, the approach was applied to service-oriented computing resulting in the design of \cows~\cite{COWS_JAL} \modif{whose basic constructs permitted to express correlations between clients and services and to deal with services failures}. (Section~\ref{sec:cows}).  Instead, to deal with autonomic computing the \SCEL language~\cite{DLPT14} was introduced \modif{that had explicit notions of agents knowledge and primitives and policies for its manipulations together with an original approach to ad hoc ensembles formation} (Section~\ref{sec:scel}).  Finally, to model and prove emergent properties of collective adaptive systems a distilled version of \SCEL named \abc~\cite{Alrahman3}  was introduced \modif{that had specific operators for  selecting communication partners using predicates on the run time value of relevant attributes of the agents forming the system} (Section~\ref{sec:abc}).} 

In the following parts of this paper, for each of these domain-specific languages, we discuss the design choices behind it, present its syntax and informal semantics, and provide an excerpt of the rules defining its formal operational semantics in terms of labelled transition systems by relying on the Structural Operational Semantics style~\cite{SOS:Plotkin04}. For each language, we also briefly describe the programming environments that have been developed to support program execution and outline some of the techniques that have been advocated for the verification of properties of the specified systems. 

Moreover, to assess the expressive power of the different formalisms and to put them at work, we show how they can be used to model a simple scenario that is instrumental to highlight distinguishing features. For each formalism, we also provide some code snippets  showing how close the specification of the model is to its underlying implementation. The scenario considers an online travel broker that, starting from specific requirements of customers, looks for hotel rooms and flights. Customers communicate their preferences to the broker and this, after some preliminary assessments, forwards the requirements to a number of hotels and air companies. Those, upon request, declare their availability and prices so that the customers can take the final choices and proceed with the booking. 

The paper ends with a summary of distinguishing features of the presented languages and with a few considerations about the lessons learnt (Section~\ref{sec:conclusion}).

\section{\textbf{\textsc{Klaim}}: Kernel language for Agents Interaction and Mobility}
\label{sec:klaim}

Network awareness indicates the ability of the software components of a distributed application to manage directly a sufficient amount of knowledge about the network environment where they are currently deployed. This capability allows components to have a highly dynamic behaviour and manage unpredictable changes of the network environment  over time. This is of great importance when programming mobile components capable of disconnecting from one node of the underlying infrastructure and of reconnecting to a different node. Programmers are usually supported with primitive constructs that enable components to communicate, and to distribute and retrieve data to and from the nodes of the underlying infrastructure. 

\klaim\ (\emph{Kernel Language for Agents Interaction and Mobility}, \cite{Klaim98}) has been specifically devised to design distributed applications consisting of several components, both stationary and mobile, deployed over the nodes of a distributed infrastructure. The \klaim\ programming model relies on a unique interface (\ie set of operations) supporting component communications and data management. 

\emph{Localities} are the basic building blocks of \klaim\ for guaranteeing network awareness. They are symbolic addresses (\ie network references) of nodes and are referred by means of identifiers. Localities can be exchanged among the computational components and are subjected to sophisticated scoping rules. They provide the naming mechanism to identify network resources and to represent the notion of administrative domain: computations at a given locality are under the control of a specific authority. This way, localities naturally support the programming of spatially distributed applications.

\klaim\ builds on Linda's notion of \emph{generative communication} through a single shared tuple space \cite{Gel85} and generalises it to multiple distributed tuple spaces. A tuple space is a multiset of tuples. Tuples are \emph{anonymous} sequences of data items and are retrieved from tuple spaces by means of an \emph{associative selection}. Interprocess communication occurs through \emph{asynchronous} exchange of tuples via tuple spaces: there is no need for producers (\ie senders) and consumers (\ie receivers) of a tuple to synchronise.

The obtained communication model has a number of properties that make it appealing for distributed computing in general (see, e.g., \cite{Gel89,davies97limbo,CCR96,deugo01}). It supports \emph{time uncoupling} (data life time is independent of the producer process life time), \emph{destination uncoupling} 
(data producers do not need to know the future use or the final destination of the data) and \emph{space uncoupling} (programmers need to know a single interface only to operate over the tuple spaces, regardless of the network node where the action will take place). 

\subsection{Syntax}

The syntax of \klaim\ is presented in Table~\ref{klaim-syntax}. We assume existence of two disjoint sets: the set of \emph{localities}, ranged over by $\sito$, and the set of \emph{locality variables}, ranged over by $\locv$, with the distinguished variable $\self$ denoting the locality of the node using it. Their union gives the set of \emph{names}, ranged over by $\llv$. We also assume three other disjoint sets: a set of \emph{value variables}, ranged over by $x$, a set of \emph{process variables}, ranged over by $X$, and a set of \emph{process identifiers}, ranged over by $A$.

\begin{table}[t]
\footnotesize
\hrule
\begin{tabular}{l@{\hspace{.2cm}}c@{\hspace{.2cm}}l@{\hspace{.5cm}}l@{\hspace{.3cm}}|}
\multicolumn{3}{l}{\textsc{Nets:}} & \\
\quad $N$ & ::= & $\sito ::_{\rho} P$ & (computational node) \\
    & $ \sep$ & $ \sito :: \tuple {et}$ & (located tuple) \\
    & $ \sep$ & $ N_1 \parallel N_2$ & (net composition) \\[.1cm]
\multicolumn{3}{l}{\textsc{Processes:}} & \\
\quad $P$ & ::= & $ \mmnil$ & (inert process) \\
    & $ \sep$ & $a.P$ & (action prefixing) \\
    & $ \sep$ & $P_1 \choice P_2$ & (choice)\\
    & $ \sep$ & $P_1 \mid P_2$ & (parallel composition) \\
    & $ \sep$ & $X$ & (process variable) \\
    & $ \sep$ & $A(\bar{p})$ & (process invocation) \\[.1cm]
\multicolumn{3}{l}{\textsc{Actions:}} & \\
\quad $a$ & ::= & $ \oap {\llv} {t} $ & (output) \\
    & $ \sep$ & $ \iap {\llv} {T} $ & (input) \\
    & $ \sep$ & $ \rap {\llv} {T} $ & (read) \\
    & $ \sep$ & $ \goap \llv P $ & (migration) \\
    & $ \sep$ & $ \nnew \locv$ & (creation) \\[.1cm]
\end{tabular}
\begin{tabular}{l@{\hspace{.2cm}}c@{\hspace{.2cm}}l@{\hspace{.5cm}}l}\vspace*{-.3cm}\\
\multicolumn{3}{l}{\textsc{Tuples:}} & \\
\quad $t$ & $ ::=$  & \multicolumn{2}{l}{$f \sep f, t$} \\[.1cm]
\multicolumn{3}{l}{\textsc{Tuple fields:}} & \\
\quad $f$ & $ ::=$  & \multicolumn{2}{l}{$e \sep \llv \sep P$} \\[.1cm]
\multicolumn{3}{l}{\textsc{Evaluated tuples:}} & \\
\quad $et$ & $ ::=$  & \multicolumn{2}{l}{$ef \sep ef, et$} \\[.1cm]
\multicolumn{3}{l}{\textsc{Evaluated tuple fields:}} & \\
\quad $ef$ & $ ::=$  & \multicolumn{2}{l}{$V \sep \sito \sep P$} \\[.1cm]
\multicolumn{3}{l}{\textsc{Templates:}} & \\
\quad $T$ & $ ::=$  & \multicolumn{2}{l}{$F \sep F, T$} \\[.1cm]
\multicolumn{3}{l}{\textsc{Template fields:}} & \\
\quad $F$ & $ ::=$  & \multicolumn{2}{l}{$f \sep !\,x \sep !\,\locv  \sep !\,X$} \\[.1cm]
\multicolumn{3}{l}{\textsc{Expressions:}} & \\
\quad $e$ & $ ::=$  & \multicolumn{2}{l}{$V \sep x \sep \ldots$} \\[.1cm]
    \\
\end{tabular}
\hrule
\caption{Klaim syntax} 
\label{klaim-syntax}
\end{table}

\textsc{Nets} are finite collections of nodes where processes and data can be placed. A \emph{computational node} takes the form $\sito ::_{\rho} P$, where $\rho$ is an {\em allocation environment} and $P$ is a process. Since processes may refer to locality variables, the allocation environment acts as a \emph{name solver} binding locality variables to specific localities.

\textsc{Processes} are the active computational units of  \klaim. Each process is obtained by composing subprocesses or the \emph{inert} process $\mmnil$ via \emph{action prefixing} ($a.P$), nondeterministic \emph{choice} ($P_1 \choice P_2$), \emph{parallel composition} ($P_1 \mid P_2$), \emph{process variable} ($X$), and parameterised \emph{process invocation} ($A(\bar{p})$). Recursive behaviours are modelled via process definitions; it is assumed that each identifier $A$ has a {\em single} defining equation $A(\bar{q}) \define P$. Lists of actual and formal parameters are denoted by $\bar{p}$ and $\bar{q}$, respectively.

The tuple space of a node consists of all the \textsc{evaluated tuples} located there. \textsc{Tuples} are sequences of \emph{actual} fields, \ie expressions, localities or locality variables, or processes. The precise syntax of \textsc{expressions} is deliberately not specified; it is just assumed that they contain, at least, basic values, ranged over by $V$, and value variables, ranged over by $x$. \textsc{Templates} are sequences of actual and formal fields, and are used as patterns to select tuples in a tuple space. \emph{Formal} fields are identified by the $!$-tag (\eg $!\,x$) and are used to bind variables to values. 

\subsection{Informal semantics}

\textsc{Nets} aggregate nodes through the \emph{composition} operator $\_ \parallel \_$, which is both commutative and associative. \textsc{Processes} are concurrently executed in an \emph{interleaving} fashion, either at the same computational node or at different nodes. They can perform operations borrowed from a unique interface which provides two categories of actions. The first one consists of the programming abstractions supporting data management. Three primitive behaviours are provided: adding (\textbf{out}), withdrawing (\textbf{in}) and reading (\textbf{read}) a tuple to/from a tuple space. Input and output actions are \emph{mutators}: their execution modifies the tuple space. The read action is an \emph{observer}: it checks the availability and takes note of the content of a certain tuple without removing it from the tuple space. The second category of actions refers to network awareness: the migration action (\textbf{eval}) activates a new process over a network node, while the creation action (\textbf{newloc}) generates a new network node. The latter action is the only one not indexed by a locality  because it acts locally; all the other actions are tagged with the (possibly remote) locality where they will take place. Note that, in principle,  each network node can provide its own implementation of the action interface. This feature can be suitably exploited to sustain different policies for data handling as done, e.g., in \metaklaim~\cite{metaklaim}.

Only evaluated tuples can be added to a tuple space and templates must be evaluated before they can be used for retrieving tuples. Tuple and template evaluation amounts to computing the values of expressions and using the local allocation environment as a name solver for mapping locality variables to localities. As a consequence, the locality variables within processes in a tuple are mapped to localities by using the \emph{local} allocation environment. Localities and formal fields are left unchanged by such evaluation. 
A \emph{pattern-matching} mechanism is then used for associatively selecting (evaluated) tuples from tuple spaces according to (evaluated) templates. 

Process variables support \emph{higher-order} communication, namely the capability to exchange (the code of) a process and possibly execute it. This is realised by first adding a tuple containing the process to a tuple space and then retrieving/withdrawing this tuple while binding the process to a process variable. 

Finally, \klaim\ offers two forms of process mobility. One is based on \emph{static scoping}: by exploiting higher-order communication, a process moves along the nodes of a net with a fixed binding of resources determined by the allocation environments of the nodes from where, from time to time, it is going to move. The other form of mobility relies on \emph{dynamic scoping}: when migrating, a process breaks the local links to resources and inherits those of the destination node. 

\subsection{A taste of the operational semantics}

The operational semantics is only defined for well-formed nets and it is given in terms of a structural congruence and a reduction relation over nets. 
A net is deemed {\em well-formed} if for each node $\sito ::_{\rho} P$ we have that $\rho(\self) = \sito$ and $\fl{P} \subseteq dom(\rho)$, and for any pair of nodes $\sito ::_{\rho} P$ and $\sito' ::_{\rho'} P'$, we have that $\sito = \sito'$ implies $\rho = \rho'$. 
Notation $dom(\rho)$ denotes the set of locality variables mapped by the allocation environment $\rho$, while $\fl{P}$ denotes the set of \emph{free variables} of process $P$. A variable is \emph{free} in $P$ if it is not bound and it is \emph{bound} in $P$ if it occurs within a formal field of $\iap {\llv} {T}$ or $\rap {\llv} {T}$, or is the argument $\locv$ of $\nnew \locv$; the scope of the binding is the process after the prefix.
Actions \textbf{out} and \textbf{eval} are not binders, but their arguments may contain variables.
For the sake of simplicity, we assume that, for the processes we consider, bound variables are all distinct and different from the free ones.


The \emph{structural congruence}, $\equiv$, identifies syntactically different nets that intuitively represent the same net. It is defined as the smallest congruence relation over nets that satisfies a given set of laws. The most significant law is $\sito ::_{\rho} (P_1 | P_2) \equiv \sito ::_{\rho} P_1 \parallel \sito ::_{\rho} P_2$ meaning that it is always possible to transform a parallel of co-located processes into a parallel over nodes. The remaining laws express that $(i)$~$\parallel$ is commutative and associative, $(ii)$~the inert process can always be safely removed/added, and $(iii)$~a process identifier can be replaced with the body of its definition.
 
The reduction relations exploits two functions: one for evaluating tuples and templates, the other for selecting tuples in a tuple space.
The \emph{evaluation function} for tuples and templates takes as parameter the allocation environment of the node where the evaluation takes place. The main clauses of its definition are given below:
\begin{center}
$\kvalt{\locv}_{\rho} = \left\{
                            \begin{array}{ll}
\rho(\locv) & \mbox{\ \ if $\locv \in dom(\rho)$}\\
\mbox{\em undef} & \mbox{\ \ otherwise}
                            \end{array}
                   \right.$
\hspace{1cm}
$\kvalt{P}_{\rho} = P \{\rho\}$
\end{center}
where $P \{\rho\}$ denotes the process term obtained from $P$ by replacing any free occurrence of a locality variable \mbox{$\locv \in dom(\rho)$} that is not within the argument of an \mmeval\ with $\rho(\locv)$. 
Two examples of process evaluation are
$\kvalt{out(P)@\ell.Q}_{\rho} =
out(\kvalt{P}_{\rho})@\rho(\ell).Q\{\rho\}$ and
$\kvalt{eval(P)@\ell.Q}_{\rho} = eval(P)@\rho(\ell).Q\{\rho\}$.
We shall write $\kvalt{t}_{\rho} = et$ to denote that evaluation of tuple $t$ using $\rho$ succeeds and returns the evaluated tuple $et$.

For selecting an evaluated tuple $et$ from a tuple space according to an evaluated template $ET$, the \emph{pattern-matching} function, \mbox{$\mumatch(ET, et) = \sigma$,} is used. This function is defined by means of a set of inference rules which intuitively state that: an evaluated template matches against an evaluated tuple if both have the same number of fields and corresponding fields do match; two values match only if they are identical, while formal fields match any value of the same type. A successful matching returns a substitution function $\sigma$ associating the variables contained in the formal fields of the template with the values contained in the corresponding actual fields of the accessed tuple.

The \emph{reduction relation}, $\redarrow{}$, is defined as the least relation induced by a given set of inference rules. It is defined over configurations of the form $\netconf L N$, where $L$ is a finite set of localities keeping track of the localities occurring free in $N$ (that is, $\fl N \subseteq L$). $L$ is needed to ensure global freshness of new (dynamically generated) network localities and is indeed omitted whenever a reduction does not generate any fresh locality.

The most significant rules are reported in Table~\ref{Klaim-opsemrules}, where we write $\rho(\llv) = \sito$ to denote that either $\llv = \sito$ or $\llv$ is a locality variable that $\rho$ maps to $\sito$.
\begin{table}[t!]
$$
\begin{array}{@{}c@{}}
\hline
\\[-8pt]
\infer[\outrule]{
    \sito ::_{\rho} \oap {\llv} {t} .P \parallel \sito' ::_{\rho'} P'
    \redarrow{}
    \sito ::_{\rho} P \parallel \sito' ::_{\rho'} P' \parallel \sito' :: \tuple{et}
}{
    \rho(\llv) = \sito' && \kvalt{t}_{\rho} = et
}
\\[.35cm]
\infer[\evalrule]{
    \sito ::_{\rho} \goap {\llv} Q.P \parallel \sito' ::_{\rho'} P'
    \redarrow{}
    \sito ::_{\rho} P \parallel \sito' ::_{\rho'} P' | Q
}{
    \rho(\llv) = \sito'
}
\\[.35cm]
\infer[\inrule]{
    \sito ::_{\rho} \iap {\llv} {T}.P \parallel \sito' :: \tuple{et}
    \redarrow{}
    \sito ::_{\rho} P \sigma \parallel \sito' :: \emptytuple
}{
    \rho(\llv) = \sito' && \mumatch(\kvalt{T}_{\rho}, et) = \sigma
}
\\[.35cm]
\infer[\readrule]{
    \sito ::_{\rho} \rap {\llv} {T}.P \parallel \sito' :: \tuple{et}
    \redarrow{}
    \sito ::_{\rho} P \sigma \parallel \sito' :: \tuple{et}
}{
    \rho(\llv) = \sito' && \mumatch(\kvalt{T}_{\rho}, et) = \sigma
}
\\[.35cm]
\infer[\newrule]{
    \netconf L {\sito ::_{\rho} \nnew{\locv}.P}
    \redarrow{}
    \netconf {L \cup \{ \sito' \}}
            {\sito ::_{\rho} \musub P {\sito'} \locv \parallel \sito' ::_{\rho[\sito'/\self]} \mmnil}
}{
   \sito' \not\in L
}
\\[.1cm]
\hline
\end{array}
$$
\vspace{-.5cm}
\caption{\klaim\ operational semantics}
\label{Klaim-opsemrules}
\end{table}
In rule \outrule, the local allocation environment is used both to determine the name of the node where the tuple must be placed and to evaluate the argument tuple. This implies that if the argument tuple contains a field with a process $P$, the corresponding field of the evaluated tuple contains the process resulting from the evaluation of its locality variables, that is $P \{\rho\}$. Hence, processes in a tuple are transmitted after the interpretation of their free locality variables through the local allocation environment. This corresponds to having a {\em static scoping} discipline for the (possibly remote) generation of tuples. \outrule\ requires existence of the target node at $\sito'$, which is left unchanged by the reduction, and that the tuple $t$ argument of $\mmout$ is evaluable. As a result of the reduction, the tuple resulting from the evaluation of $t$ is added to the tuple space at $\sito'$.
A {\em dynamic linking} strategy is adopted for the $\mmeval$ operation, rule \evalrule. In this case the locality variables of the spawned process are not interpreted using the local allocation
environment: the linking of locality variables is done at the remote node. The underlying assumption that all the equations for process definitions are available everywhere greatly simplifies rule \evalrule, because it permits avoiding mechanisms for code inspection to find the process definitions needed by $Q$. 
Rule \inrule\ requires that the template $T$ argument of $\mmin$ is evaluable and that a matching tuple at the target node exists. As a result of the reduction, the matched tuple is removed from the target tuple space and the substitution returned by the pattern-matching function is applied to the continuation of the process performing the action, in order to replace the free occurrences of the variables bound by $T$ with the corresponding values of $et$. 
Rule \readrule\ is similar, it only differs from \inrule\ just because the accessed tuple is still left in the tuple space.
Finally, in rule \newrule, the premise exploits the set $L$ to choose a fresh locality $\sito'$ for naming the new node. In the continuation of the process performing the action the locality variable $\locv$ argument of $\mmnew$ is replaced by $\sito'$, thus the new locality becomes usable for the process. Notably, $\sito'$ is not yet known to any other node in the net. Hence, it can be used by the creating process as a \emph{private} name. The allocation environment of the new node is derived from that of the creating one with the obvious update for the location variable $\self$. Therefore, the new node inherits all the bindings of the creating node.

\subsection{A travel booking scenario}
\label{sec:booking_scenario_in_klaim}
We illustrate some of the distinguishing features of the \klaim\ programming model by using the online travel booking scenario informally 
presented in the Introduction. 
The \klaim\ specification consists of a collection of \klaim\ nodes, each modelling a component of the software architecture of the scenario.
For simplicity, we focus on three main components:
\begin{itemize}
\item  the {\sc broker} component where customers enter their requests, including the date and the origin-destination of the travel;
\item the {\sc hotel} and {\sc flight} components that are in charge of selecting hotels and flights in compliance with customers' requests.
\end{itemize}

\begin{figure}[t]
\centering
\includegraphics[width=\textwidth]{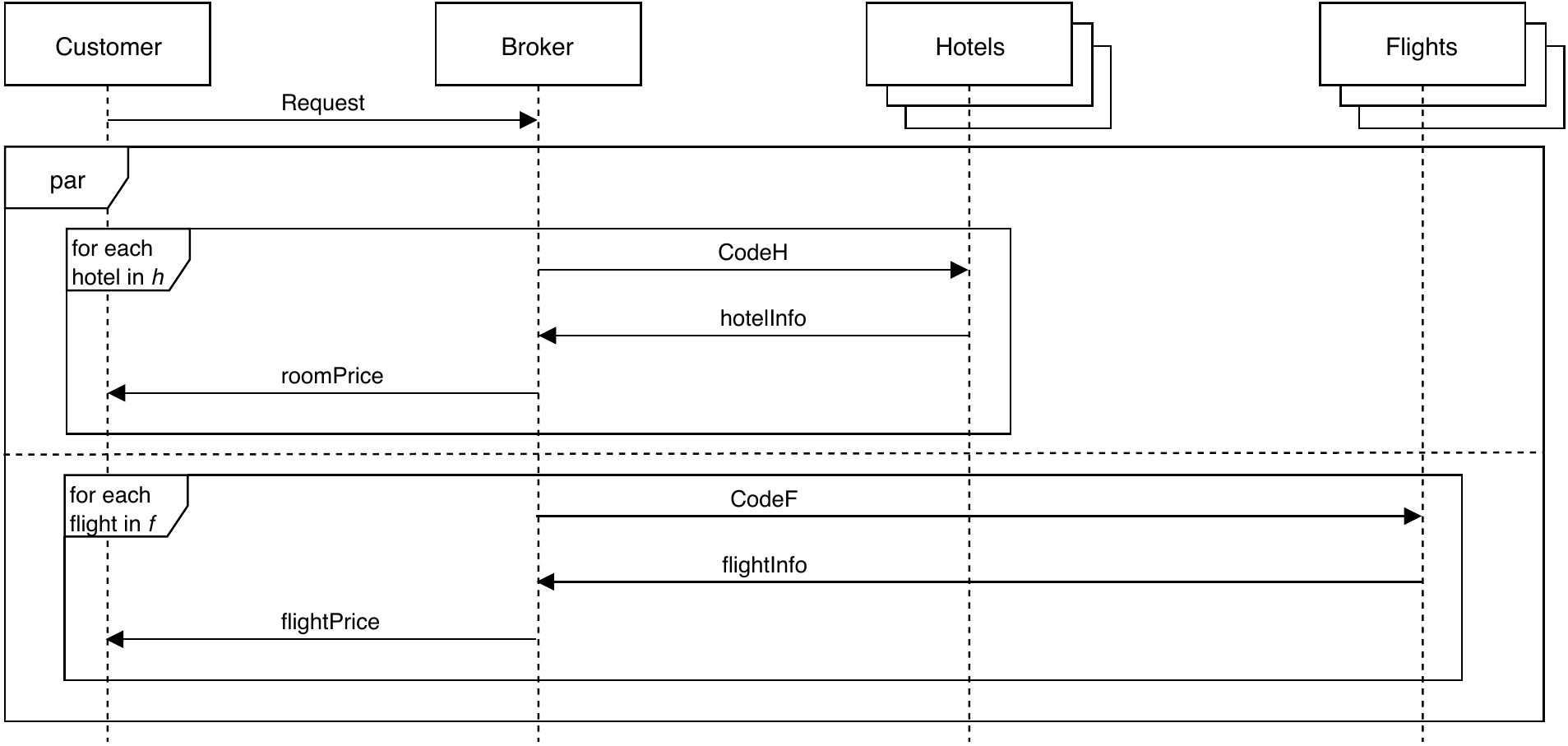}
\caption{Travel Booking Scenario in \klaim: Sequence Diagram.}
\label{fig:Klaim_seq_diag}
\end{figure}

\modif{The UML activity diagram displayed in Fig.~\ref{fig:Klaim_seq_diag} illustrates the flow of control inside the \klaim\ net implementing the travel booking scenario. The {\sc broker}, after collecting requests from customers, exploits \emph{code mobility} to activate some spider processes in the {\sc hotel} and {\sc flight} nodes. These spider processes act on behalf of the {\sc broker} to find hotels and flights matching customer's request. The exploitation of code mobility within the workflow of the application is expressed in the diagram by the $\mathit{CodeH}$ (resp. $\mathit{CodeF}$) label. Once available, the result of the search carried out by the spider processes is communicated back to the {\sc broker}. 
}

When we presented the \klaim\ programming model, we mainly focused on the linguistic primitives to \emph{structure} distributed applications  and to \emph{program} behaviour. Indeed, we have deliberately not considered primitive data types. We now show how to equip \klaim\ with simple data types. As an example, we introduce a data type for handling non-empty sequences of locations. We represent them through (the standard) square brackets comma-separated value notation: $[l_1, l_2, \ldots l_k]$. The unary function $[\: ]$ takes as input a location and yields as result the sequence consisting of the argument location only: $[\: ](l)=[l]$. The binary function $::$ (read \emph{cons}) takes as input a location and a sequence, and produces as result a new sequence whose first element is the argument location: $l :: [l_1, l_2, \ldots l_k] =  [l, l_1, l_2, \ldots l_k]$. Sequences now appear in tuples, hence pattern-matching has to be extended accordingly. For instance, the template $[!u]$ matches all sequences consisting of one location only. The template $!u:: !s$ matches all sequences having at least two elements. Hereafter, we will apply pattern-matching  to values in order to recognise the form of values and let the computation be guided accordingly. The same approach can be followed to include other data types, such as Strings, Dates, and so on. 

The structure of the nodes where the hotel and flight facilities are deployed is intuitively clear. The node hosting the hotel booking facility is  presented below:
\[
  l_h ::_{\rho_h} (\mathit{HotelManager}
  \mid T_h)
\]
The node hosts the  process $\mathit{HotelManager}$, which manages the hotel booking requests activated by the broker component. Moreover, it exposes room availability through suitable tuples stored  in the local tuple space $T_h$.

The \klaim\ node that specifies the behaviour of the {\sc broker} component is as follows:
\[
  \begin{array}{ll}
   l_{br} ::_{\rho_{br}} (\mathit{Handler} \mid \mathit{SessionManage}r \mid T_{br})  
   \end{array}
\]  
The node hosts the  processes $\mathit{Handler}$ and $\mathit{SessionManager}$ presented below, together with the local tuple space, represented by $T_{br}$. For the sake of readability, we exploit a sort of macro-like mechanism to associate a name to a piece of \klaim\ specification, e.g. we write $A \equiv P$ to indicate that the code of the process $P$ will replace the identifier $A$ each time this is encountered in the \klaim\ specification.
\[
\begin{array}{ll}
\mathit{Handler} \define \mmin(!usr, !date, !origin, !dest, !res)@\self.\\
\quad \mmout(``\mathit{Manage}", usr)@\self.\,   \mmin(usr,!sid)@\self. \\
\quad  ( (\emph{SpiderHotel}(sid,date) \mid \mmread(usr, ``\mathit{h}", !\mathit{hpref})@\self. \\
\quad \quad \mathit{ManageHotelPref}(usr, sid, res, \mathit{hpref})) \\
\quad  \mid (\emph{SpiderFlight}(sid, date, origin, dest) \mid \mmread(usr, ``\mathit{f}", 
!\mathit{fpref})@\self. \\
\quad \quad \mathit{ManageFlightPref}(usr, sid, res, \mathit{fpref})) \\
\quad \mid \mathit{Handler}) \\ \\
\mathit{SessionManager} \define \mmin(``\mathit{Manage}", !usr)@\self. \cdots .\\
\phantom{\mathit{SessionManager} \define\ }\mmout(usr,\mathit{sessionId})@\self.\,\mathit{SessionManager} \\
\end{array}
\]
The handler process $Handler$ receives the customer request, obtained by sensing in the tuple space the tuple containing the data about the customer code, the date-origin-destination of the travel, and the location of the node where the results of the request will be stored. 
The handler process then activates the manager process $\mathit{SessionManager}$, by emitting in the local tuple space a tuple tagged by $``\mathit{Manage}"$, and gets the actual session identifier, by inspecting the tuple space. We abstract from the detailed description of process $\mathit{SessionManager}$, 
since it deals with some low-level computational aspects specific for the considered application, rather than taking care of coordinating activities.
We just assume that the omitted code creates the unique session identifier for the customer's request, and associates to it the sequence of hotels and the sequence of airline companies that must be queried to satisfy the customer's request. 

The handler process exploits two recursive processes to activate the spider (mobile) processes in charge of finding hotels and flights. 
\[
\begin{array}{ll}
\mathit{SpiderHotel}(id, d) \define \mmin(id, ``h", [!u])@\self.\,\mmeval(\mathit{CodeH})@u.\,\mmnil \\
\quad +\ \mmin(id, ``h", [!u::!s])@\self.\,\mmeval(\mathit{CodeH})@u.\,\\
\qquad\, \mmout(id,``h",s)@\self.\,\mathit{SpiderHotel}(id,d) \\
\\
\mathit{CodeH} \equiv \mmout(``\mathit{check}", id, d)@\self. \\
\quad \quad (\mmin(``\mathit{avail}", id, !\mathit{info})@\self.\,\mmout(id, \mathit{info})@l_{br}.\,\mmnil\\
\quad\quad\ +\ \mmin(``\mathit{no\mathrm{-}avail}", id)@\self.\,\mmnil) \\ 
\\
\mathit{SpiderFlight}(id, d, orn, dst) \define \mmin(id,``f", [!u])@\self.\,\mmeval(\mathit{CodeF})@u.\,\mmnil \\
\quad +\  \mmin(id,``f", [!u::!s])@\self.\,\mmeval(\mathit{CodeF})@u.\,\\
\qquad\, \mmout(id,``f",s)@\self.\,\mathit{SpiderFlight}(id, d, orn, dst) \\
\\
\mathit{CodeF} \equiv \mmout(``check", id, d, orn, dst)@\self.\\
\quad \quad (\mmin(``\emph{avail}", id, !\mathit{info})@\self.\,\mmout(id, \mathit{info})@l_{br}.\,\mmnil\\
\quad\quad\  +\  \mmin(``\mathit{no\mathrm{-}avail}", id)@\self.\,\mmnil) \\ \\
\end{array}
\]
The spider processes take fully advantage of \klaim\ dynamic linking mobility through the $\mmeval$ primitive. This ensures that each spider will be spawned on the remote node {\em without} evaluating its locality variables according to the allocation environment of the broker component. This programming choice implies that when the mobile code will run in the remote node of the hotel (resp. of the flight), the location $\self$ will be bound to the actual address of the location where the hotel (resp. flight) component is deployed. The result of this search is then forwarded back to the tuple space of the broker. 

The last part of the behaviour of the broker consists in the management of the customer's preferences. This strongly depends on the data type used to store preferences. We outline the abstract specification of the facility matching customer's preferences with respect to the hotel information, as the treatment of flight information is similar. For simplicity, we assume that the hotel information is stored in tuples of the form $(\mathit{name}, \mathit{rate}, \mathit{hotel group})$. We also assume that the customer has a loyalty card for a specific group of hotels which is stored in the tuple space of the broker and that the customer will add a distinguished tuple to the tuple space of the broker to signal the termination of the hotel booking activity.
\[
\begin{array}{ll}
\mathit{ManageHotelPref}(usr, sid, res, \mathit{hpref}) \define \\
\qquad \mmin(sid, (!name, !rate, !gruop))@\self.\\ 
\qquad\qquad \mmout(``hCheck",group=\mathit{hpref})@\self.\\
\qquad\qquad (\ \mmin(``hCheck",\mathbf{true})@\self.\\
\qquad\qquad \quad \quad \mmout(sid, name, rate, ``\mathit{reduced\mathrm{-}price"})@res. \\
\qquad\qquad \quad \quad \mathit{ManageHotelPref}(usr, sid, res, \mathit{hpref}) \\
\qquad\qquad \ +\ \mmin(``hCheck",\mathbf{false})@\self.\\
\qquad\qquad \quad \quad \mmout(sid, name, rate, ``\mathit{standard\mathrm{-}price"})@res. \\
\qquad\qquad \quad \quad \mathit{ManageHotelPref}(usr, sid, res, \mathit{hpref})\ ) \\
\qquad +\  \mmin(sid, ``\mathit{endHotelBooking}")@\self.\mmnil
\end{array}
\]
The process $\mathit{ManageHotelPref}$ senses the local tuple space of the broker to identify the information about the hotels made available by the spider processes. This information is checked against the customer's preference (i.e., the hotel group) in order to report the presence of a reduced rate. Whenever information  on the hotel meets the customer's preferences, the tuple containing the hotel data is stored in the remote tuple space of the customer with a flag indicating the availability of the reduced rate.

\modif{
\paragraph{Discussion}
The \klaim\ primitive constructs for code mobility are instrumental to support the workflow of the travel booking scenario. We have seen that the spider processes exploit dynamic linking mobility. This has the additional benefit that the preferences associated with the specific customer are confined to the location of the broker. This is a simple way to obtain a suitable form of \emph{data privacy}. More sophisticated forms of security could be obtained through the use of \klaim\ types for access control or hierarchical \klaim\ nets. We refer to~\cite{DFPV00} and to~\cite{H-KLAIM} for details.
}

\subsection{Programming environment}

\XKlaim\footnote{\XKlaim is available online at \url{https://github.com/LorenzoBettini/xklaim}.} (\emph{eXtended \klaim}, \cite{BDFP98}) is an experimental programming language that extends \klaim with a high level syntax for processes. It provides variable declarations, enriched operations, assignments, conditionals, sequential and iterative process composition. The implementation of \XKlaim is based on \Klava\footnote{\Klava is available online at \url{http://music.dsi.unifi.it}.} (\emph{\klaim in \java}, \cite{BDP02}), a \java\ package that provides the run-time system for \XKlaim operations, and on a compiler, which translates \XKlaim programs into \java\ programs that use \Klava. A renewed and enhanced version of  \XKlaim is proposed in \cite{NEW_XKLAIM}. The new implementation comes together with an Eclipse-based IDE tooling, and relies on recent powerful frameworks for the development of programming languages, in particular the \xtext framework~\cite{XtextBook}.

\mbox{\XKlaim} can be used to write the higher layer of distributed applications while \Klava can be seen both as a middleware for \XKlaim programs and as a \java\ framework for programming according to the \klaim paradigm. By using \Klava directly, the programmer is able to 
implement a finer grained type of mobility.

\modif{Fig.~\ref{lis:xklaim_code} lists a significant fragment of code\footnote{\modif{The \XKlaim source code for the complete scenario can be downloaded from \url{https://bitbucket.org/tiezzi/jlamp_survey_code/src/master/Klaim/}.}} of the \XKlaim implementation of the \klaim specification of the travel booking scenario, presented in Section~\ref{sec:booking_scenario_in_klaim}. 
The \XKlaim code permits appreciating how  close a \klaim specification is to its \XKlaim implementation. 
Indeed, the syntax of the communication primitives is the same, except for the notation of formal fields
that in \XKlaim are specified as (typed) variable declarations. 
Notably, concurrent subprocesses of the $\mathit{Handler}$ process, 
composed by means of the $\mid$ operator in \klaim, are activated in 
\XKlaim using the \textbf{eval} action with target $\self$.
Finally, in the definition of the network, physical localities are expressed 
in terms of the standard TCP syntax \texttt{host:port}.

}

\begin{figure}[!t]
\lstset{language=xklaim,style=mystyle}
\begin{lstlisting}{}
proc HandlerProc() {
	while (true) {
		in(var String usr, var String date, var String origin, var String dest, 
		    val PhysicalLocality res)@self
		out("Manage",usr)@self
		in(usr,var Integer sid)@self		
		eval({eval(new SpiderHotelProc(sid,date))@self
			     read(usr,"h",var String hpref)@self
			     eval(new ManageHotelPrefProc(usr,sid,res,hpref))@self})@self		
		eval({eval(new SpiderFlightProc(sid,date,origin,dest))@self
			     read(usr,"f",var String fpref)@self
			     eval(new ManageFlightPrefProc(usr,sid,res,fpref))@self})@self
	}
}


net TravelBookingNet physical "localhost:9999" {
	node Customer logical "customer" {
		eval(new CustomerProc)@self
	}

	node Broker logical "broker"{
		...initialize broker's tuple space...
		eval(new HandlerProc)@self
		eval(new SessionManagerProc)@self
	}
	
	node Hotel1 logical "hotel1"{
		eval(new HotelManagerProc("hotel1","hotelGroupA"))@self
	}	
	. . .	
	node Flight1 logical "flight1"{
		eval(new FlightManagerProc("flight1","flightGroupA"))@self
	}	
	. . .
}
\end{lstlisting}
\vspace*{-.3cm}
\caption{The process $\mathit{Handler}$ and the network of the travel booking scenario implemented in \XKlaim.}
\label{lis:xklaim_code}
\end{figure}

\subsection{Verification techniques}

Many verification techniques have been defined for \klaim and variants thereof. Here we only mention a few of them.
In \cite{DL02:MLMA} a temporal logic
is proposed for specifying and verifying dynamic properties of mobile processes specified in \klaim. The inspiration for the proposal was the Hennessy-Milner Logic, but it needed significant adaptations due to the richer operating context of components. The resulting logic provides tools for establishing not only deadlock freedom, liveness and correctness with respect to given specifications (which are crucial properties for process calculi and similar formalisms), but also properties that relate to resource allocation, resource access and information disclosure (which are important issues for processes involving different actors and authorities).

An important topic deeply investigated for \klaim is the use of type systems for security \cite{DFPV00,DGP06,GP09}, devoted to control accesses to tuple spaces and mobility of processes. In these type systems, traditional types are generalised to \emph{behavioural types}. These are abstractions of process behaviours that provide information about \emph{processes capabilities}, namely the operations that processes can execute at a specific locality (downloading/consuming a tuple, producing a tuple, activating a process, and creating a new node). When using behavioural types, each \klaim node is equipped with a security policy, determined by a net coordinator, that specifies the execution privileges; the policy of a node describes the actions processes there located  can execute. By exploiting static and dynamic checks, type checking guarantees that only processes whose intentions match the rights granted to them by coordinators are allowed to proceed. An expressive language extension, called \MetaKlaim, equipped with a powerful type system is described in \cite{metaklaim}. \MetaKlaim is a higher order distributed process calculus equipped with staging mechanisms. It integrates \MetaML (an extension of SML for multi-stage programming) and \klaim, to permit interleaving of meta-programming activities (such as assembly and linking of code fragments), dynamic checking of security policies at administrative boundaries, and traditional computational activities on a wide area network (such as remote communication and code mobility). \MetaKlaim exploits a powerful type system (including polymorphic types \emph{\`a la} system \F) to deal with highly parameterised mobile components and to dynamically enforce security policies: types are metadata that are extracted from code at run-time and are used to express trustiness guarantees. The dynamic type checking ensures that the trustiness guarantees of wide area network applications are maintained also when computations interoperate with potentially untrusted components.

An alternative approach to control accesses to tuple spaces and mobility of processes is introduced in~\cite{DGHNNPP10}. It is based on Flow Logic and permits statically checking absence of violations. Starting from an existing type system for \klaim with some dynamic checks, the insights from the Flow Logic approach are exploited to construct a type system for statically guaranteeing secure access to tuple spaces and safe process migration for a smooth extension of \klaim. This is the first completely static type system for controlling accesses devised for a tuple space-based coordination language. 
A static  \emph{control flow analysis} that extends the one proposed in \cite{LMCS17,icissp2019}, to manage network awareness and coordination via multiple tuple spaces has been introduced in~\cite{BDFG19}. The static analysis can be used to detect where and how tuples are manipulated and how messages flow among the nodes of a \klaim\ network. This permits to identify possible security breaches in the data workflow of a distributed application. For instance, it may keep the safe paths that data inside a tuple can traverse apart from those that pass through a possible untrusted node.

\modif{
We now outline how the static methodology presented in~\cite{BDFG19} can be applied to investigate the security of \klaim\ code. We illustrate this by resorting to the \klaim\ specification of the travel booking scenario. The static methodology enables us to construct an \emph{abstract} graph-based model of the behaviour of the \klaim\ specification of the scenario. This abstract model supports a reasoning technique which permits to detect ($i$) the path in the network through which (a value in) a tuple of a specific node reaches another one, and ($ii$) the transformations which are applied to a selected value along those paths.

In the travel booking scenario, the abstract model approximates the trajectories of each piece of data. For instance, the abstract \emph{trajectory} below
\[
\begin{array}{l}
Trajectory(d_0, d_f) = 
l_u,{d_0} : l_{br}, d_1 : l_h, d_2 : l_{br,} d_3 : l_u, {d_f} \\
\hspace*{-1.5cm}\mbox{where}\\
{d_0} = \langle customer, date, origin, dest \rangle \qquad \mbox{ Customer Request} \\
{d_f} = bookingData  \qquad \qquad \qquad \qquad \quad \mbox{ Result of the search} \\
\end{array}
\]
expresses the path of the value $bookingData$ associated to the customer's request. This trajectory, made of pairs of the form $location, datum$ separated by the symbol `:', encodes the data transformations generated by each of the involved components in processing the customer's request together with the sequence of locations traversed due to the computation steps.

The abstract path above describes the capacity of the \klaim\ code to correctly manage customer's request. Instead, the following abstract path $l_u, d_0 : l_{br}, d_1 : l_h, d_2 : l_u, d_f$ detects a suspicious trajectory, namely a trajectory that by-passes the phase where the results of the spider processes are collected together. More generally, by analysing the abstract paths derived from the model it is possible to identify crucial code structures.
We refer to~\cite{BDFG19,BDFG20} for more details.
}

\subsection{Related work}

\modif{
Especially at the beginning of this century, with the manifest pervasivity of the Internet, many researchers have considered both models and implementations of network-aware formalisms that have or have been influenced by the work on \klaim and other  Linda-based models and primitives. In~\cite{BDM18}, many implementations of Linda-based models, including \klaim-based ones, to coordinate the interactions among system components are described and their efficiency is assessed. Instead, \cite{Ciatto0LOZ18} is a recent survey of coordination techniques for distributed and mobile systems, including those based on Linda and those relying on different coordination models. 
For references on network-aware programming and relation with \klaim we refer the interested reader to \cite{Klaim98,Global04}.

Among the foundational calculi aiming at capturing the key notions of network-awareness and identifying the programming abstractions  most suitable for network-aware programming, we would like to mention three different ones, namely the \emph{Distributed $\pi$-calculus} (D$\pi$)~\cite{HR02}, the \emph{Distributed Join Calculus} (DJoin)~\cite{FGLMR96,FG02}, the \emph{Ambient Calculus} (Amb)~\cite{CG00}, that were essentially proposed at the same time as \klaim.

$D\pi$ is a variant of the $\pi$-calculus enriched with explicit locations that are used to distribute processes. Interprocess communication is binary, channel-based, synchronous and local, in the sense that only processes at the same location can exchange messages. A process willing to communicate with a remote one has first to migrate to its location. 

In DJoin, a location is structured as a tree composed by the root location and its sub-locations. When a process defined at a specific location moves to a different location, the whole tree moves along with the process. Again, process communication is channel-based and  there is a unique process that can receive on each channel. To synchronise, processes rely on so-called join patterns that may require pattern matching on data and simultaneous reception of messages on different channels.

Finally, in the Ambient calculus, the key notion is that of ambient that can be thought of as a bounded environment where processes cooperate. An ambient is characterised by a name,  a collection of local agents and a collection of sub-ambients, and can be referred only through explicit naming. An agent moves together with the ambient containing it.  Communication is local to ambients and takes place through  anonymous message exchange, without resorting to channels or pattern matching. 
}

\section{\textbf{\textsc{Cows}}: Calculus for Orchestration of Web Services}
\label{sec:cows}

Since the early 2000s, the increasing success of e-business, e-learning, e-government, and other similar systems, has led the World Wide Web, initially thought of as a system for human use, to evolve towards an architecture for Service-Oriented Computing (SOC) supporting automated use. The SOC paradigm, that finds its origin in object-oriented and component-based software development, aims at enabling developers to build networks of distributed, interoperable and collaborative applications, regardless of the platform where the applications run and of the programming language used to develop them. The paradigm is based on the use of independent computational units, called \emph{services}. They are loosely coupled reusable components, that are built with little or no knowledge about clients and about other services involved in their operating environment.

One successful instantiation of the general SOC paradigm is given by the Web Service technology \cite{WS_W3C}, which exploits the pervasiveness of the Internet and related standards. Traditional software engineering technologies, however, do not neatly fit with SOC, thus hindering its full realisation in practice. The challenges come from the necessity of dealing at once with such issues as asynchronous interactions, concurrent activities, workflow coordination, business transactions, resource usage, and security, in a setting where demands and guarantees can be very different for the many involved components. 

\cows\ (\emph{Calculus for Orchestration of Web Services}, \cite{LPT07:ESOP,COWS_JAL}) is a formalism whose design has been influenced by the OASIS standard \wsbpel~\cite{WSBPEL} for orchestration of web services. In \cows, \emph{services} are computational entities capable of generating multiple instances to concurrently handle different client requests. Inter-service communication occurs through \emph{communication endpoints} and relies on pattern-matching for logically correlating messages to form an interaction session by means of their identical contents. Differently from most process calculi, and from \klaim, receive activities in \cows\ bind neither names nor variables, and this is crucial for allowing concurrent service instances to share (part of) the state. The calculus also supports service fault and termination handling by providing activities to force termination of labelled service instances and to protect service activities from a forced termination.

\subsection{Syntax}

\begin{table}[t]
\footnotesize
\hrule
\begin{tabular}{l@{\hspace{.2cm}}c@{\hspace{.2cm}}l@{\hspace{.1cm}}l@{\hspace{.3cm}}|}
\multicolumn{3}{l}{\textsc{Services:}} & \\
\quad $s$ & ::= & $\out{\xorn}{\xorn'}{\bar \expr}$ & \quad (invoke)\\
& $\sep$ & $\killing{\kl}$ & \quad (kill) \\   
& $\sep$ & $\guard$ & \quad (receive-guarded choice)\\
& $\sep$   & $s \spar s$ & \quad (parallel composition) \\
& $\sep$   & $\prot{s}$ & \quad (protection)\\
& $\sep$   & $\scope{\xnk} s$ & \quad (delimitation)\\
& $\sep$   & $\rep s$ & \quad (replication)\\[.1cm]
\end{tabular}
\begin{tabular}{l@{\hspace{.2cm}}c@{\hspace{.2cm}}l@{\hspace{.1cm}}l}
\vspace*{-.9cm}\\
\multicolumn{4}{l}{\textsc{Receive-guarded choice:}}\\
\quad $\guard$ & ::= & $\nil$ & \quad (nil) \\
& $\sep$   & $\inp{\partner}{\op}{\bar \xorv}.s$ & \quad (request processing) \\
& $\sep$   & $\guard + \guard$ & \quad (choice)\\
\\[.9cm]
\end{tabular}
\hrule
\caption{\cows\ syntax} \label{tab:syntaxCOWS}
\end{table}

The syntax of \cows\ is presented in Table~\ref{tab:syntaxCOWS}. We use three countable disjoint sets: the set of \emph{values} (ranged over by $\val$), the set of `write once' \emph{variables} (ranged over by $\var$), and set of \emph{killer labels} (ranged over by $\kl$). The set of values is left unspecified; however, we assume that it includes the set of partner and operation \emph{names} (ranged over by $\name$, $\partner$, $\op$) mainly used to represent communication endpoints. We also use a set of \emph{expressions} (ranged over by $\expr$), whose exact syntax is deliberately omitted; we just assume that expressions contain values and variables, and do not contain killer labels. As a matter of notation, $\xorv$ ranges over values and variables, $\xorn$ ranges over names and variables, and $\xnk$ ranges over \emph{elements}, i.e. killer labels, names and variables. Notation $\bar \cdot$ stands for tuples, e.g.~$\bar \var$  means $\arr{x_1,\ldots,x_n}$ (with $n \geq 0$), where variables in the same tuple are all distinct. 

\emph{Services} are structured activities built from basic
activities, i.e. the \emph{empty activity} $\nil$,
the \emph{invoke activity} $\out{\und}{\und}{\und}$,
the \emph{receive activity} $\inp{\und}{\und}{\und}$\,, 
and the \emph{kill activity} $\killing{\und}$, 
by means of \emph{prefixing} $\und\, . \und$\,, 
\emph{choice} $\und + \und$\,, 
\emph{parallel composition} $\und \spar \und$\,, 
\emph{protection} $\prot{\und\,}$\,,
\emph{delimitation} $\scope{\und} \und$
and \emph{replication} $\rep \und$\,.
We write $I \define s$ to assign a name $I$ to the term $s$.

\subsection{Informal semantics}

\emph{Invoke} and \emph{receive} are the communication activities. The former permits invoking an operation (i.e., a functionality like a method in object-oriented programming) offered by a service, while the latter permits  waiting for an invocation to arrive. Besides output and input parameters, both activities indicate an endpoint through which communication should occur. 

An \emph{endpoint} $\epc{\partner}{\op}$ can be interpreted as a specific implementation of operation $\op$ provided by the service identified by the logic name $\partner$. The names composing an endpoint can be dealt with separately, as in an asynchronous request-response interaction, where usually the service provider statically knows the name of the operation for sending the response, but not the partner name of the requesting service it has to reply to. Partner and operation names can be exchanged in communication, thus enabling many different interaction patterns among service instances. However, dynamically received names cannot form the endpoints used to receive further invocations (as in \emph{localised \pic}\ \cite{LOCALISEDPI}). In other words, endpoints of receive activities are identified statically because the syntax only allows using names and not variables for them. This design choice reflects  the current (web) service technologies that require endpoints of receive activities to be statically determined.

An invoke $\out{\partner}{\op}{\arr{\expr_{1},\ldots,\expr_{n}}}$ can proceed as soon as all expression arguments are successfully evaluated. A receive $\inp{\partner}{\op}{\arr{\xorv_{1},\ldots,\xorv_{n}}}.s$ offers an invocable operation $\op$ along with a given partner name $\partner$, thereafter the service continues as $s$. An inter-service communication between these two activities takes place when the tuple of values $\arr{\val_{1},\ldots,\val_{n}}$, resulting from the evaluation of the invoke argument, matches the template $\arr{\xorv_{1},\ldots,\xorv_{n}}$ argument of the receive. This causes a substitution of the variables in the receive template (within the scope of variables declarations) with the corresponding values produced by the invoke. 

Communication is asynchronous, as in \klaim. This results from the syntactic constraints that invoke activities cannot be used as prefixes and choice can only be guarded by receive activities (as in \emph{asynchronous \pic}\ \cite{ACS98}). Indeed, in service-oriented systems, communication is usually asynchronous, in the sense that (i) there may be an arbitrary delay between the sending and the receiving of a message, (ii) the order in which messages are received may differ from that in which they were sent, and (iii) a sender cannot determine if and when a sent message will be received.

The \emph{empty} activity does nothing, while \emph{choice} permits selecting for execution one between two alternative receives.

Execution of \emph{parallel} services is interleaved. However, if more matching receives are ready to process a given invoke, only one of the receives that generate a substitution with smallest size (in terms of number of variable-value replacements) is allowed to progress (namely, execution of this receive takes precedence over that of the others). This mechanism permits to model the precedence of a service instance over the corresponding service specification when both of them can process the same request, and enables a sort of blind-date conversation joining strategy \cite{SOCA}. 

\emph{Delimitation} is the only binding construct: $\scope{\xnk} s$ binds the element $\xnk$ in the scope $s$. According to its first argument, delimitation is used for three different purposes: (i) to regulate the range of application of substitutions produced by communication, when the delimited element is a variable; (ii) to generate fresh names, when the delimited element is a name; (iii)  to confine the effect of a kill activity, when the delimited element is a killer label. The scope of names can be dynamically extended, in order to model the communication of private names, as done with the restriction operator in \emph{\pic}\ \cite{PICALC}. Instead, killer labels cannot be dynamically extended, because the activities whose termination would be forced by the execution of a kill need to be statically determined.

The \emph{kill} activity forces immediate termination of all the concurrent activities not enclosed within the \emph{protection} operator. To faithfully model fault and termination handling of SOC applications, kill activities are executed eagerly with respect to the communication activities enclosed within the delimitation of the corresponding killer label.

Finally, the \emph{replication} construct $\rep s$ permits to spawn in parallel as many copies of $s$ as necessary. This, for example, is exploited to implement recursive behaviours and to model business process definitions, which can create multiple instances to serve several requests simultaneously.

\subsection{A taste of the operational semantics}

\begin{table}[t!]
$$
\begin{array}{@{\ }l@{\hspace*{-.5cm}}r}
\hline
\\[-8pt]
\infer[\rulelabel{inv}]{
\out{\partner}{\op}{\bar \expr} \transition{\outl{\partner}{\op\,}{\,\bar \val}} \nil
}{
\eval{\bar \expr}=\bar \val}
&
\raisebox{.4cm}{$\inp{\partner}{\op}{\bar {\xorv}}.s \transition{\inpl{\partner}{\op\,}{\,\bar{\xorv}}} s
\; \rulelabel{rec}$}
\\[.35cm]
\multicolumn{2}{c}{
\infer[\rulelabel{com}]{
s_1 \spar s_2 \transition{\substl} s_1' \spar s_2'
}{
s_1 \transition{\inpl{\partner}{\op\,}{\,\bar \xorv}} s_1'
&& s_2 \transition{\outl{\partner}{\op\,}{\,\bar \val}} s_2'
&& \matchCows{\bar \xorv}{\bar \val}\!=\!\substl}
}
\\[.35cm]
\infer[\rulelabel{del$_{com}$}]{
\scope{\var} s \transition{\substl} s' \!\cdot \substi{\assoc{\var}{\val}}
}{
\raisebox{.1cm}{$s \transition{\substl \,\uplus\, \substi{\assoc{\var}{\val}}} s'$}}
&
\infer[\rulelabel{prot}]{
\prot{s} \transition{\alpha} \prot{s'}
}{
\raisebox{.1cm}{$s \transition{\alpha} s'$}}
\\[.35cm]
\raisebox{.4cm}{$\killing{\kl} \transition{\kl} \nil \; \rulelabel{kill}$}
&
\infer[\rulelabel{par$_{kill}$}]{
s_1 \spar s_2 \transition{\kl} s_1' \spar \halt{s_2}
}{
s_1 \transition{\kl} s_1'}
\\[.35cm]
\infer[\rulelabel{del$_{kill}$}]{
\scope{\kl} s \transition{\taukill} \scope{\kl} s'
}{
\raisebox{.1cm}{$s \transition{\kl} s'$}}
&
\infer[\rulelabel{del}]{
\scope{\xnk} s \transition{\alpha} \scope{\xnk} s'
}{
s \transition{\alpha} s'
& \xnk \!\notin \dset{\alpha}
& \alpha \neq \kl,\taukill
& \nokill{s}{\xnk}}
\\[.1cm]
\hline
\end{array}
$$
\vspace*{-.5cm}
\caption{\cows\ operational semantics (selected rules)}
\label{tab:COWSsemantics}
\end{table}

The operational semantics of \cows\ is defined only for
\emph{closed} services, i.e.~services without free variables
and killer labels.
As usual, the semantics is formally given in terms of a structural congruence
and of a labelled transition relation.
The former identifies syntactically different services that intuitively represent 
the same service. Its definition is standard, except for the scope extension 
laws that permit to extend the scope of names (as in the \pic) and variables, thus enabling
possible communication, but prevent extending the scope of killer labels.

We report in Table~\ref{tab:COWSsemantics} an excerpt of 
the operational rules defining the labelled transition relation. 
We comment on the rules below. 

A service invocation can proceed only if the expressions in the
argument can be evaluated (rule \rulelabel{inv}).
To this aim, we use the evaluation function $\eval{\und}$ that
takes a closed expression and returns the corresponding value. 
This function is not explicitly defined, since the exact syntax of 
expressions is deliberately not specified.
%
%
A receive activity offers an invocable operation along a given
partner name (rule \rulelabel{rec}).
Communication can take place when two parallel services perform
matching receive and invoke activities (rule \rulelabel{com}).
We use here the partial function $\matchCows{\und\,}{\und}$ for
performing \emph{pattern-matching} on semi-structured data 
(\`a la \klaim). Pattern-matching permits to determine if a receive and 
an invoke over the same endpoint can synchronise. 
When tuples $\bar \xorv$ and $\bar
\val$ do match, $\matchCows{\bar \xorv}{\bar \val}$ returns a
substitution $\substl$ for the variables in $\bar \xorv$; otherwise, it is
undefined. Substitutions are functions mapping variables to values and are written as collections
of pairs of the form $\assoc{\var}{\val}$. Application of
substitution $\substl$ to $s$, written $s \cdot \substl$, has the
effect of replacing every free occurrence of $\var$ in $s$ with
$\val$, for each $\assoc{\var}{\val} \in \substl$.
The label of a communication transition indicates the generated substitution (for subsequent application), rather than a silent action as in most process calculi.
%
%
When the delimitation of a variable $\var$ argument of a receive
involved in a communication is encountered, i.e. the whole scope of
the variable is determined, the delimitation is removed and the
substitution for $\var$ is applied to the term (rule
\rulelabel{del$_{com}$}). Variable $\var$ disappears from the term
and cannot be reassigned a value (for this reason 
\cows 's variables are deemed `write once'). 
We use $\substl_1 \uplus \substl_2$ to denote the union of substitutions 
$\substl_1$ and $\substl_2$ when they have disjoint domains.
%
%

Execution of \emph{parallel} services is interleaved but, if more matching receives are ready to process a given invoke, only one of the receives that generate a substitution with smallest size (in terms of number of variable-value replacements) is allowed to progress (namely, execution of this receive takes precedence over that of the others). This mechanism permits to model the precedence of a service instance over the corresponding service specification when both of them can process the same request (we refer to \cite[Sec.~3.2]{COWS_JAL} for a complete account on this feature), and enables a sort of blind-date conversation joining strategy \cite{SOCA}. For the sake of presentation, we have omitted here this precedence mechanism, thus presenting a simplified version of the operational rules concerning the parallel composition operator.

Activity $\killing{\kl}$ forces termination of all unprotected parallel activities (rules \rulelabel{kill}\ and \rulelabel{par$_{kill}$}) inside 
the innermost enclosing $\scope{\kl}$. Termination of a service $s$ is achieved by means of function $\halt{s}$, which returns the service obtained by only 
retaining the protected activities inside $s$. 
The delimitation $\scope{\kl}$ stops the killing effect 
by turning the transition label $\kl$ into
$\taukill$ (rule \rulelabel{del$_{kill}$}).
Such delimitation, whose existence is ensured by the assumption that the semantics is only defined for closed services, prevents a single service to be capable to stop all the other parallel services, which would be unreasonable in a service-oriented setting (as services are loosely coupled and organized in different administrative domains).
Critical activities can be protected from killing
by putting them into a protection $\prot{\und}$; this
way, $\prot{s}$ behaves like $s$ (rule \rulelabel{prot}). Similarly,
$\scope{\xnk} s$ behaves like $s$ (rule \rulelabel{del}),
except when the transition label $\alpha$ contains $\xnk$, in which
case $\alpha$ must correspond either to a communication assigning a
value to $\xnk$ (rule \rulelabel{del$_{com}$}) or to a kill
activity for $\xnk$ (rule \rulelabel{del$_{kill}$}), or when a
free kill activity for $\xnk$ is active in $s$, in which case only
actions corresponding to kill activities can be executed.
Predicate $\nokill{s}{\xnk}$ is used to check the absence of
a free kill activity: it holds true if either $\xnk$ is
not a killer label, or $\xnk=\kl$ and $s$ cannot immediately perform
a free kill activity $\killing{\kl}$.
In this way, kill activities are executed \emph{eagerly} with
respect to the activities enclosed within the delimitation of the
corresponding killer label.

\subsection{A travel booking scenario}
\label{sec:booking_scenario_in_cows}
We provide here, in an incremental way, the \cows\ specification of our 
travel brokering scenario.


At a high level of abstraction, the travel broker service is rendered in \cows\ as:
$$
\begin{array}{rcl}
\mathit{Broker} & \define &
\rep\scope{\var_{cust},\var_{dates},\var_{dest}}
\inp{\mathit{\pTA}}{\mathit{o_{req}}}{\arr{\var_{cust},\var_{dates},\var_{dest}}} .\\
&& \hspace{3.5cm}
\out{\var_{cust}}{\mathit{o_{resp}}}{\arr{book(\var_{dates},\var_{dest})}}
\end{array}
$$
The replication operator $\rep$ is used here to specify that the service 
is \emph{persistent}, i.e.~capable of creating multiple instances 
to serve several requests simultaneously.
The delimitation operator specifies the 
scope of the variables arguments of the subsequent receive activity on operation $\mathit{o_{req}}$,
used to receive a request message from a customer. Besides dates and 
destination of the travel, this message contains the partner name that the customer 
will use to receive the response, which will be sent by the service by means of the invoke activity 
on operation $\mathit{o_{resp}}$. Booking of hotel and flight is here abstracted by
the (unspecified) expression $book(\var_{dates},\var_{dest})$.

A customer of the broker service is specified as follows:
$$
\begin{array}{rcl}
\mathit{Customer} & \define &
\out{\mathit{\pTA}}{\mathit{o_{req}}}{\arr{p_{c},\val_{dates},\val_{dest}}}
 \spar \scope{x_{travel}}\inp{p_{c}}{\mathit{o_{resp}}}{\arr{x_{travel}}}.s\\
\end{array}
$$
The customer behaviour is specular to that of the broker: it starts with an invoke and 
then waits for a response message containing the travel data. 

The overall specification of the scenario is simply the parallel composition of the two components: 
$
(\mathit{Customer} \spar \mathit{Broker})
$.
Whenever prompted by a client request, the broker service creates an instance to serve that specific request, and is immediately ready to concurrently serve other possible requests. Therefore, the resulting \cows\ term after such a computational step is the following:
$$
\begin{array}{l}
\scope{x_{travel}}\inp{p_{c}}{\mathit{o_{resp}}}{\arr{x_{travel}}}.s
\ \spar\
\mathit{Broker}
\ \spar\
\bgInstanceA{$\out{p_{c}}{\mathit{o_{resp}}}{\arr{book(\val_{dates},\val_{dest})}}$}
\end{array}
$$
The created service instance (highlighted by a grey background) is represented as a service running in parallel with the other terms. Notably, the variables of the invoke activity are instantiated (i.e., replaced) by the corresponding values exchanged in the communication. This invoke activity can now synchronise with the receive activity of the customer, whose execution will then continue as $s$ with $x_{travel}$ replaced by the value resulting from the evaluation of the $book$ expression. 

Let us now consider a more refined specification, where the role of the $book$ expression is played by the interactions with services for flights and hotels searching.
\begin{figure}[t]
\centering
\includegraphics[width=\textwidth]{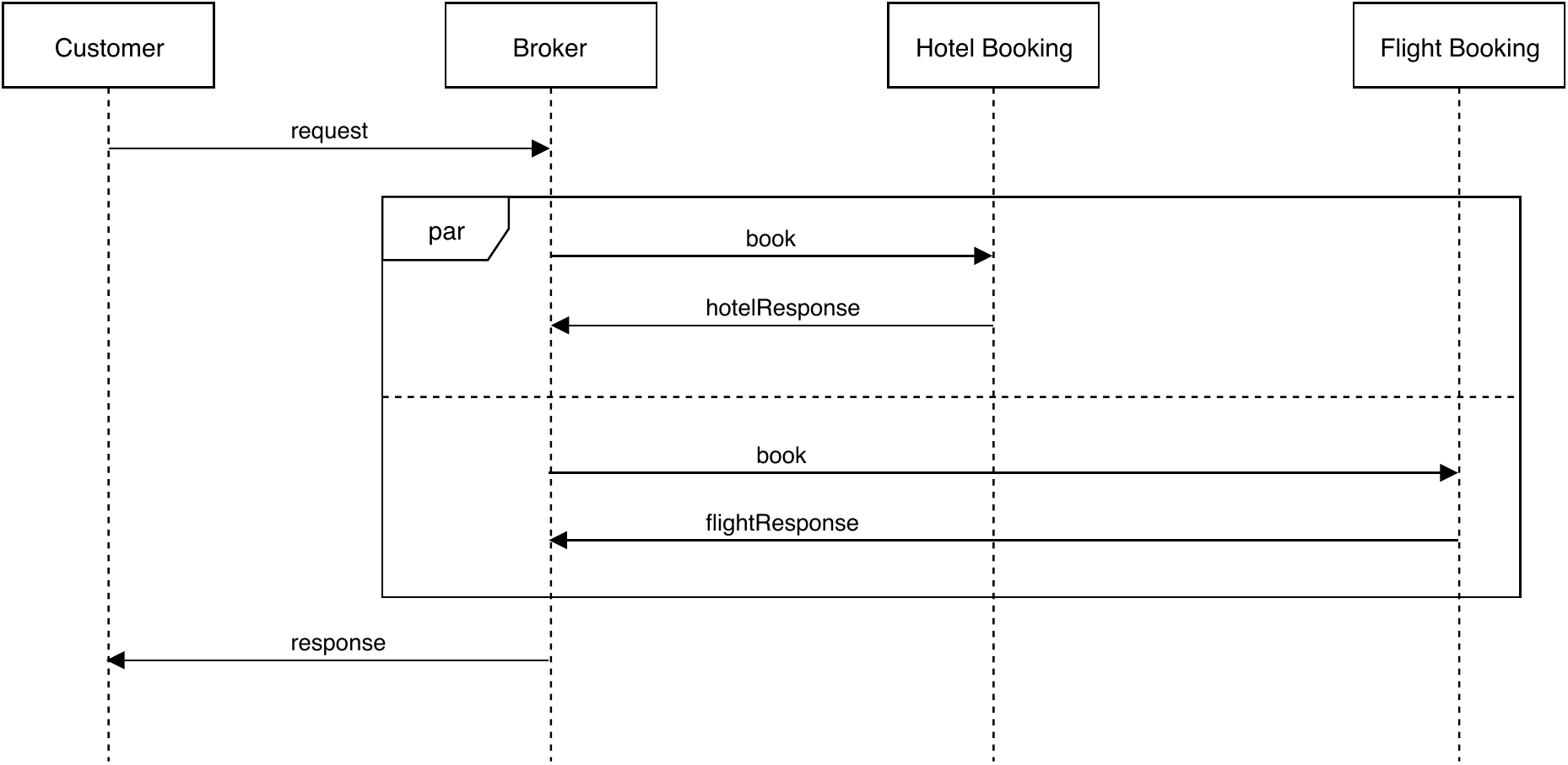}
\caption{Travel Booking Scenario in \cows: Sequence Diagram.}
\label{fig:cows_seq_diag}
\end{figure}
\modif{The interactions between a customer, the (refined) broker and the searching services
are described by the UML activity diagram in Fig.~\ref{fig:cows_seq_diag}. The figure highlights 
that the broker service interacts in parallel with the flights and hotels searching services, 
and that it replies to the customer after both parallel interactions complete.
The refined specification of the broker is the following:}
$$
\begin{array}{@{}r@{\ }c@{\ }l}
\mathit{Broker'} & \define &
\rep\scope{\var_{cust},\var_{dates},\var_{dest}}
\inp{\mathit{\pTA}}{\mathit{o_{req}}}{\arr{\var_{cust},\var_{dates},\var_{dest}}} . \\
&& \hspace{.3cm}
\scope{p,o,\var_{\mathit{flight}},\var_{\mathit{hotel}}}\\ 
&& \hspace{.3cm}
(\ (\out{\mathit{p_{flight}}}{\mathit{o_{book}}}{\arr{\mathit{\pTA},\var_{cust},\var_{dates},\var_{dest}}}\\
&& \hspace{.9cm}
\spar\, \inp{\mathit{\pTA}}{\mathit{o_{fRes}}}{\arr{\var_{cust},\var_{dates},\var_{dest},\var_{\mathit{flight}}}} .\
(\out{p}{o}{\arr{end}} \spar s_f))\\
&& \hspace{.6cm}
\spar (\out{\mathit{p_{hotel}}}{\mathit{o_{book}}}{\arr{\mathit{\pTA},\var_{cust},\var_{dates},\var_{dest}}}\\
&& \hspace{.9cm}
\spar\, \inp{\mathit{\pTA}}{\mathit{o_{hRes}}}{\arr{\var_{cust},\var_{dates},\var_{dest},\var_{hotel}}} .\
(\out{p}{o}{\arr{end}} \spar s_h))\\
&& \hspace{.6cm}
\spar\, 
\inp{p}{o}{\arr{end}}.\ \inp{p}{o}{\arr{end}}.\ 
\out{\var_{cust}}{\mathit{o_{resp}}}{\arr{\var_{\mathit{flight}},\var_{\mathit{hotel}}}}\ )
\end{array}
$$
After the reception of a customer request, 
the service contacts in parallel the two searching services (by invoking the operation $o_{book}$). When the responses from both services are available, the broker service combines them and replies to the customer. To this aim, a private endpoint $\epc{p}{o}$ is exploited: the reception of a message from a searching service triggers an $end$ signal (i.e., an internal message) along the private endpoint, and two of such signals are necessary to trigger the invoke  activity for replying to the customer. 
Suitable expression functions could be used in this last invoke activity for filtering 
the results produced by the searching services.
Notice that the scope of variable $\var_{\mathit{flight}}$ (resp. $\var_{\mathit{hotel}}$) includes not only the continuation $s_f$ (resp. $s_h$) of the service performing the receive, but also the activity for sending the response to the customer. This is different from most process calculi and accounts for easily expressing variables shared among parallel activities within the same service instance, which is a feature typically supported in SOC.

The behaviour of the above service is of particular interest when it is included in a scenario with multiple customers (the specifications of customers and searching services are omitted, we just assume that they follow the communication protocol established by the broker specification):
$$
\mathit{Customer_1} \ \spar\ \mathit{Customer_2} \ \spar\ \mathit{Broker'}
\ \spar\ \mathit{FlightBooking} \ \spar\ \mathit{HotelBooking}
$$
After a certain number of computational steps have taken place, 
we can obtain a system configuration where one instance of the broker service
is created per each customer, and both instances have sent their requests to the searching 
services and are waiting for replies. Now, to send the values resulting from the processing of the request of the first customer, the flight searching service has to perform an invoke activity of the form \mbox{$\out{\mathit{\pTA}}{\mathit{o_{fRes}}}{\arr{p_{c1},\val_{dates},\val_{dest},\val_{\mathit{flight}}}}$}. However, the broker service has two instances waiting for such message along the endpoint \mbox{$\epc{\mathit{\pTA}}{\mathit{o_{fRes}}}$}. In order to deliver the message to the proper instance, i.e. the one serving the request of the first customer, the \emph{message correlation} mechanism is used. In fact, in SOC, it is up to each single message to provide a form of context that enables services to associate the message with the appropriate instance. This is achieved by embedding values, called \emph{correlation data}, in the message itself. Pattern-matching is the mechanism used by the \cows's semantics for locating correlation data. In our example, these data are the customer's partner name, the travel dates and the destination, which have instantiated the corresponding variables in the receive activity \mbox{$\inp{\mathit{\pTA}}{\mathit{o_{fRes}}}{\arr{p_{c1},\val_{dates},\val_{dest},\var_{\mathit{flight}}}}$} within the broker instance serving $\mathit{Customer_1}$.
While the receive of the instance serving the first customer is enabled, 
the one within the other broker instance is not, as it has been instantiated 
with unmatchable values.

Finally, let us provide further details of the broker specification, in order to add fault and compensation handling activities (highlighted by a grey background):

$$
\begin{array}{@{}r@{\ }c@{\ }l}
\mathit{Broker''} & \define &
\rep\scope{\var_{cust},\var_{dates},\var_{dest}}
\inp{\mathit{\pTA}}{\mathit{o_{req}}}{\arr{\var_{cust},\var_{dates},\var_{dest}}} . \\
&& \hspace{.3cm}
\scope{p,o,\var_{\mathit{flight}},\var_{\mathit{hotel}},\bgInstanceA{$\kl$}}\\ 
&& \hspace{.3cm}
(\ (\out{\mathit{p_{flight}}}{\mathit{o_{book}}}{\arr{\mathit{\pTA},\var_{cust},\var_{dates},\var_{dest}}}\\
&& \hspace{.9cm}
\spar\, \inp{\mathit{\pTA}}{\mathit{o_{fRes}}}{\arr{\var_{cust},\var_{dates},\var_{dest},\var_{\mathit{flight}}}} .\\
&& \hspace{1.7cm}
(\out{p}{o}{\arr{end}} \spar s_f 
\\
&& \hspace{1.9cm}
\spar \bgInstanceA{$\prot{\inp{p}{o}{\arr{comp}}.\, \out{\mathit{p_{flight}}}{\mathit{o_{cancel}}}{\arr{\var_{cust},\var_{dates},\var_{dest}}} }$})\\
&& \hspace{1.2cm}
\bgInstanceA{$+\, \inp{\mathit{\pTA}}{\mathit{o_{fFault}}}{\arr{\var_{cust},\var_{dates},\var_{dest}}} .$}\\
&& \hspace{1.7cm}
\bgInstanceA{$(\killing{\kl} \spar \prot{\out{p}{o}{\arr{comp}} \spar \out{p}{o}{\arr{\mathit{fault}}} })$})\\
&& \hspace{.6cm}
\spar (\out{\mathit{p_{hotel}}}{\mathit{o_{book}}}{\arr{\mathit{\pTA},\var_{cust},\var_{dates},\var_{dest}}}\\
&& \hspace{.9cm}
\spar\, \inp{\mathit{\pTA}}{\mathit{o_{hRes}}}{\arr{\var_{cust},\var_{dates},\var_{dest},\var_{\mathit{hotel}}}} .\\
&& \hspace{1.7cm}
(\out{p}{o}{\arr{end}} \spar s_h 
\\
&& \hspace{1.9cm}
\spar \bgInstanceA{$\prot{\inp{p}{o}{\arr{comp}}.\, \out{\mathit{p_{hotel}}}{\mathit{o_{cancel}}}{\arr{\var_{cust},\var_{dates},\var_{dest}}} }$})\\
&& \hspace{1.2cm}
\bgInstanceA{$+\, \inp{\mathit{\pTA}}{\mathit{o_{hFault}}}{\arr{\var_{cust},\var_{dates},\var_{dest}}} .$}\\
&& \hspace{1.7cm}
\bgInstanceA{$(\killing{\kl} \spar \prot{\out{p}{o}{\arr{comp}} \spar \out{p}{o}{\arr{\mathit{fault}}} })$})\\
&& \hspace{.6cm}
\spar\, 
\inp{p}{o}{\arr{end}}.\ \inp{p}{o}{\arr{end}}.\ 
\out{\var_{cust}}{\mathit{o_{resp}}}{\arr{\var_{\mathit{flight}},\var_{\mathit{hotel}}}}\\ 
&& \hspace{.6cm}
\bgInstanceA{$\spar\, 
\prot{\inp{p}{o}{\arr{\mathit{fault}}}.\ 
\out{\var_{cust}}{\mathit{o_{fault}}}{\arr{}}}$ 
}\ )
\end{array}
$$
Now, when a positive  response from a searching service is received, a compensation handler is installed. This consists of an invoke activity on operation $\mathit{o_{cancel}}$, triggered by a $comp$ signal, devoted to cancel the booking. If a negative response on $\mathit{o_{fFault}}$ (resp. $\mathit{o_{hFault}}$) is received, the normal execution of the service is immediately terminated (by means of the $\mathbf{kill}$ activity), the activity compensating the hotel (resp. flight) booking is activated, if installed, and a $\mathit{fault}$ signal is emitted. This last signal triggers the execution of the fault handler, consisting of an invoke activity for notifying the customer that the request booking is failed. Notably, fault and compensation activities are enclosed within protection blocks, in order to protect them from the killing effect of the $\mathbf{kill}$ activities.
 

\modif{
\paragraph{Discussion}
Most of the distinguishing features of \cows find their full application in the final specification of the travel booking scenario. Let us focus on the $\mathit{Broker''}$ service. The replication operator is used to allow the broker service to create multiple instances. In particular, an instance is created for each received customer request. Pattern-matching (on the correlation values replacing variables $\var_{cust}$, $\var_{dates}$, and $\var_{dest}$) is then used to associate each message from the searching services to the appropriate broker instance. The delimitation operator is used for different purposes: to define the scope of the correlation variables; to make the endpoint \mbox{$\epc{\mathit{\partner}}{\mathit{\op}}$} private; to share variables $\var_{\mathit{flight}}$ and $\var_{\mathit{hotel}}$ among the parallel terms within the scope of the inner delimitation operator, and to limit the scope of the $\killing{\kl}$ actions. The protection operator, instead, is used to protect the fault and compensation handlers from the killing effect. 
}

\subsection{Programming environment}

To effectively program SOC applications, \cows, originally conceived as a process calculus, has been extended with high-level features, such as standard control flow constructs (i.e., sequentialisation, assignment, conditional choice, iteration) and a scope activity explicitly defining fault and compensation handlers. The implementation of the resulting orchestration language, called \lwsbpel\ \cite{LapadulaPT12}, is based on a software tool \cite{ACR} supporting a rapid and easy development of SOC applications via the translation of service orchestrations written in \lwsbpel\ into executable \wsbpel\ programs. More specifically, a \lwsbpel\ program given as input to this tool also includes a declarative part, containing the variable types and the physical service bindings, necessary for generating the corresponding WSDL document and the process deployment descriptor. These files, together with the one containing the \wsbpel\ code, are organised in a package that can be deployed and executed in a \wsbpel\ engine. 

\modif{In Fig.~\ref{lis:cows_code} we report the relevant code\footnote{\modif{The \lwsbpel source code for the complete scenario can be downloaded from \url{https://bitbucket.org/tiezzi/jlamp_survey_code/src/master/COWS/blite_code/}.}} of the \lwsbpel implementation of the \cows specification, presented in Section~\ref{sec:booking_scenario_in_cows}, of the travel booking scenario. 
Despite the use of a different notation, the invoke (\textbf{inv}) and receive (\textbf{rcv}) primitives of \lwsbpel acts similarly to the \cows' ones. To ease the programming task, \lwsbpel also
provides the high-level features for 
sequential (\textbf{seq} \_ ; \ldots ; \_ \textbf{qes}),
and parallel (\textbf{flw} \_ | \ldots | \_ \textbf{wlf}) composition.
These permit
avoiding the interactions along the private endpoint \mbox{$\epc{\mathit{\partner}}{\mathit{\op}}$}.
The last line of the listing shows a \emph{deployment} definition, 
which associates the correlation set {\{x\_cust, x\_dates, x\_dest\}}
to the broker service. The declarative part of this \lwsbpel program, 
specifying the configuration data necessary to produce the corresponding
\wsbpel program, is omitted. 
}

\begin{figure}[!t]
\lstset{language=blite,style=mystyle}
\begin{lstlisting}{}
s_broker ::= [ seq
                 rcv <p_br,cb_cust> o_req <x_cust,x_dates,x_dest>;
                 flw
                   seq
                     inv <p_flight,cb_p_flight> o_fBook <x_dates,x_dest,x_cust>;
                     rcv <cb_p_flight> o_fBook <x_dates,x_dest,x_cust,x_flight>	           
                   qes
                   |
                   seq
                     inv <p_hotel,cb_p_hotel> o_hBook <x_dates,x_cust,x_dest>;
                     rcv <cb_p_hotel> o_hBook <x_dates,x_cust,x_dest,x_hotel>
                   qes
                 wlf;
                 inv <cb_cust> o_req <x_flight,x_hotel>
               qes ];;

broker_service ::= {s_broker}{x_cust,x_dates,x_dest};;
\end{lstlisting}
\vspace*{-.3cm}
\caption{The service $\mathit{Broker}'$ of the travel booking scenario implemented in \lwsbpel.}
\label{lis:cows_code}
\end{figure}

%
%
%

\subsection{Verification techniques}

The main verification techniques devised for \cows\ specifications are the following: (i) a type system for checking confidentiality properties \cite{LPT07:FSEN}, which uses types to express and enforce policies for regulating the exchange of data among services; (ii) a bisimulation-based observational semantics \cite{PTY:ICALP09}, which permits to check interchangeability of services and conformance against service specifications; (iii) a verification methodology for checking functional properties specific  of SOC systems \cite{FGLMPT08:TOSEM}.

Concerning the third technique, the properties are described by means of \socLogic, a logic specifically designed to express in a convenient way distinctive aspects of services, such as, e.g., acceptance of a request, provision of a response, and correlation among service requests and responses. The verification of \socLogic\ formulae over \cows\ specifications is assisted by the on-the-fly model checker \cowsmc. This approach has been used in \cite{FGLMPT08:TOSEM,MasiPT09,GnesiPT11} to verify some properties of interest of an automotive scenario, an e-Health authentication protocol, and a finance case study, respectively. \cowsmc\ can also be used as an interpreter for \cows: it takes a \cows\ term as an input and analyses it syntactically; if the analysis succeeds, the tool allows the user to interactively explore the computations arising from the term.


\modif{Using the  \socLogic logic and the \cowsmc tool, we were able to specify and verify some 
properties of the broker service of the travel booking scenario, like:
\begin{itemize}
\item The service always gives a response to a request.
\end{itemize}
that is expressed by the \socLogic formula
\begin{center}
\begin{tabular}{l}
\texttt{AG [request(travelBooking)]}
\\
\texttt{\ \ \ AF \{responseOk(travelBooking)} 
\\
\texttt{\ \ \ \ \ \ \  or responseFail(travelBooking)\} true}
\end{tabular}
\end{center}
The above property is satisfied by the broker service, as well as the following ones:
\begin{itemize}
\item The service is permanently available, i.e. it is always available to accept new requests. 
\item It is possible to cancel a request after a successful response.
\end{itemize}
Instead, the following properties are not satisfied:
\begin{itemize}
\item The service is sequential, i.e. it will not be available at least until a response
is provided. 
\item The service is reliable, i.e. it always responds positively.
\end{itemize}
For the latter two properties, \cowsmc can show a counterexample, i.e. a clear 
and detailed explanation of the negative verification result. 
}

\subsection{Related work}

%

\modif{\cows covers typical aspects of SOC technologies, such as 
service instances and their interactions, delivery of correlated messages,
concurrent activities, multiple start activities, receive conflicts, 
long-running business transactions. 
Linguistic variants of \cows\ have been subsequently introduced to incorporate 
other aspects, initially not supported by the calculus, such as timed business process 
activities \cite{LPT07:ICTAC} and dynamic service discovery and negotiation 
mechanisms \cite{LPT08:WWV}. 
Other variants have also been devised to enable quantitative reasoning on 
service behaviours \cite{PQ07:ICSOC,PQZ08:COORDINATION}.

Many other formalisms for SOC have been defined as enrichments of 
existing process calculi with constructs inspired by those of 
\wsbpel. For example, the $\pi$-calculus has been extended in \cite{WebPi,WebPiAtWork,WEBPIINF,MAZ_LUC}  to study 
a simplified version of the `scope' construct of \wsbpel,
while CSP has been extended with compensation mechanisms in \cite{CCSP}. 
Differently from these works, when designing \cows the starting point was the 
technological perspective provided by the OASIS standard \wsbpel and the 
related web service technology. Indeed, \cows aims at conveying, in 
a distilled form, different key notions underlying SOC, to make  
a direct representation of SOC applications possible. 

Another large body of work on modelling of SOC interactions relies on the  explicit use of interaction sessions (by exploiting private channels \`a la \pic). This is the case, e.g., of the SCC \cite{SCC} and Caspis \cite{Caspis} formalisms. \cows, instead, does not provide an explicit representation of sessions, which anyway can be identified by correlating the related messages by their content. This is the approach fostered by the SOC technology (in particular by \wsbpel), as it is more robust and  fits better with the loosely coupled nature of SOC than that based on explicit session references.

The formalism closest to \cows\ is, perhaps, SOCK \cite{SOCK06}, as it also relies on a form of correlation-based communication, but \cows is more amenable to formal reasoning, as it has a much simpler operational semantics. SOCK is a three-layered calculus, which decomposes a SOC specification into three parts: the behaviour (process description), the declaration (concerning execution modalities, e.g. concurrent vs. sequential) and the composition (consisting of the parallel composition of service engines). \cows\ represents a more foundational formalism than SOCK, in that it does not explicitly consider the declaration layer. The interested reader is referred to \cite{COWS_JAL} for a description of the way  services' execution modalities can be rendered in \cows. 

For other references on service-oriented computing and relation with \cows we refer the interested reader to \cite{COWS_JAL,sensoriaBook}.
}

\modif{
More recently, a new architectural style for structuring applications as a collection of services is emerging. It is called \emph{microservices} and shares with SOC many design principles, e.g. complex distributed applications structured in terms of loosely coupled, independent and interoperable components. However, microservices are more lightweight than web services from the technological point of view, e.g. they interact asynchronously by directly using the HTTP protocol. In particular, microservices do not exploit the OASIS standard \wsbpel\ for their orchestration, which is instead at the basis of the design of \cows.
Therefore, in the first instance, \cows\ does not seem suitable for directly modelling microservices, but we leave a more thorough investigation of this issue for future work.
}

\section{\textbf{\textsc{Scel}}: Software Component Ensemble Language}
\label{sec:scel}

Developing massively distributed and highly dynamic computing systems which control and interact with the physical world is a major challenge in todays software engineering. Difficulties arise from the open-ended and dynamic nature of large-scale systems, 
the unpredictable external environment, the limited if not impossible human intervention, and the need of ensembles of components to interact and collaborate for achieving specific goals, while hiding complexity to end-users. 
A possible answer to the problems posed by such systems is to make them \emph{self-aware} and \emph{context-aware},  and able to 
\emph{self-adapt} and to \emph{self-configure}.  
These and other \emph{self-management} capabilities, like self-healing, self-optimisation, and self-protection, characterise \emph{autonomic computing}~\cite{KC03} systems. In order to achieve these goals, it is needed that these systems continuously monitor their progress and the environment they are working in, to determine the actions to perform and the components to install for better dealing with the current status of affairs.

Specific support to programming these systems is provided by \SCEL\ (\emph{Software Component Ensemble Language})~\cite{DLPT14,SCEL-ASCENSbook15} a formal language equipped with a set of linguistic abstractions for specifying the behaviour of components, the interaction among them, and the dynamic formation of their \emph{ensembles}. In \SCEL, components are computational entities that have associated \emph{knowledge} repositories and specific behavioural \emph{policies}. They also have an interface used to expose their characterising features (\emph{attributes}). Ensembles, in turn, are aggregations of interacting 
components that are determined at run-time by means of predicates over components' attributes. 

\SCEL\ components acquire information about their status (self-awareness) and about their environment (context-awareness) from knowledge repositories. Moreover, by exploiting awareness and the capability of adding processes to knowledge repositories and of dynamically activating processes and components, they can guarantee self-adaptation and self-configuration, initiate self-healing actions in presence of faults, and activate self-optimisation strategies. Finally, by using suitable policy languages, components can attain data integrity and self-protection against, e.g., unauthorised accesses or denial-of-service attacks. 

\modif{
A system of collaborative \SCEL\ components can thus monitor its state and its components, as well as the execution context, and identify relevant changes that may affect the achievement of its goals or the fulfilment of its requirements. The system can  then plan reconfigurations to meet the new functional or non-functional requirements, execute them, and monitor that its goals are achieved, possibly without any interruption. Attributes are key to support this autonomic behaviour; they are used to expose distinguishing features of \SCEL\ components and to indicate occurrence of specific events. The set of all exposed attributes constitutes the common knowledge that is updated during computations and is used to build patterns of communication to enable \SCEL\ components to dynamically organise themselves into ensembles.

A careful choice and use of (sets of) attributes permit to express autonomic behaviours in \SCEL\  in a natural way as done, e.g., in~\cite{CNPPTZ13} for modelling component- and ensemble-level adaptation patterns, and in~\cite{CCCNPTZ14} for offering \emph{self-expression}, i.e. the ability of changing at run-time the coordination pattern used in an ensemble. \SCEL\ has also been used for effectively modelling autonomic systems from different application scenarios such as, e.g., swarm robotics~\cite{FACS13,DLPT14}, cooperative e-vehicles \cite{SASOAWARENESS2013E-Mobility}, service provision and cloud-computing~\cite{SCEL-FMCO11,ATC13,SASOAWARENESS2013CloudPaper}. 
}

\subsection{Syntax}
\label{sec:scelsyntax}

\SCEL\ syntax is reported in Table~\ref{SCELtsyntax}. Five pairwise disjoint sets are used: \emph{Names} (ranged over by $\naddr$, $\naddr'$, \ldots),  \emph{Predicate names} (ranged over by $\enspred{p}$, $\enspred{p'}$~\ldots), \emph{Variables for names} (ranged over by $\vaddr$, $\vaddr'$,~\ldots), \emph{Variables for processes} (ranged over by $X$, $Y$,~\ldots), and \emph{Parameterised  process identifiers} (ranged over by $A$, $B$~\ldots). The distinguished variable $\self$ is used to denote the name of the component using it.

\textsc{Systems} are obtained by aggregating \textsc{Components} which, in turn, are obtained by aggregating \textsc{Knowledge} and \textsc{Processes}, according to some \textsc{Policies}. \textsc{Processes} specify the flow of the \textsc{Actions} that can be performed and use the same composition operators as in \klaim. \textsc{Actions} can have a \textsc{Target} to determine the components, in addition to the subject one, that are involved in that action. 

\SCEL\ is parametric with respect to some syntactic categories, namely \textsc{Policies}, \textsc{Knowledge}, \textsc{Templates} and \textsc{Items} (with the last two determining the part of \textsc{Knowledge} to be retrieved/removed or added, respectively). This choice permits integrating different approaches to policy specification and knowledge handling within \SCEL, like, e.g., the \emph{access control policies} of~\cite{ATC13} and the \emph{constraint stores} of~\cite{ccSCEL15}. 

A simple, yet expressive, instance of \SCEL, named \SCELlight, has been introduced in~\cite{SCELigth14} where policies are absent (equivalently, where any process action is authorised) and knowledge repositories are implemented as tuple spaces \emph{\'a la} \klaim. A full-fledged instance, named \PSCEL\ (\emph{Policed} \SCEL), has been introduced in~\cite{SCEL-ASCENSbook15}. In \PSCEL, knowledge repositories are again implemented as \klaim\ tuple spaces, while policies are expressed by means of a simplified version of FACPL (Formal Access Control Policy Language, \cite{FACPL:TSE}), a language for defining access control, resource usage and adaptation policies.

\begin{table}[t]
\footnotesize
\hrule
\begin{tabular}{l@{\hspace{.2cm}}c@{\hspace{.2cm}}l@{\hspace{.1cm}}l@{\hspace{.3cm}}|}
\multicolumn{3}{l}{\textsc{Systems:}} & \\
\quad $S$ & ::= & $C$ & \quad (component) \\
& $\sep$ & $S_1 \parcomp S_2$ & \quad (composition) \\
& $\sep$ & $\res \naddr S$ & \quad (name restriction) \\[.1cm]
\multicolumn{3}{l}{\textsc{Components:}} & \\
\quad $C$ & ::= & $\loccomp{{\cal I}}{\knw,\pols,P}$ & \quad (single component) \\[.1cm]
\multicolumn{3}{l}{\textsc{Processes:}} & \\
\quad $P$ & ::= & $\procnil$ & \quad (inert) \\
& $\sep$ & $a.P$ & \quad (action prefixing) \\
& $\sep$ & $P_1 \choice P_2$ & \quad (choice) \\
& $\sep$ & $\auth {P_1} {P_2}$ & \quad (composition) \\
& $\sep$ & $X$ & \quad (process variable) \\
& $\sep$ & $A(\bar{p})$ & \quad (invocation) \\[.1cm]
\end{tabular}
\begin{tabular}{l@{\hspace{.2cm}}c@{\hspace{.2cm}}l@{\hspace{.1cm}}l}
\vspace*{-.9cm}\\
\multicolumn{4}{l}{\textsc{Actions:}}\\
\quad $a$ & ::= & $\acgetp T \nvaddr$ & \quad (withdraw) \\
& $\sep$ & $\acreadp T \nvaddr$ & \quad (retrieve) \\
& $\sep$ & $\acputp t \nvaddr$ & \quad (addition)\\
& $\sep$ & $\acfreshp n$ & \quad (scope) \\
& $\sep$ & $\acnewp {\interf{I},\knw,\pols,P}$ & \quad (new) \\[.1cm]
\multicolumn{4}{l}{\textsc{Targets:}}\\
\quad $\nvaddr$ & ::= & $n$ & \quad (name) \\
& $\sep$ & $x$ & \quad (variable) \\
& $\sep$ & $\self$ & \quad (self) \\
& $\sep$ & $\enspred{P}$ & \quad (predicate) \\
& $\sep$ & $\enspred{p}$ & \quad (pred. name) \\[.1cm]
\end{tabular}
\hrule
\caption{\SCEL\ syntax (\textsc{Policies} $\pols$, \textsc{Knowledge} $\knw$, \textsc{Templates} $T$, and \textsc{Items} $t$ are parameters of the language)}
\label{SCELtsyntax}
\end{table}

\subsection{Informal semantics}
\label{sec:scelsemop}

A \SCEL\ \textsc{system} in an aggregation of \textsc{components} built by means of the \emph{composition} operator $\_ \parcomp \_\,$, which is both commutative and associative. Within a system, it is also possible to restrict to a subsystem the scope of a name, say $n$, by using the name \emph{restriction} operator $\res \naddr \_$\,. Thus, in a system of the form $S_1 \parcomp \res \naddr S_2$, the effect of the operator is to make name $\naddr$ invisible from $S_1$. 

A \SCEL\  \textsc{component} $\loccomp{{\cal I}}{\knw,\pols,P}$ consists of:
\begin{itemize}
\item An \emph{interface} \interf{I} publishing and making available information about the component itself in the form of \emph{attributes}, i.e. names acting as references to information stored in the component's knowledge repository. Among them, attribute $id$ is mandatory and is bound to the name of the component. 
\item A \emph{knowledge repository} $\knw$ managing both \emph{application data} and \emph{awareness data}, equipped with specific handling mechanisms. Application data are used for determining the progress of components' computations, while awareness data provide information about the environment in which the components are running or about the status of the component itself. 
\item A set of \emph{policies} $\pols$ regulating the interaction between the different processes of the component and the interaction with other components. 
\item A \emph{process} $P$, together with a set of process definitions that can be dynamically activated. 
\end{itemize}

\textsc{Processes} are the active computational units. Each process is obtained by composing subprocesses or the \emph{inert} process $\procnil$ via \emph{action prefixing} ($a.P$), nondeterministic \emph{choice} ($P_1 \!\choice\! P_2$), controlled \emph{composition} ($\auth {P_1} {P_2}$), \emph{process variable} ($X$), and parameterised process \emph{invocation} ($A(\bar{p})$). The semantics of the construct $\auth{P_1} {P_2}$ is another parameter of \SCEL. It can be instantiated  to capture various forms of \emph{parallel composition} commonly used in process calculi. For example, in \PSCEL, it corresponds to the 
\emph{interleaved} execution of the two involved processes. Communication can be \emph{higher-order}, as in \klaim. 
We assume that $A$ ranges over a set of parameterised {\em process identifiers} that are used in (possibly recursive) process definitions. It is also assumed that each process identifier $A$ has a {\em single} definition of the form $A(\bar{f}) \define P$. Lists of actual and formal parameters are denoted by $\bar{p}$ and $\bar{f}$, respectively.

Processes can perform five different kinds of \textsc{actions}. Actions $\acgetp T \nvaddr$, $\acreadp T \nvaddr$ and $\acputp t \nvaddr$ are used to manage shared knowledge repositories by \emph{withdrawing/retrieving/adding} information items from/to the knowledge repositories identified by the target $\nvaddr$. These actions exploit templates $T$ as patterns to select knowledge items $t$ from the repositories. They depend on the chosen kind of knowledge repository (a parameter of \SCEL, as pointed out earlier) and are implemented by invoking the provided knowledge handlers. Action $\acfreshp n$ introduces a \emph{scope} restriction for the name $n$ to guarantee that this name is \emph{fresh}, i.e., different from any other previously used name. Action $\acnewp {\interf{I},\knw,\pols,P}$ creates a \emph{new} component $\loccomp{{\cal I}}{\knw,\pols,P}$.

Actions $\acget$/$\acread$ may block the executing process to wait for the expected item, in case it is not (yet) available in the knowledge repository. The two actions differ for the fact that $\acget$ removes the found item from the knowledge repository while $\acread$ leaves the target repository unchanged. Actions $\acput$, $\acfresh$ and $\acnew$ are instead immediately executed (provided that their execution is allowed by the policies in force).

To indicate the target $\nvaddr$ of an action, in addition to names and variables for names, the process performing the action can also use the distinguished variable $\self$, to refer to the name of the hosting component, or a \emph{predicate} $\enspred{P}$ (or the name $\enspred{p}$ of a predicate), to specify the \emph{ensemble} of all those components with which the process wants to interact. Predicates are boolean-valued expressions obtained by logically combining relations involving attributes.
Thus, e.g., actions $\acputp t n$ and $\acputp t {\enspred{P}}$ give rise to two different primitive forms of communication: the former is a \emph{point-to-point} communication, while the latter is a sort of \emph{group-oriented} communication. 

It is worth noticing that the group-oriented variant of action $\acput$ is used to insert a knowledge item in the repositories of \emph{all} components belonging to the ensemble identified by the target predicate. Differently, the group-oriented variants of actions $\acget$ and $\acread$ withdraw and retrieve, respectively, an item from a \emph{single} component non-deterministically selected among those satisfying the target predicate.

\subsection{A taste of the operational semantics}
\label{sec:scelexcerpt}

The semantics of \SCEL\ is defined by means of a few (Labeled) Transition Systems. In this section we introduce the inference rules, shown in Table~\ref{SCELtsystemsopsem1}, of the labeled transition relation defining the behaviour of all the variants of the action \acput: that for \emph{point-to-point} communication, which can be either local (rule \rulelabel{lput}) or remote (rule \rulelabel{ptpput}), and that for group-oriented communication (rules \rulelabel{grput} and \rulelabel{engrput}). The semantics of the other actions is defined similarly. We refer the interested reader to \cite{DLPT14,SCEL-ASCENSbook15} for a full account of \SCEL\ operational semantics.

\begin{table}[t!]
$$
\begin{array}{@{}c@{}}
\hline
\\[-8pt]
\infer[\rulelabel{put}]{
\loccomp{I}{\knw,\pols,P}
\ensarrow{\interf{I}:\doPut{t'}{\evtarg}} 
\loccomp{I}{\knw,\bullet,P'}
}{
P \procarrow{\acputp t \nvaddr} P' &&
\valt{t}{\interf{I}}=t' &&
\valn{c}{\interf{I}}=\evtarg 
}
\\[.35cm]
\infer[\rulelabel{accput}]{
\loccomp{J}{\knw,\pols,P} \ensarrow{\interf{I}:\acceptPut{t}{\interf{J'}}} \loccomp{J}{\knw',\pols',P}
}{
\pols \allow \interf{I}:\acceptPut{t}{\interf{J}},\pols' &&
\interf{J'} = \interf{J}\replace{\interf{J}.\pi}{\pols'} &&
\knw' = \knw\oplus t
}
\\[.3cm]
\infer[\rulelabel{lput}]{
C \ensarrow{\tau} C''
}{
C \ensarrow{\interf{I}:\doPut{t}{\naddr}} C' &&
\naddr = \interf{I}.\texttt{id} &&
C'[\interf{I}.\pi/\bullet] \ensarrow{\interf{I}:\acceptPut{t}{\interf{I}}} C'' 
}
\\[.25cm]
\infer[\rulelabel{ptpput}]{
S_1\parcomp S_2\ensarrow{\tau} S_1'{[\pols'/\bullet]}\parcomp S_2'
}{
S_1\ensarrow{\interf{I}:\doPut{t}{n}} S_1' &&
S_2\ensarrow{\interf{I}:\acceptPut{t}{\interf{J}}} S_2' &&
\interf{J}.\texttt{id} = \naddr &&
{\interf{I}.\pi \allow \interf{I}:\acceptPut{t}{\interf{J}},\pols'}
}
\\[.35cm]
\infer[\rulelabel{grput}]{
S_1\parcomp S_2\ensarrow{\interf{I}\replace{\interf{I}.\pi}{\pols'}:\doPut{t}{\enspred{P}}} S_1'\parcomp S_2'
}{
S_1\ensarrow{\interf{I}:\doPut{t}{\enspred{P}}} S_1' &&
S_2\ensarrow{\interf{I}:\acceptPut{t}{\interf{J}}} S_2' &&
\interf{J} \models \enspred{P} &&
\interf{I}.\pi \allow \interf{I}:\acceptPut{t}{\interf{J}},\pols'
}
\\[.35cm]
\infer[\rulelabel{engrput}]{
S\parcomp \loccomp{J}{\knw,\pols,P} \ensarrow{\interf{I}:\doPut{t}{\enspred{P}}} S'\parcomp \loccomp{J}{\knw,\pols,P}
}{
S\ensarrow{\interf{I}:\doPut{t}{\enspred{P}}} S' & \ 
(\interf{J} \not\models \enspred{P}
\,\ \vee \,\ 
\pols \not\allow \interf{I}:\acceptPut{t}{\interf{J}},\pols'
\,\ \vee \,\ 
\interf{I}.\pi \not\allow \interf{I}:\acceptPut{t}{\interf{J}},\pols')
}
\\[.1cm]
\hline
\end{array}
$$
\vspace*{-.5cm}
\caption{Inference rules for the action \acput\ ($\interf{I}$ and $\interf{J}$ range over interfaces)}
\label{SCELtsystemsopsem1}
\end{table}

The execution of actions \acput\ by a component, which is indicated by the transition label $\interf{I}:\doPut{t}{\evtarg}$ generated by rule \rulelabel{put}, requires an appropriate synchronisation with one or more target components allowing addition of the argument item to their local repository. These components perform a fictitious \emph{authorisation action}, which is indicated by the transition label \mbox{$\interf{I}\!:\acceptPut{t}{\interf{J'}}$} generated by rule \rulelabel{accput}. Eventually, the simultaneous execution of these complementary actions gives rise to a \emph{computation step}, which is denoted by a $\tau$-labelled transition generated, e.g., by rules \rulelabel{lput} and \rulelabel{ptpput}.

More specifically, rule \rulelabel{put}\footnote{Actually, this rule is an instantiation and a simplification of the original rule \rulelabel{pr-sys}~\cite{DLPT14,SCEL-ASCENSbook15} for only taking into account actions \acput, as done in the present context.}\ indicates the intention of component $\interf{I}$ to perform an action \acput\ corresponding to a  commitment of the process $P$ running locally which becomes $P'$ after execution of the action. Function $\valp{\cdot}{\interf{I}}$ denotes the evaluation of terms with respect to the interface $\interf{I}$, causing the replacement of the attributes occurring therein by the corresponding information and the replacement of $\self$ by the component name (i.e. $\interf{I}.\texttt{id}$). Therefore, $t'$ is an evaluated item and $\evtarg$ is either the name $n$ of a single component or a predicate $\enspred{P}$ indicating a set of components. Notably, the policy in force at the component performing the action might change. Indeed, after the transition, the component contains a placeholder $\bullet$ in place of the policy; it will be replaced by an actual (possibly new) policy when the target $\evtarg$ of the action is determined (see the rules \rulelabel{lput} and \rulelabel{ptpput}).

Rule \rulelabel{accput} indicates that the policy $\pols$ of component $\interf{J}$ allows component $\interf{I}$ to add the item $t$ to $\interf{J}$'s repository $\knw$ thus getting the repository $\knw'$ (this is denoted by $\knw\oplus t=\knw'$). This control is done by the so called \emph{authorisation predicate}\footnote{The actual definition of this predicate depends on the policy language, which is one of the parameters of \SCEL.}\ which is the first premise of the rule. An effect of this transition is the possible update of policy $\pols$ which becomes $\pols'$, both in the component after the transition and in the generated label (in fact, $\interf{J'}$ is $\interf{J}$ where $\pols'$ replaces the local policy, denoted by $\interf{J}.\pi$). 

In case of point-to-point communication, action \acput\ adds an item either to the local repository  (rule \rulelabel{lput}) or to a remote repository (rule \rulelabel{ptput}). Anyhow, this transition corresponds to a computation step. Notably, the transition labelled by ${\interf{I}:\acceptPut{t}{\interf{I}}}$ in the premise of \rulelabel{lput} can only be produced by rule \rulelabel{accput}; it thus ensures that the component $\interf{I}$ authorises the local addition of the item $t$ and that the component's knowledge and policy are updated accordingly. When the target of the action denotes a specific remote repository \rulelabel{ptput}, the action is only allowed if $\naddr$ is the name of the component $\interf{J}$ simultaneously willing to accept addition of the item to its local repository and if the request to perform the action at $\interf{J}$ is authorised by the local policy (this is checked by the authorisation predicate occurring as last premise). Notably, if there are multiple components with the same name $\naddr$ willing to accept the item, one of them is non-deterministically chosen.

In case of group-oriented communication, in a single transition action $\acput$ adds an item to the repository of each component satisfying its target predicate $\enspred{P}$ and willing to accept it. Indeed, rule \rulelabel{grput} permits the execution of a \acput\ for group-oriented communication when there is a parallel component, say $\interf{J}$, satisfying the target of the action and whose policy authorises this remote access. Relation $\interf{J} \models \enspred{P}$ states that predicate $\enspred{P}$ holds true when evaluated after replacement of the attributes occurring therein with the information to which these attributes are bound within the interface $\interf{J}$. The exact definition of such relation depends on the used predicates. Of course, the action \acput\ must be authorised to use $\interf{J}$ as a target also by the policy in force at the component performing the action (the policy is updated after each evaluation of the authorisation predicate). Notably, the label of the inferred transition is yet that of a \acput\ for group-oriented communication, thus further authorisation actions performed by other parallel components satisfying the target of the action can be simultaneously executed. In other words, the rule implies that multiple components can be delivered the same item in a single transition. Instead, rule \rulelabel{engrput} means that the capability of a component to perform a \acput\ for group-oriented communication is not affected by those other components not satisfying predicate $\enspred{P}$, \ie not belonging to the ensemble, or not authorised by the executing component or not authorising the action. Thus, when there is a component performing a \acput\ for group-oriented communication, by repeatedly applying rules \rulelabel{grput} and \rulelabel{engrput}, in one way or another all system components get involved in the transition, which can then be turned into a computation step (this is done by a rule of a different transition system not shown here).

\subsection{A travel booking scenario}
\label{sec:scel-trag}
\label{sec:booking_scenario_in_scel}


Our specification of the scenario is written in \PSCEL, which instantiates knowledge repositories, items and templates of the \SCEL\ syntax as shown in Table~\ref{SCEL-TS}. Notably, knowledge \textsc{items} are \emph{tuples}, i.e.\ sequences of values, while \textsc{templates} are sequences of values and formal fields binding variables. Values can be either targets $\nvaddr$, or processes $P$ or, more generally, the result of the evaluation of an expression $e$. We assume that expressions may contain attribute names, \emph{boolean}, \emph{integer}, \emph{float} and \emph{string} values and variables, and the standard operators over them. A \klaim-like pattern-matching mechanism is used to withdraw a tuple from a tuple space when a given template is specified.

\begin{table}[t]
\footnotesize
\hrule
\begin{tabular}{@{\hspace{0cm}}l@{\hspace{.75cm}}l@{\hspace{.75cm}}l}
\textsc{Knowledge:}
& \textsc{Items:}
& \textsc{Templates:}
\\[.1cm]
$\knw$  ::=  $\emptyset\!\sep\! \tuple{t} \!\sep\! \knw_1 \parcomp \knw_2$
& 
$t$  ::=  $\bexpr \!\sep\! \nvaddr \!\sep\! P \!\sep\! t_1,t_2$
&
$T$  ::= $\bexpr \!\sep\! \nvaddr \!\sep\! \form\,\vaddr \!\sep\! \form\,X \!\sep\! T_1,T_2$
\\[1pt]
\end{tabular}
\hrule
\vspace*{-.2cm}
\caption{Tuple space syntax ($e$ is an \textsc{expression})}
\label{SCEL-TS}
\end{table}

Regarding \PSCEL\ policies\footnote{We refer the interested reader to~\cite{SCEL-ASCENSbook15} for the presentation of the syntax of the policy language and of the operational semantics of \PSCEL.}, for the purposes of this example it is sufficient to know that they are hierarchically structured lists of elements containing controls on the value of attributes that should be provided by access requests generated when executing process actions. Together with permit or deny decisions, policies specify the conditions for their applicability, the combination algorithms to be used in their evaluation and the obligations for the enforcement process. Due to space limitations, we only explicitly show the policy of the customers, being those of the other components similar.

\begin{figure}[t]
\centering
\includegraphics[width=.75\textwidth]{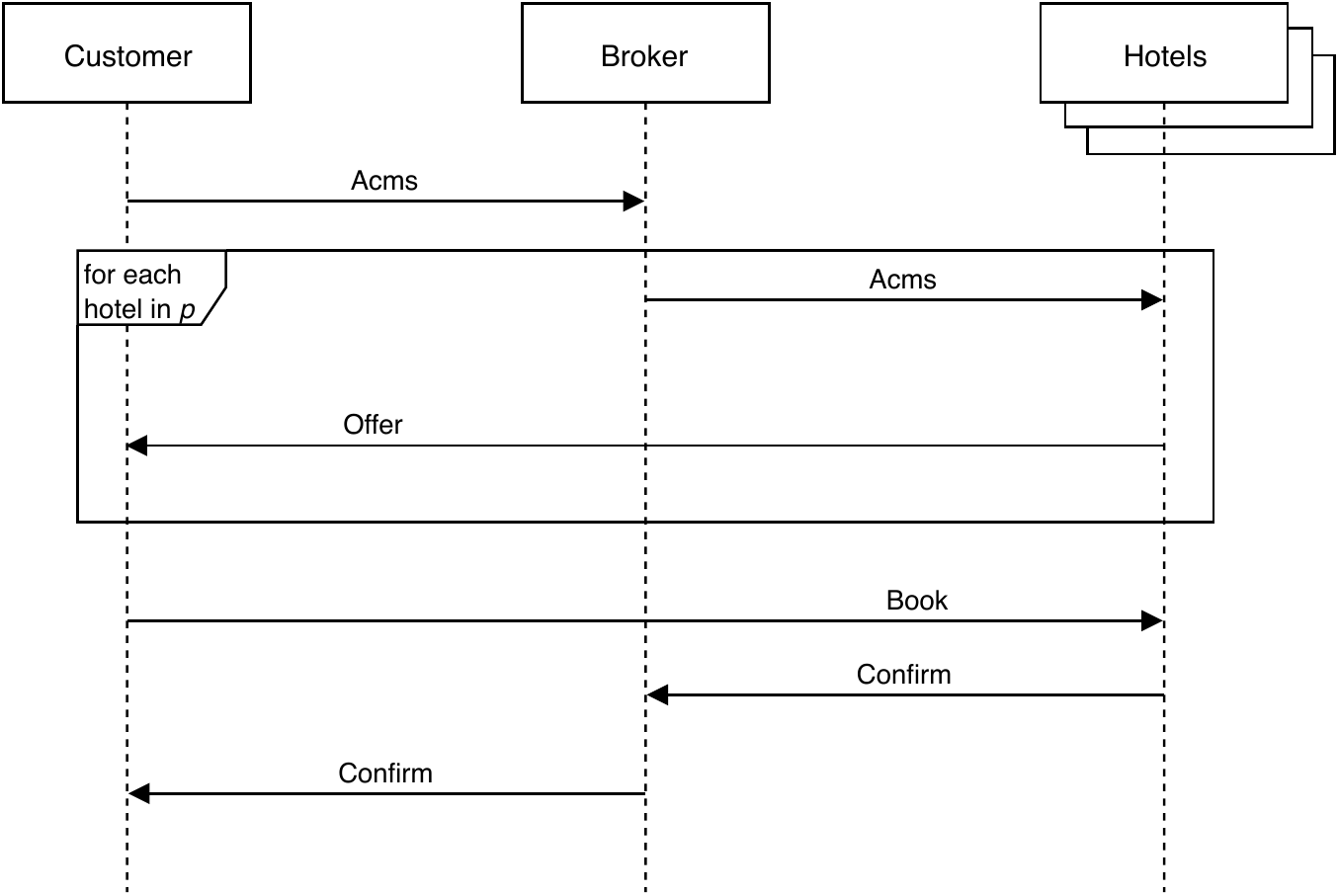}
\caption{Travel Booking Scenario in \SCEL: Sequence Diagram.}
\label{fig:SCEL_seq_diag}
\end{figure}

The scenario\footnote{We focus on hotel booking as flight booking is really similar.}\ is modelled by the following system $S$ including two customer components $\loccompAlt{{\cal I}_{c}}{\knw_{c},\pols_{c},P_{c}}$ and $\loccompAlt{{\cal I}_{d}}{\knw_{d},\pols_{d},P_{d}}$, a broker component $\loccompAlt{{\cal I}_{b}}{\knw_{b},\pols_{b},P_{b}}$, and $n$ hotel components $\loccompAlt{{\cal I}_{h_i}}{\knw_{h_i},\pols_{h_i},P_{h_i}}$:
\[
\begin{array}{@{}rcl}
S & \define &
\loccompAlt{{\cal I}_{c}}{\knw_{c},\pols_{c},P_{c}}
\parcomp
\loccompAlt{{\cal I}_{d}}{\knw_{d},\pols_{d},P_{d}}
\parcomp
\loccompAlt{{\cal I}_{b}}{\knw_{b},\pols_{b},P_{b}}
\parcomp
\\
&&
\loccompAlt{{\cal I}_{h_1}}{\knw_{h_1},\pols_{h_1},P_{h_1}}
\parcomp
\cdots
\parcomp
\loccompAlt{{\cal I}_{h_n}}{\knw_{h_n},\pols_{h_n},P_{h_n}}\, .
\end{array}
\]
\modif{
A specific UML activity diagram for 
$S$ is reported in Fig.~\ref{fig:SCEL_seq_diag}. 
Here, customer and hotel components can directly interact; this simplifies the model as it avoids the broker component to select room offers and make reservations on behalf of the customers. The figure also shows that the broker component forwards customers' requests to all those hotel components satisfying a given predicate $\enspred{p}$, which can then send offers directly to the requesting customer. The next room booking message is instead only directed to one of the hotels, which finally sends a confirmation to the customer through the broker component. In the rest of this section, we describe the different components of the system.}


\paragraph{Customer}
The following process $P_{c}$ runs at the customer component ${{\cal I}_{c}}$:
\[
\begin{array}{@{}r@{\ }c@{\ }l}
P_{c} & \define &
\acfreshp{\mathit{key}}.\\
&&
\acputp 
{\mathrm{``Acms"},\mathit{loc}, \mathit{day},\mathit{rating},\mathit{price}, \self,\mathit{key}}{({\cal I}_{b}.\texttt{id})}. \\
&&
P_{c}'(\mathit{day},\mathit{key})\\[.4cm]
P_{c}'(d,k) & \define &
\acgetp 
{\mathrm{``Offer"},d,\form f,\form p,\form h,k}
{\self}. \\
&&
(\ \acputp 
{\mathrm{``Book"},d,f,k}
{h}.\,
\acgetp 
{\mathrm{``Confirm"},d,\form X,k}
{\self}.
P_{c}''(X)\\
&&
\ \ +
\ P_{c}'(d,k)\ )
\end{array}
\]
This process firstly creates a name, denoted by $\mathit{key}$, which is guaranteed to be unique in the whole system. This name is a reference key for identifying the inquiry, thus permitting to distinguish possibly simultaneous inquiries generated by the two customers and to correlate responses to the corresponding inquiry. Then, the process sends a room reservation inquiry to the broker component by exploiting remote point-to-point communication. In addition to the string $\mathrm{``Acms"}$, 
the inquiry specifies: destination locality, day of stay, minimum star-rating, and the maximum room-price that the customer is willing to pay. The inquiry also contains the customer's identity and the name identifying the request. The continuation process $P_{c}'(d,k)$ elaborates the received room offers. By exploiting the pattern-matching mechanism, the process firstly withdraws from the local repository an offer (if any) corresponding to the inquiry done. The offer specifies features and price of the room, and the hotel's identity. 
Then, non-deterministically, either it sends a message to the hotel for reserving the room and waits for a confirmation, or (discards the offer and) continues recursively by withdrawing another offer.

In a real case study, the actual hotel would be selected after comparing price and features of the room and the related services. Here, for the sake of simplicity, we only model the selection as a non-deterministic choice. Also, we do not consider the possibility that a customer gets stuck due to lack of offers and we do not model removal of offers corresponding to finalised reservations.

In case room reservation is confirmed, by means of higher-order communication, the customer receives a process that is bound to variable $X$ and can be, later on, sent for execution by $P_{c}''$, e.g. to take advantage of some complimentary services provided by the broker. These services, together with special rates agreed between the broker and the contacted hotels, represent the advantage for the customer in booking hotels through the broker rather than directly.

The policy $\pols_{c}$ in force at the customer ${{\cal I}_{c}}$ is defined as follows:
\[
\small
\begin{array}{@{}r@{\ }c@{\!}l}
\pols_{c} & \define &
\begin{array}[t]{r@{\ }l}
\multicolumn{2}{l}{
\langle\, \denyUnless 
}\\
\ \ \x{rules:} \ 
\ruleOpt{\permit\ \ \x{target:} & \match{\x{equal}}{\this.\texttt{id}}{subject/id}\ \wedge\ \\
& \match{\x{not}\textrm{-}\x{equal}}{\acfresh}{act/action\textrm{-}id}\, }\\
\ruleOpt{\permit\ \ \x{target:} & \match{\x{equal}}{\this.\texttt{id}}{subject/id}\ \wedge\ \\
& \match{\x{equal}}{\acfresh}{act/action\textrm{-}id}\\
\x{obl:} & [ 
\acputp{{{\cal I}_{c}}.\texttt{key}, act/arg}{\self}
] \, }\\
\ruleOpt{\permit\ \ \x{target:} & \match{\x{equal}}{\acput}{act/action\textrm{-}id}\ \wedge\ \\
& \match{\x{equal}}{\this.\texttt{id}}{object/id}\ \wedge\ \\
& ( \match{\x{pattern}\textrm{-}\x{match}}{(\mathrm{``Offer"},\_,\_,\_,{{\cal I}_{c}}.\texttt{key})}{act/arg}\ \vee\ \\
& \ \match{\x{pattern}\textrm{-}\x{match}}{(\mathrm{``Confirm"},\_,\_,{{\cal I}_{c}}.\texttt{key})}{act/arg} )
\, }
\rangle
\end{array}
\end{array}
\]
The policy consists of three rules combined through the algorithm named $\denyUnless$ whose effect is to disallow any action that is not explicitly allowed by the rules. 
The first rule allows the local process (that is, $P_{c}$) to perform any action different from a \acfresh. In fact, the keyword $\this$ denotes the component where the policy is in force (i.e. the customer component), while the name $subject$ identifies the component performing the action which has to be authorised. The target of the rule requires these two components to coincide. The second rule allows the local process to perform an action \acfresh, but with the obligation of immediately performing a so called \emph{obligation action} for setting the component attribute $\texttt{key}$ to the argument of \acfresh\ (we assume that at the outset no value is bound to $\texttt{key}$ and that components attributes are managed by using the same mechanisms used for data). This means that, in general, policy evaluation can affect the system by adding further behaviours, other than simply (dis-)allowing some behaviours. The last rule allows other components to perform actions \acput\ having the local component as a target (i.e. $object$) provided that the argument of the action is a tuple with a very specific structure, where  the last field is the name purposely created to identify the specific inquiry. This check on the tuples 
aims at protecting the integrity of the repository; it ensures that the tuples added to the repository have a specific structure, with their last field containing the value of the attribute $\texttt{key}$. Indeed, the check succeed only after a value is bound to $\texttt{key}$, that is after the local process has performed the action \acfresh\ that triggers the obligation action \acput\ assigning a value to $\texttt{key}$. 

It is worth noticing that, since \SCEL\ features asynchronous communication, customers' inquiries are directly placed in the broker' repository, as well as hotels' offers are placed in the customers' repository, without requiring any synchronisation with other process actions. However, components can use policies to authorise such actions or not and, possibly, to oblige the executing process to perform some actions, like the one in the previous policy about setting the value of the attribute $\texttt{key}$.

\paragraph{Broker}
The following process $P_{b}$ runs at the broker component ${{\cal I}_{b}}$: 
\[
\begin{array}{@{}r@{\ }c@{\ }l}
P_{b} & \define &
\acgetp 
{\mathrm{``Acms"},\form l,\form d,\form r,\form p,\form c,\form k}
{\self}.\\
&&
(\auth 
{P_{b}}
{\acputp 
{\mathrm{``Acms"},d,c,k}
{\enspred{p}}. 
P_{b}'(l,d,c,k)}
)\\[.4cm]
P_{b}'(l,d,c,k) & \define &
\acgetp 
{\mathrm{``Confirm"},d,\form f,\form h,k}
\self. 
\\
&&
\acputp 
{\mathrm{``Confirm"},d,Q(l,d,f,h),k}
{c}.
\procnil\\
\end{array}
\]
where the predicate name $\enspred{p}$ is defined as 
\[
\begin{array}{@{}rcl}
\enspred{p} & \define & \texttt{serviceType}=\mathrm{``hotel"}\ \wedge\ \texttt{locality}=l\ \wedge
\\
&& 
\texttt{starRating} \geq r\ \wedge\ \texttt{roomPrice} \leq p 
\end{array}
\]
The process $P_{b}$ is triggered by a customer inquiry withdrawn from the local repository. Whenever this happens, the process continues by spawning, for parallel execution, a copy of itself, to serve other inquiries, and by making a group-oriented \acput\ using the data of the customer's inquiry, partly in the target, to filter the interacting partners, and in part in the forwarded inquiry itself. This means that there is no predefined list of hotels to interact with. Rather, a combination of predicates on the attributes named $\texttt{serviceType}$, $\texttt{locality}$, $\texttt{starRating}$, and $\texttt{roomPrice}$ is used at run-time to identify the ensemble made of all hotels which are relevant to deal with the given customer inquiry. Thus, by taking advantage of group-oriented communication, the broker is able to dynamically identify an ensemble of hotel components that can potentially provide the service requested by the customer. All the hotel components meeting these requirements get an excerpt of the original customer inquiry, specifying the day of stay, the customer's identity, and the unique inquiry's reference key.
The continuation process $P_{b}'$ waits for a confirmation of room reservation, including the features of the room, from a hotel after which it forwards the message to the customer along with a process $Q$ enabling the customer to use some complimentary services provided by the broker and depending on the destination, the day of stay, the paid price, the chosen hotel, etc.

\paragraph{Hotel}
The following process $P_{h_i}$ runs  at the hotel component ${{\cal I}_{h_i}}$:

\[
\begin{array}{@{}rcl}
P_{h_i} & \define &
\acgetp 
{\mathrm{``Acms"},\form d,\form c, \form k}
{\self}.\\
&&
(\auth 
{P_{h_i}}
\auth 
{P_{h_i}^{o} (d, c, k)}
{P_{h_i}^{b}(d,k)})
\\[.4cm]
P_{h_i}^{o}(d, c, k) & \define &
\acreadp 
{\mathrm{``Room"}, d,\form f}
{\self}.\\
&&
\acputp 
{\mathrm{``Offer"}, d, f, \mathit{p}(f,{\cal I}_{b}.\texttt{id}), \self, k}
{c}.
\procnil
\\[.4cm]
P_{h_i}^{b}(d,k) & \define &
\acgetp 
{\mathrm{``Book"},d,\form f,k}
\self. 
\acgetp 
{\mathrm{``Room"}, d, f}
\self. \\
&&
\acputp 
{\mathrm{``Confirm"},d,f,\self,k}
{({\cal I}_{b}.\texttt{id})}.
\procnil
\end{array}
\]
The process $P_{h_i}$, when triggered by a room reservation inquiry withdrawn from the local repository, spawns for parallel execution a copy of itself to serve other, possibly simultaneous, inquiries, elaborates the specific inquiry, and manages the room reservation request. The elaboration of the specific inquiry is done by the process $P_{h_i}^{o}$ which retrieves from the local repository information about room availability for the day of stay and hotel features, and sends back to the customer offers including room price. This price is calculated by function $\mathit{p}()$ that, on the basis of room features and broker identity, can return special rates depending on the specific agreements between the hotel and the broker. The process $P_{h_i}^{b}$, when triggered by a room reservation request, removes the corresponding room from the availability list for the day of stay, and sends a confirmation to the broker.

\modif{
\paragraph{Discussion}
The specification $S$ of the scenario takes advantage both of \SCEL\ general features and specific ones of the used dialect. Among the former, attributes and predicates over them are used to implement group-oriented communication, to enable the broker component to dynamically select an ensemble of hotels that could provide the service requested by customer components. Moreover, dynamic creation of unique names is used to unambiguously identify customer requests and all correlated messages subsequently exchanged, while process variables are used for exchanging and activating the processes that model the complimentary services provided by the broker.
Among the 
features of the used dialect, \klaim-like pattern-matching is exploited to selectively access items in the knowledge repositories, while policies are used both to disallow behaviours which could undermine the integrity of the repositories and to generate new behaviours through obligation actions. In particular, obligations have been used for setting some attributes when specific events occur (in this case, creation of fresh unique names).
}

\subsection{Programming environment}

\SCEL\ systems can be executed and simulated in \jresp\footnote{\jresp\ website: \url{http://jresp.sourceforge.net/}.} (Java Runtime Environment for \SCEL\ Programs), which offers specific software tools to develop and support \SCEL\ systems. In particular, \jresp\ provides an API that permits enriching \java\ programs with the \SCEL's linguistic constructs. The API is instrumental to assist programmers in the implementation of autonomic systems, which thus turns out to be definitely simpler than using ``pure'' Java. Moreover, \jresp\ provides a set of classes enabling execution of \emph{virtual components} on top of a simulation environment that can control component interactions and collect relevant simulation data. 

\modif{In Fig.~\ref{lis:scel_code} we report a significant fragment of code\footnote{\modif{The \jresp source code for the complete scenario can be downloaded from \url{https://bitbucket.org/tiezzi/jlamp_survey_code/src/master/SCEL/}.}} of the \jresp implementation of the \SCEL specification of the travel booking scenario, presented in Section~\ref{sec:booking_scenario_in_scel}. 
The \java\ classes of the broker process reported here show how programmers can directly use \SCEL\ communication primitives in Java, thus resulting in a very compact code.
A \SCEL process is rendered in \jresp as an instance of class \texttt{Agent}, which specifies the agent behaviour within the method \texttt{doRun()}. In the case of the broker, this method first performs a \texttt{get} action to retrieve an \emph{Acms} tuple, and then spawns another instance of the broker agent in the broker node, to immediately serve other possible customer requests. 
The method \texttt{get()} takes as parameters an instance of class \texttt{Template} and a target (in this case it is the local component referred by \texttt{Self.SELF}), and returns a matching tuple. 
In the subsequent \texttt{put} action, instead, the target is an ensemble (instance of class \texttt{Group}), defined by a predicate (instance of class \texttt{GroupPredicate}). Finally, in the last \texttt{put} action the target is the customer address, dynamically retrieved from the \emph{Acms} tuple initially received.  
}

\begin{figure}[!t]
\lstset{language=java,style=mystyleSmall}
\begin{lstlisting}{}
public static class BrokerAgent extends Agent {
	private Node brokerNode;

	public BrokerAgent(Node node) {
		super("BrokerProcess");
		brokerNode=node;
	}

	protected void doRun() {
		try {
			Tuple acms = get(new Template(new ActualTemplateField("Acms"),
						          new FormalTemplateField(String.class),
						          new FormalTemplateField(String.class),
						          new FormalTemplateField(Integer.class),
						          new FormalTemplateField(Double.class),
						          new FormalTemplateField(Target.class),
						          new FormalTemplateField(String.class)), Self.SELF);
			        String l = acms.getElementAt(String.class,1);
			        String d = acms.getElementAt(String.class,2);
			        Integer r = acms.getElementAt(Integer.class,3);
			        Double p = acms.getElementAt(Double.class,4);
			        Target c= acms.getElementAt(Target.class,5);
			        String k = acms.getElementAt(String.class,6);
			        
			        Agent B = new BrokerAgent(brokerNode);
			        brokerNode.addAgent(B);
			        
			        GroupPredicate p1 = new HasValue("serviceType","hotel");
			        GroupPredicate p2 = new HasValue("locality",l);
			        GroupPredicate p3 = new IsGreaterOrEqualThan("starRating", r);
			        GroupPredicate p4 = new IsLessOrEqualThan("roomPrice", p);
			        GroupPredicate predicate = new And(new And(new And(p1,p2),p3),p4);
			        put(new Tuple("Acms",d,c,k), new Group(predicate));
			        
			        Tuple confirmation = get(new Template(new ActualTemplateField("Confirm"),
								          			new ActualTemplateField(d),
								          			new FormalTemplateField(String.class),
								          			new FormalTemplateField(Target.class),
								          			new ActualTemplateField(k)), Self.SELF);	
			        String f = confirmation.getElementAt(String.class,2);
			        Target h= confirmation.getElementAt(Target.class,3);
			        
			        put(new Tuple("Confirm",d,f,h,k),c); 			
			} catch (Exception e) { e.printStackTrace(); }
		}
	}
\end{lstlisting}
\vspace*{-.3cm}
\caption{The process $P_{b}$ of the travel booking scenario implemented in \jresp.}
\label{lis:scel_code}
\end{figure}

\subsection{Verification techniques}

\begin{figure}[t]
\centering
\includegraphics[scale=.4]{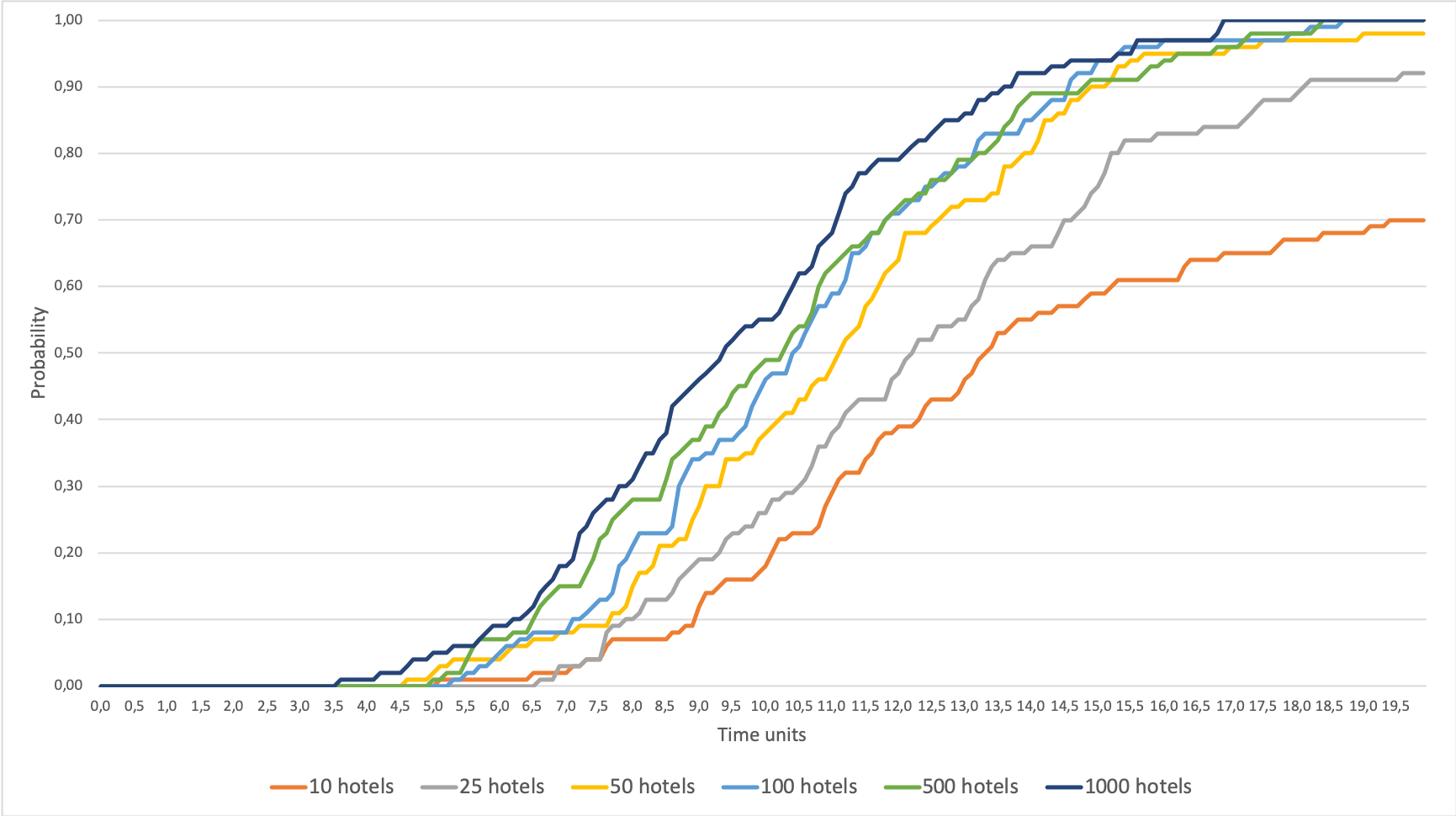}
\caption{Statistical model-checking of the Travel Booking scenario using \jresp.}
\label{fig:jresp_smc}
\end{figure}

Qualitative properties of \SCELlight\ specifications have been verified
through the \spin\ model checker~\cite{Hol}. The verification relies on a 
translation from \SCELlight\ into Promela, the input language of \spin. This approach has been used in~\cite{SCELigth14} to verify properties of interest for a service provision scenario, like absence of deadlock, server overload and responsiveness, and in~\cite{SCEL-ASCENSbook15} to verify similar properties for a swarm robotics scenario.

\SCEL's operational semantics has also been implemented by using the \maudel\ framework~\cite{Maude-book}. The outcome, named \misscel\ (\maudel\ Interpreter and Simulator for \SCEL), focuses on \SCELlight\ and exploits the rich \maudel\ toolset to perform, among other things, qualitative analysis
via \maudel's invariant and LTL model checkers, and statistical model checking via \multivesta~\cite{multivesta} (as done in~\cite{MisscelAndPirlo} for a robotic collision avoidance scenario). A further advantage of \misscel\ is that \SCEL\ specifications can be intertwined with (very expressive) raw \maudel\ code. This permits to obtain sophisticated specifications in which \SCEL\ is used to model behaviours, aggregations, and knowledge handling, while scenario-specific details are specified with \maudel. 

A prototype framework for statistical model-checking \cite{SCEL-ASCENSbook15} has been developed by relying on the simulation data provided by \jresp. 
\modif{The tool, by relying on a randomised algorithm, allows one to verify whether the implementation of a system satisfies a given property with a certain degree of confidence, depending on the number of simulations for a given experiment.}


\modif{To illustrate in practice the effectiveness of the statistical model-checking facility of \jresp\ for the verification of \SCEL specifications, we have applied it to the \SCEL model of the travel booking scenario. In particular, this verification activity aimed at evaluating the probability of reaching within a given deadline a successful execution state, where the customer receives an offer acceptable according to his/her desiderata and completes the booking process with the corresponding hotel.
Fig.~\ref{fig:jresp_smc} shows the probability that the successful state is reached within 20 units of time when the number of hotels varies among 10, 25, 50, 100, 500 and 1000. For a given number of hotels, the corresponding experiment has been repeated 100 times, thus ensuring adequate confidence for the precision of the results. 
The diagram shows that the successful state can be reached only after 3.5 units of time and that, as expected, the success probability grows with the number of hotels, although with more than 50 hotels the system performance does not improve significantly.}

%
%
%
\subsection{Related work}

%

\modif{
\SCEL\ combines the idea of dynamically forming ensembles of agents -- getting together to cooperatively work towards some collective goals -- which has been extensively analysed \cite{Durfee89,KluG02} in the area of distributed artificial intelligence and multi-agent systems \cite{ZamO04}, with other concepts that have emerged from different research fields of Computer Science and Engineering. Indeed, it borrows from software engineering the importance of component-based design and separation of concerns \cite{ComposingAdaptiveSoftware04}, from multi-agent systems the relevance of knowledge handling and of spatial representation \cite{AgentSpeak96,Jason05,JACK05,JADE07,2APL08},
from middleware and network architectures the importance of flexibility in communication \cite{MP06:logicalNeighborhoods,TSmiddleware09,TOTA09,Haggle09,MP12}, from distributed systems' security the role of policies \cite{NIST}, from actors and process algebras the importance of minimality and formality \cite{Agha90,CCS89}. 

In the area of concurrency theory, calculi such as those defined in \cite{BRF04} and in \cite{AK09b}, relying on the (bio)chemical programming paradigm, have been proposed for the specification of autonomic systems. Some other formalisms, like those introduced in \cite{MS06} and in \cite{SRS10}, aiming at modelling dynamically changing network topologies, 
also offer linguistic primitives for specifying autonomic systems. Compared to these proposals, \SCEL\ allows one to provide high-level abstract descriptions of systems that nevertheless have a direct correspondence with their implementation. 

Summing it up, the main distinctive aspect of \SCEL\ is the actual choice of the specific programming abstractions for autonomic computing and their reconciliation under a single roof with a uniform formal semantics. 
We refer the interested reader to~\cite{DLPT14,SCEL-ASCENSbook15,ascensBook} for a more complete account of autonomic computing and its relations with \SCEL.
}

\section{\abc: Attribute-based communication}
\label{sec:abc}

Collective-Adaptive Systems (CAS)~\cite{ferscha2015collective} are emerging computational systems, consisting of a massive number of components, featuring complex interaction mechanisms. These systems are usually distributed, heterogeneous, decentralised and interdependent, and are operating in dynamic and often unpredictable environments. CAS components combine their behaviours, by forming collectives, to achieve specific goals depending on their attributes, objectives, and functionalities. CAS are inherently scalable and their boundaries are fluid in the sense that components may enter or leave the collective at any time; so they need to dynamically adapt to their environmental conditions and contextual data.

\abc (Attribute-based Communication calculus, \cite{AbCSAC15,Alrahman3,AlrahmanNL19}) is a process calculus specifically designed to deal with CAS. It has been heavily inspired by \SCEL, but has been designed to reduce complexity and keep the set of linguistic primitives to a minimum. Indeed, it was originally designed as a trimmed version of \SCEL\ that was obtained by ignoring the parts relative to policies and knowledge and  concentrating only on behaviours and interfaces. In this respect, \abc has similar aims to \SCELlight, but the underlying communication paradigm is very different; explicit message passing for the former and shared memory \emph{\`a la} \klaim for the latter.

Indeed, the original goal of \abc was to assess the impact of the new message passing paradigm based on attributes and compare it with standard approaches that handle the interaction between distributed components by relying on identities (Actors~\cite{Agha:1986:AMC:7929}), or channels ($\pi$-calculus), or broadcast (B-$\pi$-calculus~\cite{prasad1991calculus}). In all these formalisms, messages exchanges rely on names or addresses of the involved components and are independent of their status and capabilities.  This makes it hard to program, coordinate, and adapt complex behaviours that highly depend on run-time modifications of components.

In \abc, the attribute-based system is however more than just the enabler of the parallel composition of interacting partners; it is also a tool for parameterising system components with respect to the environment or the space where they are executed. Indeed, the environment has a great impact on components behaviour and provides a new means of indirect communication, that allows components to mutually influence each other, also unintentionally.

\subsection{Syntax}

Table~\ref{tab:syntax} contains the  syntax of \abc.  The top-level entities of the calculus are {\sc components}. A component, $\COMPAbC{\Gamma}{I}{P}$, is a process $P$ associated with an {attribute environment} $\Gamma$, and an {interface} $I$. The \emph{attribute environment} provides a collection of attributes whose values represent the status of the component and influence its run-time behaviour. Formally, $\Gamma\!:\! \Attributes \pto \Values$ is a partial map from attribute identifiers ($a \in \Attributes$) to values ($v \in \Values$), i.e., to numbers, strings, tuples, \ldots The \emph{interface} $I\subseteq \Attributes$ contains the \emph{public attributes} of a component (the attributes in  $dom(\Gamma)-I$ being \emph{private}). Composed  components $C_1\|C_2$ are built by using the parallel operator.

\begin{table}[t]
\footnotesize
\hrule
\begin{tabular}{l@{\hspace{.1cm}}c@{\hspace{.1cm}}l@{\hspace{-.1cm}}l@{\hspace{.1cm}}|@{}}
\multicolumn{3}{l}{\textsc{Components:}} & \\
\quad $C$ & ::= & $\COMPAbC{\Gamma}{I}{P}$ & \quad (component)\\
& $\sep$ & $C_1\|C_2$ & \quad (composition)
\\[.1cm]
\multicolumn{3}{l}{\textsc{Processes:}} & \\
\quad $P$ & ::= & $\abcnil$ & \quad (inaction)\\
& $\sep$ & $\abcin{\Pi}{\tilde{x}}.U$ & \quad (attribute-based input)\\
& $\sep$ & $\abcout{\tilde{E}}{\Pi}.U$ & \quad (attribute-based output)\\
& $\sep$ & $\abcmatch{\Pi} P$ & \quad (context awareness)\\
& $\sep$ & $P_1 + P_2$ & \quad (choice)\\
& $\sep$ & $P_1 | P_2$ & \quad (parallel composition)\\
& $\sep$ & $K$ & \quad (process identifier)
\\[.1cm]
\multicolumn{3}{l}{\textsc{Updates:}} & \\
\quad $U$ & ::= & $\abcupdate{a}{E}U$ & \quad (attribute update)\\
& $\sep$ & $P$ & \quad (process)
\\[.1cm]
\end{tabular}
\begin{tabular}{@{}l@{\hspace{.1cm}}c@{\hspace{.2cm}}l@{\hspace{.1cm}}l}
\vspace*{-.5cm}\\
\multicolumn{4}{@{\,}l}{\textsc{Predicates:}}\\
\quad $\Pi$ & ::= & $\mathrm{tt}$ & \quad (true) \\
& $\sep$   & $\mathrm{ff}$ & \quad (false) \\
& $\sep$   & $E_1 \bowtie E_2$ & \quad (comparison) \\
& $\sep$   & $p(\tilde{E})$ & \quad (atomic predicate) \\
& $\sep$   & $\Pi_1 \land \Pi_2$ & \quad (conjunction) \\
& $\sep$   & $\Pi_1\lor \Pi_2$ & \quad (disjunction) \\
& $\sep$   & $\neg \Pi$ & \quad (negation)
\\[.1cm]
\multicolumn{3}{@{\,}l}{\textsc{Expressions:}} & \\
\quad $E$ & ::= & $v$ & \quad (value)\\
& $\sep$ & $x$ & \quad (variable)\\
& $\sep$ & $a$ & \quad (attribute identifier)\\
& $\sep$ & $\this.a$ & \quad (local reference)\\
& $\sep$ & $f(\tilde{E})$ & \quad (operator)\\
\end{tabular}
\hrule
\caption{The syntax of the \abc calculus}
\label{tab:syntax}
\end{table}

A {\sc process}  $P$ can be: the \emph{inactive} process $\abcnil$; the \emph{input-prefixed} process $\abcin{\Pi}{\tilde{x}}.U$ or the \emph{output-prefixed} process $\abcout{\tilde{E}}{\Pi}.U$,  where $U$ is a process preceded by a (possibly empty) sequence of \emph{attribute updates}; a \emph{context aware} process, $\abcmatch{\Pi}{P}$, where $\Pi$ is a {\sc predicate} built from boolean constants and from atomic predicates, based on {\sc expressions} over attributes, by using standard boolean operators; a \emph{nondeterministic choice} between two processes, $P_1 + P_2$; a \emph{parallel composition} of two processes, $P_1|P_2$; or a process call with an identifier $K$ used in a unique process definition $K \triangleq P$.

An expression $E$ may be a constant value $v$, a variable $x$, an
attribute name $a$, or a reference $this.a$ to attribute $a$ in the
local environment. Predicate $\Pi$ can be either $\mathrm{tt}$, or can
be built using comparison operators $\bowtie$ between two expressions and
 logical connectives $\wedge$, $\neg$, $\ldots$. Both expressions and
predicates can take more complex forms, of which we deliberately
omit the precise syntax; we just refer to them as n-ary operators on
subexpressions, i.e., $f(\tilde{E})$ and $p(\tilde{E})$.

\subsection{Informal semantics}

Attribute-based actions for sending and receiving messages permit to establish communication links between different components according to specific predicates over their attributes.

Specifically, \emph{attribute-based output} $\abcout{\tilde{E}}{\Pi}$ sends the result of the evaluation of the sequence of expressions $\tilde{E}$ to the components whose attributes satisfy the predicate $\Pi$. Notably, together with the computed values, also the portion of the attribute environment of the sending component that can be perceived by the context is sent; this is obtained from the local environment by limiting its domain to the attributes in the component interface. This information is needed to allow receivers to determine whether they are interested in the sent message.

Instead, \emph{attribute-based input} $\abcin{\Pi}{\tilde{x}}$ specifies receipt of messages from a component satisfying predicate $\Pi$; the sequence $\tilde{x}$ acts as a placeholder for received values. A message can be received when two \emph{communication constraints} are satisfied: the public local attribute environment satisfies  the predicate used by the sender to identify potential receivers, and the sender environment and the communicated message are such that the receiving predicate is satisfied. In this case, attribute updates are performed under the generated substitution.

An \emph{attribute update}  $\abcupdate{a}{E}$ assigns the value of $E$ to the attribute identifier $a$. This action is used to change the values of the attributes according to contextual conditions and to adapt component's behaviour. 

The \emph{awareness construct} $\abcmatch{\Pi}{P}$ blocks the execution of $P$ until predicate $\Pi$ is satisfied when using the local attribute environment, possibly after a change of state by a component.  This construct permits to collect awareness data and take decisions based on the changes in the attribute environment.

Attribute updates and awareness predicates are local to components and their executions are atomic with the associated
communication action.

\subsection{A taste of the operational semantics}
\label{sec:abcexcerpt}

The operational semantics of $\abc$ is based on two relations.
The transition relation $\rMapsto{\ \ }{}$ that describes the behaviour of individual components and the transition relation  $\rTo{\ \  }$ that relies on $\rMapsto{\ \ }{}$ and describes system behaviours.

Relation $\rMapsto{\  \  }{}$  defines the local behaviour of a component whose labels $\alpha$ have the following format:
$$
 \alpha ::= \abclabout{\Gamma}{\tilde{v}}{\Pi}
\quad |\quad
	\abclabin{\Gamma}{\tilde{v}}{\Pi} \quad |\quad \abclabdin{\Gamma}{\tilde{v}}{\Pi}
\qquad\qquad\qquad
$$
%
%
The $\alpha$-labels $\abclabout{\Gamma}{\tilde{v}}{\Pi}$ and $\abclabin{\Gamma}{\tilde{v}}{\Pi}$ are used to denote \abc output and input actions, respectively. The former contains the sender's predicate $\Pi$, that specifies the expected communication partners, the transmitted values $\tilde{v}$, and the portion of the sender \emph{attribute environment} $\Gamma$ that can be perceived by receivers. The latter label is just the complementary label selected among all the possible ones that the receiver may accept. The $\alpha$-labels $\abclabdin{\Gamma}{\tilde{v}}{\Pi}$ describe the actions exhibited by  a component to discard undesired input messages. The contextual label $\Gamma$ indicates the environment in which the components operate and is instrumental to determine whether actual communications can take place at the system level.

The basic rules for components interaction are reported in Table~\ref{tab:compsem1}. 
\begin{table}[t!]
$$
\begin{array}{@{\qquad}c@{\qquad}}
\hline
\\[-8pt]
\infer[\rulelabel{Brd}]{
\COMP{I}{\abcout{\tilde{E}}{\Pi_1}.U}
\rMapsto{\abclabout{\Gamma\downarrow I}{\tilde{v}}{\Pi}}{}
\llbrace\COMP{I}{U}\rrbrace
}{
\llbracket \tilde{E}\rrbracket_{\Gamma}=\tilde{v} &
\{ \Pi_1 \}_{\Gamma}=\Pi 
}
\\[.4cm]
\COMP{I}{\abcout{\tilde{E}}{\Pi}.U}
\rMapsto{\abclabdin{\Gamma'}{\tilde{v}}{\Pi'}}{}
\COMP{I}{\abcout{\tilde{E}}{\Pi}.U}
\; \rulelabel{FBrd}
\\[.4cm]
\infer[\rulelabel{Rcv}]{
\COMP{1}{_I}{\abcin{\Pi_1}{\tilde{x}}.U}
\rMapsto{\abclabin{\Gamma'}{\tilde{v}}{\Pi}}{}
\llbrace\COMP{1}{_I}{U[\tilde{v}/\tilde{x}]}\rrbrace
}{
\Gamma'\models \{\Pi_1[\tilde{v}/\tilde{x}]\}_{\Gamma_1} &
\Gamma_1\downarrow I\models \Pi 
}
\\[.4cm]
\infer[\rulelabel{FRcv}]{
\COMP{1}{_I}{\abcin{\Pi}{\tilde{x}}.U}
\rMapsto{\abclabdin{\Gamma'}{\tilde{v}}{\Pi'}}{}
\COMP{1}{_I}{\abcin{\Pi}{\tilde{x}}.U}
}{
\Gamma'\not\models \{\Pi[\tilde{v}/\tilde{x}]\}_{\Gamma} \vee
\Gamma_1\downarrow I\not \models \Pi' 
}
\\[.1cm]
\hline
\end{array}
$$
\vspace*{-.5cm}
\caption{Operational Semantics for Components' Communications}
\label{tab:compsem1}
\end{table}
The behaviour of \emph{attribute-based output} is defined by rule \rulelabel{Brd} in Table~\ref{tab:compsem1}. It states that when an output is executed, the sequence of expressions $\tilde{E}$ is evaluated, say to $\tilde{v}$, and the \emph{closure} $\Pi$ of predicate $\Pi_1$ under $\Gamma$ is computed. Hence, these values are sent to other components together with $\Gamma\downarrow I$, i.e., the portion of the \emph{attribute environment} that can be perceived by the context, obtained from $\Gamma$ by limiting its domain to the attributes in the interface $I$. It has to be noted that rule \rulelabel{Brd} is not sufficient to fully describe output; rule \rulelabel{FBrd} is also needed to model the fact that all incoming messages are \emph{discarded} when only output actions are possible.
Rule \rulelabel{Rcv} governs the execution of input actions. It states that a message can be received when two \emph{communication constraints} are satisfied: the local attribute environment restricted to interface $I$ ($\Gamma_1\downarrow I$) satisfies  $\Pi$, the predicate used by the sender to identify potential receivers; and the sender environment $\Gamma'$ satisfies the receiving predicate $\{\Pi_1[\tilde{v}/\tilde{x}]\}_{\Gamma_1}$. When these two constraints are satisfied the input action is performed and the update $U$ is applied under the substitution $[\tilde{v}/\tilde{x}]$. Rule \rulelabel{FRcv} states that an input is \emph{discarded} when the local attribute environment does not satisfy the \emph{sender's predicate}, or the \emph{receiving predicate} is not satisfied by the sender's environment.

The behaviour of an \abc system is described by means of  the transition relation $\rTo{\  \  }{}$
whose labels $\lambda$ are generated by the following grammar:
$$
\lambda  \  ::= {\overline{\Pi}(\tilde{v})}
\quad |\quad   {\Pi}(\tilde{v})
$$

The main semantics rules of \abc systems are reported in Table~\ref{abccom}.


\begin{table}[t!]
$$
\begin{array}{@{\ }l@{\hspace*{1cm}}r}
\hline
\\[-8pt]
\infer[\rulelabel{iComp}]{
\COMP{}{P}\rTo{\lambda}{}
\Gamma':{P'}
}{
\COMP{}{P}
\rMapsto{\lambda}{}
\Gamma':{P'}
}
&
\infer[\rulelabel{fComp}]{
\COMP{}{P}\rTo{{\Pi}(\tilde{v})}{}
\COMP{}{P}
}{
\COMP{}{P}
\rMapsto{\widetilde{\Pi(\tilde{v})}}{}
\COMP{}{P}
}\\[.35cm]
\infer[\rulelabel{Com}]{
C_1\ \|\ C_2\rTo{\overline{\Pi}(\tilde{v})} C_1'\ \|\ C_2'
}{
C_1 \rTo{\overline{\Pi}(\tilde{v})} C_1'
\ \
C_2 \rTo{ \Pi(\tilde{v})} C_2' \ \ 
}
&
\infer[\rulelabel{Sync}]{
C_1\ \|\ C_2\rTo{ \Pi(\tilde{v})} C_1'\ \|\ C_2'
}
{
C_1 \rTo{ \Pi(\tilde{v})} C_1'
\quad
C_2 \rTo{ \Pi(\tilde{v})} C_2'
}
\\[.1cm]
\hline
\end{array}
$$
\vspace*{-.5cm}
\caption{\abc Communication Rules}
\label{abccom}
\end{table}

Rule \rulelabel{iComp} states that a component evolves with a send $\overline{\Pi}(\tilde{v})$ or receive ${\Pi}(\tilde{v})$, action (generically denoted by $\lambda$) if its internal behaviour (denoted by the relation $\mapsto$) allows it. Rule \rulelabel{fComp} states that a component can discard a message ${\Pi}(\tilde{v})$ if its internal behaviour does not allow the reception of this message by generating the discarding label $\widetilde{\Pi(\tilde{v})}$. Rule \rulelabel{Com} states that if $C_1$ evolves to $C'_1$ by sending a message $\overline{\Pi}(\tilde{v})$ then this message should be delivered to $C_2$ which evolves to $C'_2$ as a result. Note that $C_2$ can be also a parallel composition of different components. Thus, rule \rulelabel{Sync} states that multiple components can be delivered the same message in a single transition.

The semantics of the parallel composition operator, in rules \rulelabel{Com} and \rulelabel{Sync} in Table \ref{abccom}, abstracts from the underlying coordination infrastructure that mediates the interactions between components and thus the semantics assumes atomic message exchange. This implies that no component can evolve before the sent message is delivered to all components executing in parallel. Individual components are in charge of using or discarding incoming messages. Message transmission is non-blocking, but reception is not. For instance, a component can still send a message even if there is no receiver (i.e., all the target components discard the message); a receive operation can, instead, only take place through synchronisation with an available message.

The original semantics of \abc outlined above has been formulated in a way that
when a component sends a message, that message is delivered to all
 components in the system in a single move. Atomically, each individual receivers decide
whether to use the message or to discard it.
This semantics imposes a restriction on the
ordering of message delivery because only one component can send its
message at a time.
Even if this approach is useful to describe the \abc{} one-to-many interactions in an abstract way, it is too strong when large scale distributed systems are considered.

To relax the \emph{total ordering} requirement of the original \abc{} semantics, we are currently working on an alternative semantics that takes into account  the \emph{infrastructure} responsible of message dispatching.
 \abc{} \emph{system} are built by using \emph{servers} of the form $\{ \cdot \}^{\iota,\omega}$ that are responsible to manage a set of \emph{components}. Each server is equipped with an \emph{input queue} $\iota$ and an \emph{output queue} $\omega$. The former is the queue of messages that, coming from the environment, the server must deliver to the managed components. The latter is the queue of messages that have been generated locally and that the server must forward to the enclosing system.

\subsection{A travel booking scenario}
In this section we consider the travel agency scenario
and outline its  \abc model. We do restrict attention to the part of the scenario concerned with hotel booking, with customers interacting with a broker for room booking. Customers contact the broker that in turn contacts those hotels that expose attributes meeting customers expectations. After receiving pricing and availability information, the broker forwards the best options to customers that then proceeds with the booking directly with the hotel.

The \abc specification of the above scenario relies on three types of components, namely Customer, Broker and Hotel.
$$Cust_1 \parallel \ldots \parallel Cust_n \parallel Broker \parallel Hotel_1 \parallel \ldots \parallel Hotel_p$$
where each component has the following form:
$$Cust_i \triangleq \Gamma_{i} : P_C \;, Broker \triangleq \Gamma: P_B \;, Hotel_k \triangleq \Gamma_{k} : P_H$$
with the same behaviour and the same set of attributes names for each type.  
\begin{figure}[t]
\centering
\includegraphics[width=.75\textwidth]{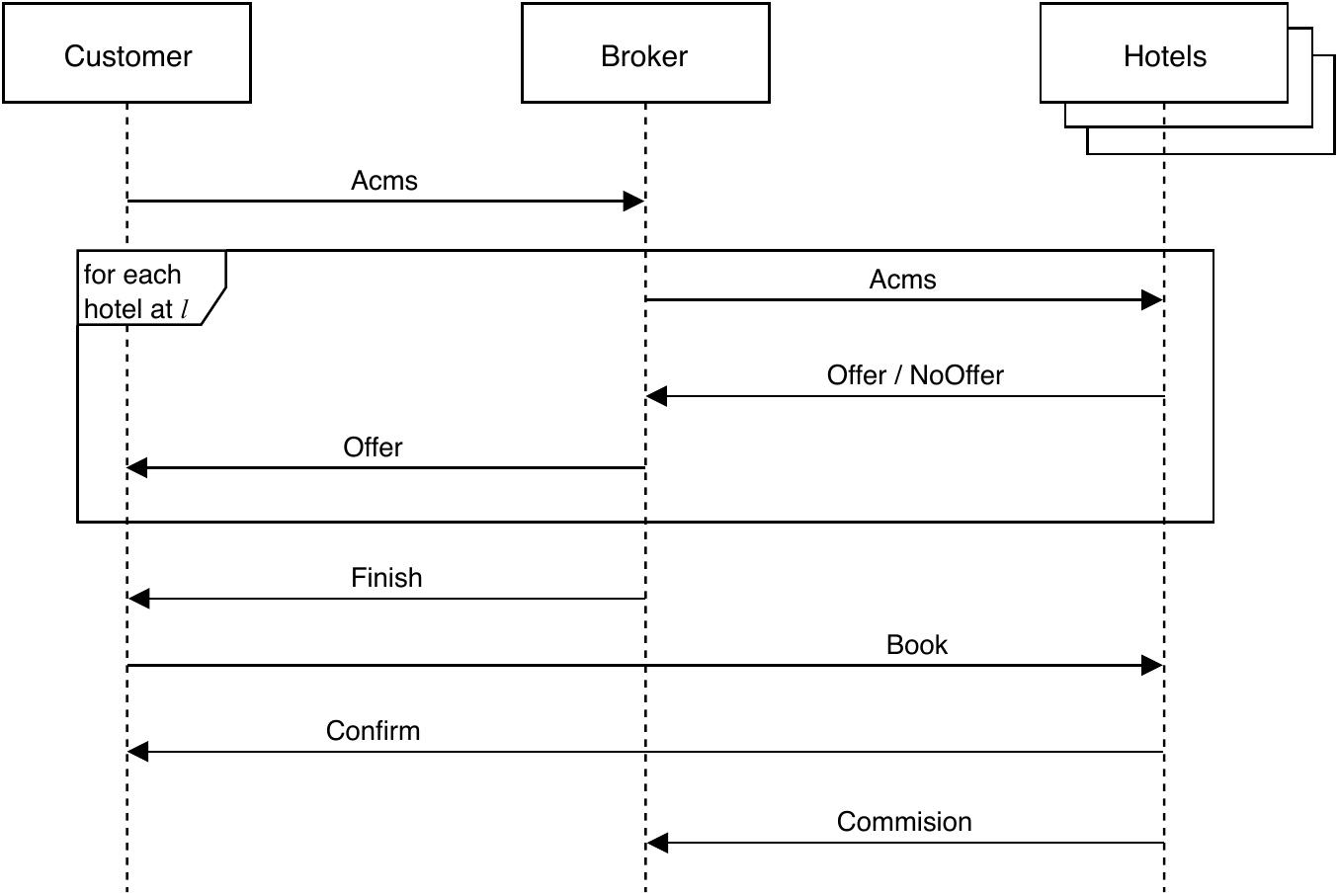}
\caption{Travel Booking Scenario in \abc: Sequence Diagram.}
\label{fig:abc_seq_diag}
\end{figure}
\modif{A UML activity diagram for the \abc specification of the travel booking scenario  is reported in Fig.~\ref{fig:abc_seq_diag}. 
The figure highlights that the broker acts as a mediator between the customer and the hotels 
and filters the offers according to the customer's preferences. 
Instead, once the customer has selected the desired hotel, the interactions between 
them, for booking and confirmation, are not mediated by the broker. }
In what follows, we provide the specification of the individual components.

\paragraph{Customer} The main attributes of a customer are:
\texttt{id, loc, day, price, dist} that represent the customer id, the
 preferred location, the chosen day, the maximum price,
the tolerated distance between the ideal location \texttt{loc} and the actual
location  of the hotel.
In addition, customers use other internal attributes, namely
\texttt{favh} - the favorite hotel, \texttt{ref} - the reference broker, and a flag \texttt{send} to control when to send out an inquiry
message.
The behaviour of a customer is encoded by:
$$P_C \triangleq F \; \vert \; A $$
Process $F$, when enabled by the flag \texttt{send}, proceeds by
getting the customer preferences via different getter functions - $\mathsf{get\_*()}$ - 
to be used to set up user's preferences.
 These attribute updates are triggered by exploiting an empty send
(a sort of ``fake'' output action). After setting the preferences, $F$ sends an inquiry to the broker,
specified in the sending predicate by (type = ``Broker''). The
message to be sent includes a tag ``acms'' for accommodation, the location, the booking date and the
maximum price the client is prepared to pay. The flag \texttt{send} may be enabled again for
another try, if the customer does not find a room at the wanted condition.
{\small
  \begin{align*}
    F \triangleq  &\; \langle send \rangle()@(\mathrm{ff}).\\
    & \; [\mathrm{day} := \mathsf{get\_day()}, \mathrm{price} := \mathsf{get\_price()},\mathrm{dist} := \mathsf{get\_dist()}, \mathrm{loc} := \mathsf{get\_loc()}]\\
                  & \; (\mathrm{``acms"},this.id,this.\mathrm{loc},this.\mathrm{day},this.\mathrm{price})@(type = \mathrm{``Broker"}).[\mathrm{send := ff}]F
  \end{align*}
}%
Process $A$, instead, handles the replies from the broker. The first branch
deals with `offer' messages associated with hotels located within the
preferred \texttt{dist}ance and at a good \texttt{price}. In such case, a possibly
better price $p$ offered by a hotel $h$ made available by the broker
with id $b$ are stored in the relevant attributes for later use.  The
second branch waits for a ``finish'' message indicating that there is
no more offers to be considered, in which case process $A$ continues
as $B$.  {\small
  \begin{align*}
  A \triangleq & \;(x = \mathrm{``offer"} \wedge this.price \ge p \wedge  \mathsf{diff}(this.loc,l) \le this.dist)(x,h,l,p,b).\\
                          & \; [\mathrm{price := p, favh := h, ref := b}]A \\
                      & \; +  \\
               & \; (x = \mathrm{``finish"})(x).B
  \end{align*}
}%
We assume here the availability of function $\mathsf{diff}$ that computes the distance between two coordinates.
  
Upon receipt of the ``finish'' message, process $B$ makes a decision about how to proceed with the actual booking. Two possibilities are considered:\\ 
1. There is no hotels that meets customer's requirement, i.e., (favh = $\bot$), then the first branch of $B$
sets `send'  to true for another try.  {\small
  \begin{align*}
    B \triangleq  &\; \langle \mathrm{favh = \bot} \rangle()@(\mathrm{ff}).[\mathrm{send := tt}]A\\
                  & \; + \\
                  & \; \langle \mathrm{favh \neq \bot} \rangle(\mathrm{``book"}, this.id, this.day, this.price, this.ref)@(id = this.favh). \\
    & \quad (\; (x = \mathrm{``confirm"})(x).0 \\
                  & \quad  + \\
    & \quad\; (x = \mathrm{``toolate"})(x).[\mathrm{send := tt}]A )
\end{align*}
}%
\noindent
2. A  good hotel is found, i.e., (favh $\neq \bot$), the
process sends a `book' message attaching the customer id, the wanted
date, the paid price and the reference broker to the chosen hotel. If
it receives a confirmation afterwards, the session is considered as
finished, otherwise, a `toolate', may arrive meaning that the
preferred room in that hotel has already been taken. In that case, a
customer may retry again with a new set of preferences by enabling
\texttt{send}.

\paragraph{Broker} The set of attributes of an online broker, $P_B$
are \texttt{id, type}, besides the internal attributes
\texttt{nh,cnt} needed for local computation. $P_B$, waits for
request messages from its customers and, for each inquiry, creates a
handler $H$ to process the inquiry, and spawns a copy of itself for
serving future requests:
{\small
  \begin{align*}
    P_B \triangleq & \; (x = \mathrm{``acms"})(x,c,l,d,p).[nh_{l} := \mathsf{get\_hotels}(l), cnt_{c} : = 0](H | P_B) \\[.4cm]
    H \triangleq & \; (\mathrm{``acms"},c, d, this.id)@(type = ``Hotel" \wedge locality = l).(A | U)
\end{align*}
}%
Please notice that process $H$ can use the bound variables
\textit{c,l,d,p} and that two internal attributes $\mathtt{nh_l}$
keeping the number of hotels at locality $l$ and $\mathtt{cnt_c}$
counting the number of replies from hotels are initialised for this
session. These attributes can be structured as dynamic
  vectors, and their slot $k$ can be used for storing relevant
  information of the session.

Process $H$ forwards the inquiry to all hotels (type = ``Hotel'') at
locality $l$. The message contains the reference customer identity, the
required date $d$ and the broker identity \texttt{id}.
$H$ then continues as two parallel processes: $A$ dealing with valid
offers, $U$ with invalid offers, in addition to sending a termination
message.

Process $A$ receives the price-wise acceptable offers from hotels and
forwards them to the customer: an offer is forwarded
only if the offered price is not greater than the maximum stated
price. Furthermore, to handle other messages in parallel, $A$ also
spawns a new instance of itself.
{\small
  \begin{align*}
    A \triangleq & \; \langle cnt_c < nh_l \rangle(x = \mathrm{``offer"} \wedge c = cust \wedge op \le p)(x,cust,h,l,op).(S \vert A)\\[.4cm]
    S \triangleq &  \; (\mathrm{``offer"},h,l,op,this.id)@(id = c).[cnt_c := cnt_c + 1]0 
\end{align*}
}%
Process U discards ``expensive'' offers and ignores `nooffer'
messages; in both cases it increases counter $cnt_c$ by 1. The last
branch of $U$ waits until all hotels at the locality $l$ have
replied in order to send a `finish' message to the customer.
{\small
  \begin{align*}
    U \triangleq &  \; \langle cnt_c < nh_l \rangle(x = \mathrm{``offer"} \wedge c = cust \wedge op > p)(x,cust,h,l,op). \\
    & \;  [cnt_c := cnt_c + 1]U \\
    & \; + \\
                 & \; \langle cnt_c < nh_l \rangle(x = \mathrm{``nooffer"} \wedge c = cust)(x,cust).[cnt_c := cnt_c + 1]U \\
    & \; + \\
                 & \; \langle cnt_c = nh_l \rangle(\mathrm{``finish"})@(id = c).0
\end{align*}
}%

\paragraph{Hotel} The Hotel component has attributes \texttt{id},
\texttt{type}, \texttt{locality}, a list of \texttt{room}s and their
associated \texttt{price}, and a list of trusted brokers
\texttt{blist}. Its behaviour is the parallel composition
of two subprocesses for dealing with messages from the Broker and Customers components.
$$P_H \triangleq B_H \vert C_H$$
On receiving a message from the broker, process $B_H$ gets the
number of rooms available for the required date, i.e., $room_d$, and
the corresponding price of the rooms, taking into account the broker
identity $b$, i.e., $price^{b}_{d}$. $B_H$ also replicates itself for intercepting
 other messages while processing the current one using process $A$.
{\small
  \begin{align*}
    B_H \triangleq & \; (x = \mathrm{``acms"} \wedge b \in this.blist)(x,c,d,b).(A \vert B_H)\\[.4cm]
    A \triangleq & \; \langle room_d > 0 \rangle(\mathrm{``offer"},c,this.id,this.locality,this.price^{b}_d)@(id = b).0 \\
                  & + \\
                   & \; \langle room_d = 0 \rangle(\mathrm{``nooffer"},c)@(id = b).0
  \end{align*}
}%
$A$ can either reply to the inquiry broker a ``offer'' or ``nooffer'' message,
depending on the availability of the hotel rooms at day $d$. The ``offer''
message contains also relevant information such as hotel's id,
location and the room price. Moreover, the replies messages carry also
a reference of the original request, in our case it is the
customer id - $c$.

A hotel may receive booking messages directly from some
customers. This type of messages is handled by process
$C_H$. Specifically, $C_H$ reacts to a booking request by using
process $R$, in addition to spawn its own copy. A ``book'' message is processed
if it contains a valid broker identifier $b$, i.e.,
$b \in this.blist$, which may  be used by the hotel to
pay a commission. If there are still rooms available at day $d$, the
process sends a confirmation to the customer, atomically decrease the number of
rooms. In the other case it sends a ``toolate'' message to the sender of the request and terminates.
{\small
  \begin{align*}
    C_H \triangleq & \; (x = \mathrm{``book"} \wedge b \in this.blist)(x,c,d,p,b).(R \vert C_H)\\[.4cm]
    R \triangleq   & \; \langle room_d > 0\rangle(\mathrm{``confirm"})@(id = c).[room_d := room_d - 1]\\
                   &\; (\mathrm{``comission"}, p * 10\%)@(id = b).0 \\
                   &  \; + \\
                   & \; \langle room_d = 0\rangle(\mathrm{``toolate"})@(id = c).0
\end{align*}
}%
\modif{\paragraph{Discussion}
In our view, the above specification shows that the possibility of using primitives for comparing customer preferences with the features offered by providers greatly simplifies system descriptions and give rise to compact and elegant programs. This also permits to natural handle dynamic changes of demands and offers. Moreover, the minimality of the language and its relatively simple operational semantics makes the specification amenable to formal verification.}

\subsection{Programming environment}

Basing the interaction on the values of run-time attributes is indeed a nice idea, but it needs to be supported by a middleware that provides efficient ways for distributing messages, checking attribute values, and updating them. A typical approach is to rely on a centralised broker that keeps track of all components, intercepts every message and forwards it to registered components. It is then the responsibility of each component to decide whether to receive or discard the message. This is the solution proposed in~\cite{Alrahman-Isola} where a Java implementation of \abc is provided, that however suffers of serious performance problems. Two additional implementations of \abc have thus been considered, which are built on the top of two well-established programming languages largely used for concurrent programming, namely $\mathit{Erlang}$ and Go, guaranteeing better scalability. The two implementations are called $\mathit{AErlang}$, for Attribute $\mathit{Erlang}$, and \goat, for Go with attributes.

$\mathit{AErlang}$~\cite{rocco} is a middleware enabling attribute-based communication among programs in $\mathit{Erlang}$~\cite{Erlang}, a concurrent functional programming language originally designed for building telecommunication systems and recently successfully adapted to broader contexts, such as large-scale distributed messaging platforms like Facebook and WhatsApp. $\mathit{AErlang}$  lifts $\mathit{Erlang}$'s send and receive communication primitives to attribute-based reasoning.  In $\mathit{Erlang}$, the send primitive requires an explicit destination address while in $\mathit{AErlang}$ processes are not aware of the presence and identity of each other, and communicate using predicates over attributes. $\mathit{AErlang}$ has two main components: (i) a process registry that keeps track of process details, such as the process identifier and the current status, and (ii) a message broker that undertakes the delivery of outgoing messages. The \emph{Process registry} is a generic server that accepts requests regarding process (un)registration and internal updates. It stores process identifiers and all the information used by the message broker to deliver messages. The \emph{Message broker} is responsible for delivering messages between processes. It is implemented as an $\mathit{Erlang}$ server process listening for interactions from attribute-based send. To address potential bottlenecks arising in the presence of a very large number of processes, the message broker can be set up to run in multiple parallel threads. Like the Java implementation for \abc presented in~\cite{Alrahman-Isola}, the message broker is still centralised, however, to avoid broadcasts, the broker has an attribute registry where components register their attribute values and the broker is now responsible for message filtering. Different distribution policies have been implemented that can be used by taking into account dynamicity of attributes and of predicates.

\goat\footnote{\goat codes and examples can be retrieved from \url{https://giulio-garbi.github.io/goat/}.} extends Go~\cite{godoc}, the language introduced by Google to handle massive computation clusters, and to make working in these environments more productive. Go has an intuitive and lightweight concurrency model with a well-understood semantics and extends the CSP model~\cite{hoare1978communicating} with channel mobility, like in $\pi$-calculus. It also supports buffered channels, to provide mailboxes \emph{\`a la} $\mathit{Erlang}$. The \emph{Attribute-based API} for Go offers the possibility of using the \abc primitives to program the interaction of CAS applications directly in Go. The actual implementation faithfully models the formal semantics of \abc and it is parametric with respect to the infrastructure that mediates interactions. The \goat API offers the possibility of  using three different distributed coordination infrastructures for message exchange, namely cluster, ring, and tree. For all three infrastructures, it has been proved that the message delivery ordering is the same as the one required by the original formal semantics of \abc~\cite{Goat18}. An Eclipse plugin permits programming in a high-level syntax, which can be analysed via formal methods by relying on the operational semantics of \abc. Once the code has been analysed, the \goat plugin will generate formally verifiable Go code. Examples available from \goat 's site permit to appreciate how intuitive it is to program a complex variant of the well-known problem of Stable Allocation in Content Delivery Network~\cite{akamai}. 

\modif{
\abel is a recently developed programming framework for \abc, implemented in $\mathit{Erlang}$, that offers a set of APIs with a direct correspondence to the operators used in the \abc syntax. This direct correspondence allows us to automatically translate \abc specifications into \abel terms for actual executions. More details about \abel can be found in \cite{AbelCoord19}. To provide evidence of such a direct correspondence, below we provide some \abel code snippets that are obtained from the \abc processes modelling the customer component of the travel booking scenario.

The specification of the process $F$

{\footnotesize
  \begin{align*}
    F \triangleq  &\;  \bgInstanceA{
    $\langle send \rangle()@(\mathrm{ff}).[\mathrm{day} := \mathsf{get\_day()}, \mathrm{price} := \mathsf{get\_price()},\mathrm{dist} := \mathsf{get\_dist()}, \mathrm{loc} := \mathsf{get\_loc()}]$}\\
                  & \; 
                  \bgInstanceB{
                  $(\mathrm{``acms"},this.id,this.\mathrm{loc},this.\mathrm{day},this.\mathrm{price})@(type = \mathrm{``Broker"}).[\mathrm{send := ff}]F$}
  \end{align*}
}%
\noindent
is translated into \abel code\footnote{\modif{The \abel source code for the complete scenario can be downloaded from \url{https://bitbucket.org/tiezzi/jlamp_survey_code/src/master/AbC/}.}} in terms of two function definitions that make use of the $\mathit{prefix}$ command, as shown in Fig.~\ref{lis:abc_code}.}
\begin{figure}[!t]
{\footnotesize
  \begin{align*}
\bgInstanceA{
{f(C,V)}} & \rightarrow \\
                    & \mathrm{G} = \{\mathrm{fun(L) \rightarrow att(send,L) == true \; end}\},\\
                    & \mathrm{M} =  \{\}, \\
                    & \mathrm{S} =  \{\mathrm{fun(L) \rightarrow false \; end}\}, \\
                    & \mathrm{U} = [\{\mathrm{day, fun(L) \rightarrow get\_day() \; end}\}, \{\mathrm{price, fun(L) \rightarrow get\_price() \; end, \ldots}\} ]\},\\
                    & \mathrm{Act_o} = \{\mathrm{G, M,S,U}\},\\
                    & 
                  \bgInstanceA{{prefix(C,V,\{$\mathrm{Act_o}$, fun(\_V) $\rightarrow$ \; f1(C,\_V) \; end\})}.} \\
   \bgInstanceB{
{f1(C,V)}} & \rightarrow \\
                    & \mathrm{M} =  \{\textrm`acms\textrm',\mathrm{fun(L) \rightarrow att(id,L) \; end, fun(L) \rightarrow att(loc,L) \; end}, \ldots \}, \\
                    & \mathrm{S = fun(L,R) \rightarrow att(type,R) == 'Broker' \; end}, \\
                    & \mathrm{U} = [\{\mathrm{send, fun(L) \rightarrow false \; end}\}]\},\\
                    & \mathrm{Act_o} = \{\mathrm{G, M,S,U}\},\\
                    & 
                    \bgInstanceB{{prefix(C,V,\{$\mathrm{Act_o}$, fun(\_V) $\rightarrow$ f(C,\_V) end\})}}.
  \end{align*}
}
\vspace*{-.7cm}
\caption{The custmer process $\mathit{F}$ of the travel booking scenario implemented in \abel.}
\label{lis:abc_code}
\end{figure}
\modif{
An \abc process definition is represented in \abel through one or more
function definitions. As required by the framework, two parameters,
namely the address of the hosting component C and a set of bound
variables V of the executing process are associated with the definitions.

In the code snippet, the definition of
$\mathtt{f}$ encodes the first part of the process $F$ (the grey area)
while that of $\mathtt{f1}$ corresponds to the remaining one (the blue
area). Both definitions have the same structure: first they model an output
action and then invoke the prefix command. The output action is
characterised by 4 elements: an awareness predicate $G$, a
message $M$, a sending predicate $S$ and an attribute update
$U$. Their implementation follows the specification of \abc actions
and respects the conventional style of \abel. Thus, a sending
predicate S is implemented as a (boolean-valued) function
parametrised with the local (L) and remote (R) environments. The
\texttt{prefix} command takes as input an action and a continuation
process that states how to proceed upon action termination. 
In the example, the continuation of $\mathtt{f}$ refers to
$\mathtt{f1}$, and implements the sequential composition of the two actions,
while the continuation of $\mathtt{f1}$ refers to $\mathtt{f}$, thus realising  
a recursive behaviour.
}

 \subsection{Verification techniques}

Some work has now started to verify properties of \abc programs. On the one hand, it is under investigation the use of the generic tools that have been designed for verifying properties of $\mathit{Erlang}$ and Go programs. On the other hand, tools are under development to prove directly properties of the \abc specifications. The second alternative is under consideration because in some cases the correspondence between the actual \abc specifications and the running programs may not be immediate, and the difference would reduce the effectiveness of the effort.

A novel approach to the analysis of concurrent systems modelled as \abc terms has been introduced in~\cite{AbCBrinksma17}. It relies on the UMC model checker, a tool based on modelling concurrent systems as communicating UML-like state machines~\cite{UMC11}. A structural translation from \abc specifications to the UMC internal format is used as the basis for program analysis. This permits identifying emerging properties of systems and unwanted behaviours. Indeed, we have used the tool outlined in~\cite{AbCBrinksma17} and presented in \cite{tan-thesis} to translate the \abc specification  of the travel agent scenario into a formalism which is accepted as input by UMC~\cite{UMC11}. To make our model concrete we initialised the scenario with two customers, one broker and three hotels. Using the  logics ACTL supported by UMC, we were able to specify and verify some orchestration and liveness properties, like:
\begin{itemize}
\item A customer that sends an ``acms'' message will eventually receives a ``finish'' message 
\end{itemize}
that is expressed by the ACTL formula
\begin{center}
\texttt{AG [send(Cust, ``acms'')] AF \{receive(Cust,``finish'')\}}
\end{center}
Also other properties like:
\begin{itemize}
\item A customer that sends a ``book'' message will eventually receives either a ``confirm'' or a ``toolate'' message
\item If a customer receives a ``toolate'' message, it will react by sending another ``acms'' message
\item If a customer receives a ``confirm'' message, then the broker will receive a ``commission'' message
\end{itemize}
have been verified and have indeed been used to interactively refine the actual specification of the case study.


\modif{
\subsection{Related work}
In this section, we report on related works concerning languages and calculi with primitives that either model multiparty interactions or provide interesting ways to establish interactions.


Many calculi have been proposed to provide tools for specifying and reasoning about  communicating systems. Psi-calculus~\cite{psi} and its broadcast version~\cite{broadcastpsi} are the calculi closest to \abc. 
Psi-calculus is an extension of the $\pi$-calculus that aims at serving as a meta-theory for process calculi in general.  The environment/knowledge is encoded as a special process, named assertion, which influences the behaviour of the process within its scope. The evolution of knowledge is rather complex in Psi-calculus, while the clean separation between knowledge and behaviour in \abc avoids dependencies between components and enhances readability, compositional reasoning, and maintainability. 
Broadcast Psi~\cite{broadcastpsi} is an extension of Psi-calculus with broadcast primitives whose main communication rule requires that interacting agents take into account the knowledge of each other. Modelling reconfiguration  is not easy; it requires modelling different connectivity configurations using assertions and relies on the generation of assertions tagged by a fresh generation number; only the most recent generation is used. A generation becomes obsolete 
when composed with an assertion from later generation.  In \abc, components can choose to discard messages based on their run time status through the discard rules.

Other calculi, namely CBS~\cite{prasad1995calculus}, $b\pi$~\cite{ene2001broadcast},
CPC~\cite{given2010concurrent}, attribute $\pi$-calculus~\cite{john2010attributed}, imperative $\pi$-calculus~\cite{john2009dynamic},  Set-Pi~\cite{bruni} and  Broadcast Quality Calculus of~\cite{Vigo2013}, share similarities with \abc. We refer the interested reader to \cite{AlrahmanNL19} for a more detailed account of their relationships with \abc. Indeed, \abc combines the lessons learnt  from the above mentioned calculi, in the sense that \abc strives for expressivity while aiming at minimality and simplicity. The dynamic settings of attributes and the possibility of inspecting/modifying the environment give \abc greater flexibility and expressivity while keeping models as much natural as possible.

}

\section{Concluding remarks}
\label{sec:conclusion}

This paper surveyed four domain-specific coordination languages supporting the engineering of different classes of modern distributed systems. These languages have been developed in the last twenty years by the authors (three of which have been working for quite a while in the Concurrency and Mobility Group at University of  Florence) and other collaborators. Within the coordination community other research groups have followed a similar methodology, however relying on different specification models, \eg coalgebras \cite{ArbabR02}, actors \cite{Rebeca2004} or automata \cite{BaierSAR06}, rather than process algebras.

Below, we summarise the programming abstractions introduced with the different formalisms, by also highlighting their main differences, and the lessons learned when designing and using languages for
\begin{enumerate}
\item Network-Aware Programming,
\item Service-Oriented Computing,
\item Autonomic Computing,
\item Collective Adaptive Systems Programming.
\end{enumerate}

The design of \klaim\ has shown that network awareness in distributed systems can be achieved by the explicit use of localities as first-order citizens of the language. Localities, indeed, identify network nodes, where computation takes place and data is stored. Network awareness relies on the notion of (multiple) tuple spaces, which can be accessed via a unique interface to insert and retrieve data. Communication is thus asynchronous, anonymous and associative, pattern matching plays a crucial role and guarantees high expressive power. Network awareness paves the way for different kinds of optimisations that take advantage of code mobility, which in \klaim\ can rely on both static and dynamic scoping disciplines for the interpretation of the locality variables occurring within the mobile code.

%
%
%

\modif{
From \cows, we learnt that SOC typically abstracts from the structure of the underlying network and from data distribution, both of which become transparent to the programmer. Therefore, differently from \klaim, \cows\ does not provide code mobility. 
Although the \cows\ interaction model has some commonalities with \klaim's one (both are asynchronous and based on pattern-matching) some of their features are significantly different.
Indeed, \cows interactions are triggered by invocations of services along with communication endpoints and use pattern-matching for supporting message correlation. Besides, in SOC a service invocation results either in delivering the message to the corresponding instance or in 
creating
a new instance, but only when no proper instance exists for handling the message. This means that at run time, generally, concurrent instances of the same service can share (part of) their state. The implementation of these mechanisms in \cows relies on the combined use of suitable binder operators and non-standard receive activities that, differently from the \klaim ones, bind neither names nor variables and exploit pattern-matching to enforce priority among concurrent activities. 
Two additional distinguishing features of SOC are service persistence, and service fault and termination handling. \cows uses the standard process replication operator for modelling the former, and for modelling the latter relies on the combination of some ingenious constructs to either enforce termination or protect activities in case of termination of concurrent activities .
}

\modif{
\SCEL\ was designed by leveraging the experience gained with \klaim and completely differs from \cows. However, while the syntax of \SCEL\ processes may resemble that of \klaim\ ones, their semantics rely on very different mechanisms. In \klaim, the actions for data management are tagged with the locality where they will take place. In \SCEL, instead, they are tagged with a predicate over the attributes in the interface of the components that specifies the ensemble of all those components where the action will take place. Thus, process communication in \klaim\ (like in \cows) is always point-to-point, while in \SCEL\ it can also be group-oriented. This feature can be exploited for the formation of components ensembles, which are dynamically created opportunistically and transparently. 
Moreover, \SCEL components are equipped with knowledge repositories that generalise \klaim 's tuple spaces by supporting different knowledge representations and handling mechanisms. Self- and context-awareness make these components capable to adapt their behaviour to the evolving needs and the environmental changes.
Differently from \klaim, \SCEL does not have constructs for process migration (although a form of mobility can be realised through higher-order communication) but it permits to define policies for regulating processes behaviour which, although they are a parameter of the language, can be fully integrated with its operational semantics. 
}

\modif{
Finally, \abc is strongly inspired by \SCEL. It refines the group-oriented communication model of \SCEL  to convey in a distilled form the attribute-based communication paradigm to model, program and verify properties of Collective Adaptive Systems. The result of this synthesis effort is a compact calculus, suitable for studying the theoretical impact of the novel communication paradigm and for building new programming frameworks on the top of well-established programming languages, such as Java, Erlang and Go. Compared with \SCEL, the knowledge representation in \abc is abstract and is not designed for detailed reasoning during system evolution. This reflects the different objectives of \SCEL and \abc; while \SCEL focuses on programming issues, \abc concentrates on devising a minimal set of primitives to study the effectiveness of attribute-based communication.
}

To recap, we think that the engineering methodology we presented, as witnessed by the four instantiations we have illustrated, provides a uniform linguistic approach, based on formal methods techniques, for ensuring the trustworthiness of the considered classes of systems and possibly of the other ones that may emerge in the near future. In this respect, we plan to consider the Aggregate Programming \cite{aggregate} domain, where the abstraction level in designing distributed systems further increases. In such an engineering approach, data and devices are aggregated via `under-the-hood' coordination mechanisms. Although these aggregations resemble the notions of ensemble and collectives discussed in this paper, they mainly focus on distributed computation rather than on communication mechanisms.


\paragraph{\textbf{Acknowledgements}} This work would not have been possible without the contribution of our collaborators that have helped us in shaping the four languages we have introduced. They are too many to be listed, but their names could be inferred from the bibliography below. However, we would like to make an exception and explicitly thank Michele Loreti. Michele has been a driving force for most of the results we have presented, he is not among the authors only because he is one of the PC chairs of the conference to which the work was submitted. The research has been partially supported by the MIUR project PRIN 2017FTXR7S ``IT-MaTTerS'' (Methods and Tools for Trustworthy Smart Systems), and by the MIUR project PRIN 2017TWRCNB ``SEDUCE'' (Designing Spatially Distributed Cyber-Physical Systems under Uncertainty).

\end{document}